\documentclass[12pt]{article}
\usepackage{amssymb}
\usepackage{amsmath}

\allowdisplaybreaks

\begin{document}

\author{C. Bizdadea\thanks{%
e-mail address: bizdadea@central.ucv.ro}, E. M. Cioroianu\thanks{%
e-mail address: manache@central.ucv.ro}, D. Cornea\thanks{%
e-mail address: dcornea@central.ucv.ro}, \and E. Diaconu\thanks{%
e-mail address: ediaconu@central.ucv.ro}, S. O. Saliu\thanks{%
e-mail address: osaliu@central.ucv.ro}, S. C. Sararu\thanks{%
e-mail address: scsararu@central.ucv.ro} \\
Faculty of Physics, University of Craiova,\\
13 Al. I. Cuza Str., Craiova 200585, Romania}
\title{Interactions for a collection of spin-two fields intermediated by a
massless $p$-form}
\maketitle

\begin{abstract}
Under the general hypotheses of locality, smoothness of interactions in the
coupling constant, Poincar\'{e} invariance, Lorentz covariance, and
preservation of the number of derivatives on each field, we investigate the
cross-couplings of one or several spin-two fields to a massless $p$-form.
Two complementary cases arise. The first case is related to the standard
interactions from General Relativity, but the second case describes a new,
special type of couplings in $D=p+2$ spacetime dimensions, which break the
PT-invariance. Nevertheless, no consistent, indirect cross-interactions
among different gravitons with a positively defined metric in internal space
can be constructed.

\end{abstract}

\section{Introduction}

Theories involving one or several spin-two fields have raised a constant
interest over the last thirty years, especially at the level of direct or
intermediated graviton interactions~\cite%
{cutwald1,wald2,ovrutwald,ancoann,gupta54,kraich55,wein65,deser70,boul75,fangfr,berburgdam,wald86,hatffeyn,multi,noijhepdirac,noiRS,boulcqg,boulcqgexotic}%
. In this context more results on the impossibility of consistent
cross-couplings among different gravitons have been obtained, either without
other fields~\cite{multi} or in the presence of a scalar field~\cite{multi},
a Dirac spinor~\cite{noijhepdirac}, or respectively of a massive
Rarita-Schwinger field~\cite{noiRS}. All these no-go results have been
deduced under some specific hypotheses, always including the preservation of
the derivative order of each field equation with respect to its free limit
(derivative order assumption). Through their implications, these findings
support the common belief that the only consistent interactions in graviton
theories require a single spin-two field and are subject to the standard
prescriptions of General Relativity (meaning diffeomorphisms for the gauge
transformations of the graviton and diffeomorphism algebra for the gauge
algebra of the interacting theory). This idea is also strengthened by the
confirmation of the uniqueness of Einstein-Hilbert action~\cite{multi}
having the Pauli-Fierz model as its free limit or the uniqueness of $N=1$, $%
D=4$ SUGRA action~\cite{boulcqg} allowing for a Pauli-Fierz field and a
massless Rarita-Schwinger spinor as the corresponding uncoupled limit.
Indirect arguments are thus presented in favour of ruling out $N>8$ extended
supergravity theories since they require more than one spin-two field. It is
nevertheless known that the relaxation of the derivative order condition may
lead to exotic couplings for one or a collection of spin-two fields~\cite%
{boulcqgexotic}, which are no longer mastered by General Relativity.

Our paper submits to the same topic, of constructing spin-two field(s)
couplings, initially in the presence of a massless vector field and then of
a $p$-form, with $p>1$. We employ a systematic approach to the construction
of interactions in gauge theories~\cite{def,def1,def2}, based on the
cohomological reformulation of Lagrangian BRST symmetry~\cite%
{batvilk1,batvilk2,batvilk3,henproc,hencarte}. In this approach interactions
result from the analysis of consistent deformations of the generator of the
Lagrangian BRST symmetry (known as the solution of the master equation) by
means of specific cohomological techniques, relying on local BRST cohomology~%
\cite{gen1,gen11}. The emerging deformations, and hence also the
interactions, are constructed under the general hypotheses of locality,
smoothness in the coupling constant, Poincar\'{e} invariance, Lorentz
covariance, and derivative order assumption. In this specific situation the
derivative order assumption requires that the interaction vertices contain
at most two spacetime derivatives of the fields, but does not restrict the
polynomial order in the undifferentiated fields either in the Lagrangian or
in the gauge symmetries. Our analysis envisages three steps, which introduce
gradually the situations under investigation, according to the complexity of
their cohomological content.

Initially, we consider the case of couplings between a single
Pauli-Fierz field~\cite{pf,pf1} and a massless vector field. In this
setting we compute the coupling terms to order two in the coupling
constant $k$ and find two distinct solutions. The first solution
leads to the full cross-coupling Lagrangian in all $D>2$
\begin{eqnarray*}
\mathcal{L}_{\mathrm{I}}^{(\mathrm{int})} &=&-\frac{1}{4}\sqrt{-g}g^{\mu \nu
}g^{\rho \lambda }\bar{F}_{\mu \rho }\bar{F}_{\nu \lambda }+k\left(
q_{1}\delta _{3}^{D}\varepsilon ^{\mu _{1}\mu _{2}\mu _{3}}\bar{V}_{\mu _{1}}%
\bar{F}_{\mu _{2}\mu _{3}}\right. \\
&&\left. +q_{2}\delta _{5}^{D}\varepsilon ^{\mu _{1}\mu _{2}\mu _{3}\mu
_{4}\mu _{5}}\bar{V}_{\mu _{1}}\bar{F}_{\mu _{2}\mu _{3}}\bar{F}_{\mu
_{4}\mu _{5}}\right) ,
\end{eqnarray*}%
which respects the standard rules of General Relativity. The second
solution is more unusual: it `lives' only in $D=3$, produces
polynomials of order two in the coupling constant (and not series,
like in the first case), and the the couplings are mixing-component
terms that can be written in terms of a deformed field strength (of
the massless vector-field) as
\begin{equation*}
\mathcal{L}_{\mathrm{II}}^{(\mathrm{int})}=-\frac{1}{4}F^{\prime \mu \nu
}F_{\mu \nu }^{\prime },\qquad F_{\mu \nu }^{\prime }=F_{\mu \nu
}+2k\varepsilon _{\mu \nu \rho }\partial ^{\lbrack \theta }h_{\ \ \theta
}^{\rho ]}.
\end{equation*}%
By contrast to General Relativity, where all the gauge symmetries
are deformed, here only those of the vector field are modified by
terms of order one in the coupling constant that involve the
Pauli-Fierz gauge parameters, while the spin-two field keeps its
original gauge symmetries, namely the linearized version of
diffeomorphisms. To our knowledge, \emph{this is the first situation
where the linearized version of the spin-two field allows for
non-trivial couplings, other than those subject to General
Relativity, which fulfill all the working hypotheses, including that
on the derivative order}.

Next, we focus on the investigation of cross-interactions among different
gravitons intermediated by a massless vector field. In view of this, we
start from a finite sum of Pauli-Fierz actions with a positively defined
metric in internal space and a massless vector field. The cohomological
analysis reveals again two cases. The former is related to the standard
graviton-vector field interactions from General Relativity and exhibits no
consistent cross-interactions among different gravitons (with a positively
defined metric in internal space) in the presence of a massless vector
field. At most one graviton can be coupled to the vector field via a
Lagrangian similar to $\mathcal{L}_{\mathrm{I}}^{(\mathrm{int})}$, while
each of the other spin-two fields may interact only with itself through an
Einstein-Hilbert action with a cosmological term. The latter case seems to
describe some new type of couplings in $D=3$, which appear to allow for
cross-couplings among different gravitons. The coupled Lagrangian is, like
in the case of a single graviton, a polynomial of order two in the coupling
constant, obtained by deforming the vector field strength
\begin{equation*}
\mathcal{\hat{L}}_{\mathrm{II}}^{(\mathrm{int})}=-\frac{1}{4}\hat{F}^{\mu
\nu }\hat{F}_{\mu \nu },\qquad \hat{F}^{\mu \nu }=F^{\mu \nu }+2k\varepsilon
^{\mu \nu \rho }\sum_{A=1}^{n}\left( y_{3}^{A}\partial _{\lbrack \theta
}h_{\rho ]}^{A\ \theta }\right) ,
\end{equation*}%
where $y_{3}^{A}$ are some arbitrary, nonvanishing real constants.
Nevertheless, these cross-couplings can be decoupled through an orthogonal,
linear transformation of the spin-two fields, in terms of which $\mathcal{%
\hat{L}}_{\mathrm{II}}^{(\mathrm{int})}$ becomes nothing but $\mathcal{L}_{%
\mathrm{II}}^{(\mathrm{int})}$, with $h_{\mu \nu }$ replaced for instance by
the first transformed spin-two field from the collection. In consequence,
these case also leads to no indirect cross-couplings between different
gravitons.

Then, we show that all the new results obtained in the case a massless
vector field can be generalized to an arbitrary $p$-form. More precisely, if
one starts from a free action describing an Abelian $p$-form and a single
Pauli-Fierz field, then \emph{it is possible to construct some new
deformations in }$D=p+2$\emph{\ that are consistent to all orders in the
coupling constant and are not subject to the rules of General Relativity}.
It is important to remark that all the working hypotheses, including the
derivative order assumption, are fulfilled. There are several physical
consequences of these couplings, such as the appearance of a constant
linearized scalar curvature if one allows for a cosmological term or the
modification of the initial $\left( p+1\right) $-order conservation law for
the $p$-form by terms containing the spin-two field. Regarding a collection
of spin-two fields, we find that the deformed Lagrangian does not allow for
cross-couplings between different gravitons intermediated by a $p$-form,
either in the setting of General Relativity or in the special, $\left(
p+2\right) $-dimensional situation.

This paper is organized in seven sections. In Section \ref{free} we
construct the BRST symmetry of a free model with a single
Pauli-Fierz field and one massless vector field. Section \ref{brief}
briefly addresses the deformation procedure based on the BRST
symmetry. In Sections \ref{spintwoem} and \ref{manyspintwoem} we
compute the deformations corresponding to a vector field and one or
respectively several spin-two fields, and emphasize the
Lagrangian formulation of the resulting theories. Section \ref%
{comm} discusses the generalization of the previous results to the
case of couplings between one or several gravitons and an arbitrary
$p$-form gauge field. Section \ref{conc} ends the paper with the
main conclusions.

\section{BRST symmetry of the free model\label{free}}

Our starting point is represented by a free Lagrangian action, written as
the sum between the linearized Hilbert-Einstein action (also known as the
Pauli-Fierz action) and Maxwell's action in $D>2$ spacetime dimensions
\begin{eqnarray}
S_{0}^{\mathrm{L}}[h_{\mu \nu },V_{\mu }] &=&\int d^{D}x\left[ -\frac{1}{2}%
\left( \partial _{\mu }h_{\nu \rho }\right) \partial ^{\mu }h^{\nu \rho
}+\left( \partial _{\mu }h^{\mu \rho }\right) \partial ^{\nu }h_{\nu \rho
}\right.  \notag \\
&&\left. -\left( \partial _{\mu }h\right) \partial _{\nu }h^{\nu \mu }+\frac{%
1}{2}\left( \partial _{\mu }h\right) \partial ^{\mu }h-\frac{1}{4}F_{\mu \nu
}F^{\mu \nu }\right]  \label{a1} \\
&\equiv &\int d^{D}x\left( \mathcal{L}_{0}^{\left( \mathrm{PF}\right) }+%
\mathcal{L}_{0}^{\left( \mathrm{vect}\right) }\right) .  \notag
\end{eqnarray}%
The restriction $D>2$ is required by the spin-two field action, which is
known to reduce to a total derivative in $D=2$. Throughout the paper we work
with the flat metric of `mostly plus' signature, $\sigma _{\mu \nu }=\left(
-+\ldots +\right) $. In the above $h$ denotes the trace of the Pauli-Fierz
field, $h=\sigma _{\mu \nu }h^{\mu \nu }$, and $F_{\mu \nu }$ represents the
Abelian field-strength of the massless vector field ($F_{\mu \nu }\equiv
\partial _{\lbrack \mu }V_{\nu ]}$). The theory described by action (\ref{a1}%
) possesses an Abelian and irreducible generating set of gauge
transformations
\begin{equation}
\delta _{\epsilon }h_{\mu \nu }=\partial _{(\mu }\epsilon _{\nu )},\qquad
\delta _{\epsilon }V_{\mu }=\partial _{\mu }\epsilon ,  \label{a2}
\end{equation}%
with $\epsilon _{\mu }$ and $\epsilon $ bosonic gauge parameters. The
notation $\left[ \mu \ldots \nu \right] $ (or $\left( \mu \ldots \nu \right)
$) signifies antisymmetry (or symmetry) with respect to all indices between
brackets without normalization factors (i.e., the independent terms appear
only once and are not multiplied by overall numerical factors).

In order to construct the BRST symmetry for action (\ref{a1}), it is
necessary to introduce the field/ghost and antifield spectra%
\begin{eqnarray}
\Phi ^{\alpha _{0}} &=&(h_{\mu \nu },V_{\mu }),\qquad \Phi _{\alpha
_{0}}^{\ast }=(h^{\ast \mu \nu },V^{\ast \mu }),  \label{a4a} \\
\eta _{\alpha _{1}} &=&(\eta _{\mu },\eta ),\qquad \eta ^{\ast \alpha
_{1}}=(\eta ^{\ast \mu },\eta ^{\ast }).  \label{a4b}
\end{eqnarray}%
The fermionic ghosts $\eta _{\alpha _{1}}$ are associated with the gauge
parameters $\epsilon _{\alpha _{1}}=\left\{ \epsilon _{\mu },\epsilon
\right\} $ respectively and the star variables represent the antifields of
the corresponding fields/ghosts. (According to the standard rule of the BRST
method, the Grassmann parity of a given antifield is opposite to that of the
corresponding field/ghost.) Since the gauge generators are field-independent
and irreducible, it follows that the BRST differential decomposes into
\begin{equation}
s=\delta +\gamma ,  \label{a3}
\end{equation}%
where $\delta $ is the Koszul-Tate differential and $\gamma $ denotes the
exterior longitudinal derivative. The Koszul-Tate differential is graded in
terms of the antighost number ($\mathrm{agh}$, $\mathrm{agh}\left( \delta
\right) =-1$, $\mathrm{agh}\left( \gamma \right) =0$) and enforces a
resolution of the algebra of smooth functions defined on the stationary
surface of field equations for action (\ref{a1}), $C^{\infty }\left( \Sigma
\right) $, $\Sigma :\delta S_{0}^{\mathrm{L}}/\delta \Phi ^{\alpha _{0}}=0$.
The exterior longitudinal derivative is graded in terms of the pure ghost
number ($\mathrm{pgh}$, $\mathrm{pgh}\left( \gamma \right) =1$, $\mathrm{pgh}%
\left( \delta \right) =0$) and is correlated with the original gauge
symmetry via its cohomology in pure ghost number zero computed in $C^{\infty
}\left( \Sigma \right) $, which is isomorphic to the algebra of physical
observables for this free theory. These two degrees of the BRST generators
are valued as
\begin{eqnarray}
\mathrm{agh}(\Phi ^{\alpha _{0}}) &=&\mathrm{agh}(\eta _{\alpha
_{1}})=0,\qquad \mathrm{agh}(\Phi _{\alpha _{0}}^{\ast })=1,\qquad \mathrm{%
agh}(\eta ^{\ast \alpha _{1}})=2,  \label{a5} \\
\mathrm{pgh}(\Phi ^{\alpha _{0}}) &=&0,\qquad \mathrm{pgh}(\eta _{\alpha
_{1}})=1,\qquad \mathrm{pgh}(\Phi _{\alpha _{0}}^{\ast })=\mathrm{pgh}(\eta
^{\ast \alpha _{1}})=0.  \label{a6}
\end{eqnarray}%
The overall degree that grades the BRST complex is named ghost number ($%
\mathrm{gh}$) and is defined like the difference between the pure ghost
number and the antighost number, such that $\mathrm{gh}\left( s\right) =%
\mathrm{gh}\left( \delta \right) =\mathrm{gh}\left( \gamma \right) =1$. The
actions of the operators $\delta $ and $\gamma $ (taken to act as right
differentials) on the BRST generators read as
\begin{eqnarray}
\delta h^{\ast \mu \nu } &=&2H^{\mu \nu },\qquad \delta V^{\ast \mu
}=-\partial _{\nu }F^{\nu \mu },  \label{a7} \\
\delta \eta ^{\ast \mu } &=&-2\partial _{\nu }h^{\ast \nu \mu },\qquad
\delta \eta ^{\ast }=-\partial _{\mu }V^{\ast \mu },  \label{a8} \\
\delta \Phi ^{\alpha _{0}} &=&0,\qquad \delta \eta _{\alpha _{1}}=0,
\label{a9} \\
\gamma \Phi _{\alpha _{0}}^{\ast } &=&0,\qquad \gamma \eta ^{\ast \alpha
_{1}}=0,  \label{a10} \\
\gamma h_{\mu \nu } &=&\partial _{(\mu }\eta _{\nu )},\qquad \gamma V_{\mu
}=\partial _{\mu }\eta ,  \label{a11} \\
\gamma \eta _{\mu } &=&0,\qquad \gamma \eta =0.  \label{a12}
\end{eqnarray}%
In the above $H^{\mu \nu }$ is the linearized Einstein tensor
\begin{equation}
H^{\mu \nu }=K^{\mu \nu }-\frac{1}{2}\sigma ^{\mu \nu }K,  \label{a13}
\end{equation}%
with $K^{\mu \nu }$ and $K$ the linearized Ricci tensor and the linearized
scalar curvature respectively, both obtained from the linearized Riemann
tensor%
\begin{equation}
K_{\mu \nu |\alpha \beta }=-\frac{1}{2}(\partial _{\mu }\partial _{\alpha
}h_{\nu \beta }+\partial _{\nu }\partial _{\beta }h_{\mu \alpha }-\partial
_{\nu }\partial _{\alpha }h_{\mu \beta }-\partial _{\mu }\partial _{\beta
}h_{\nu \alpha }),  \label{a14}
\end{equation}%
from its trace and double trace respectively
\begin{equation}
K_{\mu \alpha }=\sigma ^{\nu \beta }K_{\mu \nu |\alpha \beta },\qquad
K=\sigma ^{\mu \alpha }\sigma ^{\nu \beta }K_{\mu \nu |\alpha \beta }.
\label{a15}
\end{equation}%
The BRST differential is known to have a canonical action in a structure
named antibracket and denoted by the symbol $\left( ,\right) $ ($s\cdot
=\left( \cdot ,\bar{S}\right) $), which is obtained by considering the
fields/ghosts conjugated respectively to the corresponding antifields. The
generator of the BRST symmetry is a bosonic functional of ghost number zero,
which is solution to the classical master equation $\left( \bar{S},\bar{S}%
\right) =0$. The full solution to the master equation for the free model
under study reads as
\begin{equation}
\bar{S}=S_{0}^{\mathrm{L}}[h_{\mu \nu },V_{\mu }]+\int d^{D}x\left( h^{\ast
\mu \nu }\partial _{(\mu }\eta _{\nu )}+V^{\ast \mu }\partial _{\mu }\eta
\right)  \label{a16}
\end{equation}%
and encodes all the information on the gauge structure of the theory (\ref%
{a1})--(\ref{a2}).

\section{Brief review of the deformation procedure\label{brief}}

We begin with a \textquotedblleft free\textquotedblright\ gauge theory,
described by a Lagrangian action $S_{0}^{\mathrm{L}}\left[ \Phi ^{\alpha
_{0}}\right] $, invariant under some gauge transformations $\delta
_{\epsilon }\Phi ^{\alpha _{0}}=\bar{Z}_{\;\;\alpha _{1}}^{\alpha
_{0}}\epsilon ^{\alpha _{1}}$, i.e. $\frac{\delta S_{0}^{\mathrm{L}}}{\delta
\Phi ^{\alpha _{0}}}\bar{Z}_{\;\;\alpha _{1}}^{\alpha _{0}}=0$, and consider
the problem of constructing consistent interactions among the fields $\Phi
^{\alpha _{0}}$ such that the couplings preserve the field spectrum and the
original number of gauge symmetries. This matter is addressed by means of
reformulating the problem of constructing consistent interactions as a
deformation problem of the solution to the master equation corresponding to
the \textquotedblleft free\textquotedblright\ theory~\cite{def,def1,def2}.
Such a reformulation is possible due to the fact that the solution to the
master equation contains all the information on the gauge structure of the
theory. If an interacting gauge theory can be consistently constructed, then
the solution $\bar{S}$ to the master equation associated with the
\textquotedblleft free\textquotedblright\ theory, $\left( \bar{S},\bar{S}%
\right) =0$, can be deformed into a solution $S$%
\begin{equation}
\bar{S}\rightarrow S=\bar{S}+kS_{1}+k^{2}S_{2}+\cdots =\bar{S}+k\int
d^{D}x\,a+k^{2}\int d^{D}x\,b+\cdots  \label{a17}
\end{equation}%
of the master equation for the deformed theory
\begin{equation}
\left( S,S\right) =0,  \label{a18}
\end{equation}%
such that both the ghost and antifield spectra of the initial theory are
preserved. The projection of equation (\ref{a18}) on the various orders in
the coupling constant $k$ leads to the equivalent tower of equations%
\begin{eqnarray}
\left( \bar{S},\bar{S}\right) &=&0,  \label{a19} \\
2\left( S_{1},\bar{S}\right) &=&0,  \label{a20} \\
2\left( S_{2},\bar{S}\right) +\left( S_{1},S_{1}\right) &=&0,  \label{a21} \\
&&\vdots  \notag
\end{eqnarray}%
Equation (\ref{a19}) is fulfilled by hypothesis. The next equation requires
that the first-order deformation of the solution to the master equation, $%
S_{1}$, is a co-cycle of the \textquotedblleft free\textquotedblright\ BRST
differential $s$, $sS_{1}=0$. However, only cohomologically nontrivial
solutions to (\ref{a20}) should be taken into account, since the BRST-exact
ones can be eliminated by some (in general nonlinear) field redefinitions.
This means that $S_{1}$ pertains to the ghost number zero cohomological
space of $s$, $H^{0}\left( s\right) $, which is nonempty because it is
isomorphic to the space of physical observables of the \textquotedblleft
free\textquotedblright\ theory. It has been shown (by of the triviality of
the antibracket map in the cohomology of the BRST differential) that there
are no obstructions in finding solutions to the remaining equations, namely (%
\ref{a21}), etc. However, the resulting interactions may be nonlocal and
there might even appear obstructions if one insists on their locality. The
analysis of these obstructions can be done with the help of cohomological
techniques.

\section{Consistent interactions between the spin-two field and a massless
vector field\label{spintwoem}}

\subsection{Standard material: basic cohomologies\label{stand}}

The aim of this section is to investigate the cross-couplings that
can be introduced between the spin-two field and a massless vector
field. This matter is addressed in the context of the antifield-BRST
deformation procedure described in the above and relies on computing
the solutions to equations (\ref{a20})--(\ref{a21}), etc., with the
help of the BRST cohomology of the free theory. The deformations are
obtained under the following (reasonable) assumptions: smoothness in
the deformation parameter, locality, Lorentz covariance,
Poincar\'{e} invariance, and the presence of at most two derivatives
in the coupled Lagrangian. `Smoothness in the deformation parameter'
refers to the fact that the deformed solution to the master
equation, (\ref{a17}), is smooth in the coupling constant $k$ and
reduces to the original solution, (\ref{a16}), in the free limit
$k=0$. The hypothesis on the deformed theory to be Poincar\'{e}
invariant means that one does not allow an explicit dependence on
the spacetime coordinates into the deformed solution to the master
equation. The requirement concerning the maximum number of
derivatives allowed to enter the deformed Lagrangian is frequently
imposed in the literature; for instance, see the case of couplings
between the Pauli-Fierz and the massless Rarita-Schwinger
fields~\cite{boulcqg} or of cross-interactions for
a collection of Pauli-Fierz fields~\cite{multi}. If we make the notation $%
S_{1}=\int d^{D}x\,a$, then equation (\ref{a20}), which controls the
first-order deformation, takes the local form
\begin{equation}
sa=\partial _{\mu }m^{\mu },\qquad \mathrm{gh}\left( a\right) =0,\qquad
\varepsilon \left( a\right) =0,  \label{a23}
\end{equation}%
for some local current $m^{\mu }$. It shows that the nonintegrated density
of the first-order deformation pertains to the local cohomology of the free
BRST\ differential in ghost number zero, $a\in H^{0}\left( s|d\right) $,
where $d$ denotes the exterior spacetime differential. The solution to (\ref%
{a23}) is unique up to $s$-exact pieces plus divergences
\begin{equation}
a\rightarrow a+sb+\partial _{\mu }n^{\mu },  \label{a24}
\end{equation}%
with $\mathrm{gh}\left( b\right) =-1$, $\varepsilon \left( b\right) =1$, $%
\mathrm{gh}\left( n^{\mu }\right) =0$, and $\varepsilon \left( n^{\mu
}\right) =0$. At the same time, if the general solution of (\ref{a23}) is
found to be completely trivial, $a=sb+\partial _{\mu }n^{\mu }$, then it can
be made to vanish, $a=0$.

In order to analyze equation (\ref{a23}) we develop $a$ according to the
antighost number
\begin{equation}
a=\sum\limits_{i=0}^{I}a_{i},\qquad \mathrm{agh}\left( a_{i}\right)
=i,\qquad \mathrm{gh}\left( a_{i}\right) =0,\qquad \varepsilon \left(
a_{i}\right) =0,  \label{a25}
\end{equation}%
and assume, without loss of generality, that decomposition (\ref{a25}) stops
at some finite value of $I$. This can be shown for instance like in Appendix
A of~\cite{multi}. Replacing decomposition (\ref{a25}) into (\ref{a23}) and
projecting it on the various values of the antighost number by means of (\ref%
{a3}), we obtain that (\ref{a23}) is equivalent with the tower of equations
\begin{eqnarray}
\gamma a_{I} &=&\partial _{\mu }m_{I}^{\mu },  \label{a26} \\
\delta a_{I}+\gamma a_{I-1} &=&\partial _{\mu }m_{I-1}^{\mu },  \label{a27}
\\
\delta a_{i}+\gamma a_{i-1} &=&\partial _{\mu }m_{i-1}^{\mu },\qquad 1\leq
i\leq I-1,  \label{a28}
\end{eqnarray}%
where $\left( m_{i}^{\mu }\right) _{i=\overline{0,I}}$ are some local
currents, with $\mathrm{agh}\left( m_{i}^{\mu }\right) =i$. Moreover,
according to the general result from~\cite{multi} in the absence of
collection indices, equation (\ref{a26}) can be replaced in strictly
positive antighost numbers by
\begin{equation}
\gamma a_{I}=0,\qquad I>0.  \label{a29}
\end{equation}%
Due to the second-order nilpotency of $\gamma $ ($\gamma ^{2}=0$), the
solution to (\ref{a29}) is unique up to $\gamma $-exact contributions%
\begin{equation}
a_{I}\rightarrow a_{I}+\gamma b_{I},\qquad \mathrm{agh}\left( b_{I}\right)
=I,\qquad \mathrm{pgh}\left( b_{I}\right) =I-1,\qquad \varepsilon \left(
b_{I}\right) =1.  \label{a30}
\end{equation}%
Meanwhile, if it turns out that $a_{I}$ reduces to $\gamma $-exact terms, $%
a_{I}=\gamma b_{I}$, then it can be made to vanish, $a_{I}=0$. In other
words, the nontriviality of the first-order deformation $a$ is translated at
its highest antighost number component into the requirement that $a_{I}\in
H^{I}\left( \gamma \right) $, where $H^{I}\left( \gamma \right) $ denotes
the cohomology of the exterior longitudinal derivative $\gamma $ in pure
ghost number equal to $I$. So, in order to solve equation (\ref{a23})
(equivalent with (\ref{a29}) and (\ref{a27})--(\ref{a28})), we need to
compute the cohomology of $\gamma $, $H\left( \gamma \right) $, and, as it
will be made clear below, also the local cohomology of $\delta $, $H\left(
\delta |d\right) $.

Using the results on the cohomology of $\gamma $ in the Pauli-Fierz sector~%
\cite{multi} as well as definitions (\ref{a10})--(\ref{a12}), we can state
that $H\left( \gamma \right) $ is generated on the one hand by $\Phi
_{\alpha _{0}}^{\ast }$, $\eta ^{\ast \alpha _{1}}$, $F_{\mu \nu }$, and $%
K_{\mu \nu \alpha \beta }$, together with their spacetime derivatives and,
on the other hand, by the undifferentiated ghosts $\eta $ and $\eta _{\mu }$
as well as by their antisymmetric first-order derivatives $\partial _{[ \mu
}\eta _{\nu ]}$. (The spacetime derivatives of $\eta $ are $\gamma $-exact,
in agreement with the latter definition from (\ref{a11}), and the same is
valid for the derivatives of $\eta _{\mu }$ of order two and higher.) So,
the most general (and nontrivial) solution to (\ref{a29}) can be written, up
to $\gamma $-exact contributions, as
\begin{equation}
a_{I}=\alpha _{I}([F_{\mu \nu }],[K_{\mu \nu \rho \lambda }],[\Phi _{\alpha
_{0}}^{\ast }],[\eta ^{\ast \alpha _{1}}])e^{I}(\eta ,\eta _{\mu },\partial
_{[ \mu }\eta _{\nu ]}),  \label{a31}
\end{equation}%
where the notation $f\left( \left[ q\right] \right) $ means that $f$ depends
on $q$ and its derivatives up to a finite order, while $e^{I}$ denotes the
elements of a basis in the space of polynomials with pure ghost number $I$
in $\eta $, $\eta _{\mu }$, and $\partial _{[ \mu }\eta _{\nu ]}$. The
objects $\alpha _{I}$ (obviously nontrivial in $H^{0}\left( \gamma \right) $%
) were taken to have a finite antighost number and a bounded number of
derivatives, and therefore they are polynomials in the antifields, in the
linearized Riemann tensor $K_{\mu \nu \alpha \beta }$, and in the
field-strength $F_{\mu \nu }$ as well as in their subsequent derivatives.
They are required to fulfill the property $\mathrm{agh}\left( \alpha
_{I}\right) =I$ in order to ensure that the ghost number of $a_{I}$ is equal
to zero. Due to their $\gamma $-closeness, $\gamma \alpha _{I}=0$, and to
their polynomial character, $\alpha _{I}$ will be called invariant
polynomials. In antighost number zero the invariant polynomials are
polynomials in the linearized Riemann tensor, in the field-strength of the
Abelian field, and in their derivatives. The result that one can replace
equation (\ref{a26}) with (\ref{a29}) is a consequence of the triviality of
the cohomology of the exterior spacetime differential in the space of
invariant polynomials in strictly positive antighost numbers. For more
details, see subsection A.1 from~\cite{multi}.

Inserting (\ref{a31}) in (\ref{a27}), we obtain that a necessary (but not
sufficient) condition for the existence of (nontrivial) solutions $a_{I-1}$
is that the invariant polynomials $\alpha _{I}$ are (nontrivial) objects
from the local cohomology of the Koszul-Tate differential $H\left( \delta
|d\right) $ in antighost number $I>0$ and in pure ghost number zero
\begin{equation}
\delta \alpha _{I}=\partial _{\mu }j_{I-1}^{\mu },\qquad \mathrm{agh}\left(
j_{I-1}^{\mu }\right) =I-1,\qquad \mathrm{pgh}\left( j_{I-1}^{\mu }\right)
=0.  \label{a32}
\end{equation}%
We recall that the local cohomology $H\left( \delta |d\right) $ is
completely trivial in both strictly positive antighost \emph{and} pure ghost
numbers (for instance, see Theorem 5.4 from~\cite{gen1} and also~\cite{gen11}%
). Using the fact that the Cauchy order of the free theory under study is
equal to two, the general results from~\cite{gen1} and~\cite{gen11},
according to which the local cohomology of the Koszul-Tate differential in
pure ghost number zero is trivial in antighost numbers strictly greater than
its Cauchy order, ensure that
\begin{equation}
H_{J}\left( \delta |d\right) =0,\qquad J>2,  \label{a33}
\end{equation}%
where $H_{J}\left( \delta |d\right) $ denotes the local cohomology of the
Koszul-Tate differential in antighost number $J$ and in pure ghost number
zero. It can be shown that any invariant polynomial that is trivial in $%
H_{J}\left( \delta |d\right) $ with $J\geq 2$ can be taken to be trivial
also in $H_{J}^{\mathrm{inv}}\left( \delta |d\right) $. ($H_{J}^{\mathrm{inv}%
}\left( \delta |d\right) $ denotes the invariant characteristic cohomology
in antighost number $J$ --- the local cohomology of the Koszul-Tate
differential in the space of invariant polynomials.) Thus:%
\begin{equation}
\left( \alpha _{J}=\delta b_{J+1}+\partial _{\mu }c_{J}^{\mu }, \mathrm{agh}%
\left( \alpha _{J}\right) =J\geq 2\right) \Rightarrow \alpha _{J}=\delta
\beta _{J+1}+\partial _{\mu }\gamma _{J}^{\mu },  \label{a34}
\end{equation}%
with both $\beta _{J+1}$ and $\gamma _{J}^{\mu }$ invariant polynomials.
Results (\ref{a34}) and (\ref{a33}) yield the conclusion that the invariant
characteristic cohomology is trivial in antighost numbers strictly greater
than two%
\begin{equation}
H_{J}^{\mathrm{inv}}\left( \delta |d\right) =0,\qquad J>2.  \label{a35}
\end{equation}%
By proceeding in the same manner like in~\cite{multi} and~\cite{knaep1}, it
can be proved that the spaces $H_{2}\left( \delta |d\right) $ and $H_{2}^{%
\mathrm{inv}}\left( \delta |d\right) $ are spanned by
\begin{equation}
H_{2}\left( \delta |d\right) ,H_{2}^{\mathrm{inv}}\left( \delta |d\right)
:\left( \eta ^{\ast },\eta ^{\ast \mu }\right) .  \label{a36}
\end{equation}%
In contrast to the groups $\left( H_{J}\left( \delta |d\right) \right)
_{J\geq 2}$ and $\left( H_{J}^{\mathrm{inv}}\left( \delta |d\right) \right)
_{J\geq 2}$, which are finite-dimensional, the cohomology $H_{1}\left(
\delta |d\right) $ in pure ghost number zero, known to be related to global
symmetries and ordinary conservation laws, is infinite-dimensional since the
theory is free. Fortunately, it will not be needed in the sequel.

The previous results on $H\left( \delta |d\right) $ and $H^{\mathrm{inv}%
}\left( \delta |d\right) $ in strictly positive antighost numbers are
important because they control the obstructions of removing the antifields
from the first-order deformation. Based on formulas (\ref{a33})--(\ref{a35}%
), one can eliminate all the pieces of antighost number strictly greater
than two from the nonintegrated density of the first-order deformation by
adding only trivial terms. Consequently, one can take (without loss of
nontrivial objects) $I\leq 2$ into the decomposition (\ref{a25}). (The proof
of this statement can be realized like in subsection A.3 from~\cite{multi}.)
In addition, the last representative reads as in (\ref{a31}), where the
invariant polynomial is necessarily a nontrivial object from $H_{2}^{\mathrm{%
inv}}\left( \delta |d\right) $ if $I=2$ and from $H_{1}\left( \delta
|d\right) $ if $I=1$ respectively.

\subsection{\label{deformareaI}Computation of first-order deformations}

Assuming $I=2$, the nonintegrated density of the first-order deformation (%
\ref{a25}) becomes
\begin{equation}
a=a_{0}+a_{1}+a_{2}.  \label{a37}
\end{equation}%
We can further decompose $a$ in a natural manner as
\begin{equation}
a=a^{\left( \mathrm{PF}\right) }+a^{\left( \mathrm{int}\right) }+a^{\left(
\mathrm{vect}\right) },  \label{a38}
\end{equation}%
where $a^{\left( \mathrm{PF}\right) }$ contains only
fields/ghosts/antifields from the Pauli-Fierz sector, $a^{\left( \mathrm{int}%
\right) }$ mixes both fields, and $a^{\left( \mathrm{vect}\right) }$
involves only the vector field sector. The component $a^{\left(
\mathrm{PF}\right) }$ is completely known~\cite{multi} and satisfies
by itself an equation of the type (\ref{a23}). It admits a
decomposition similar to (\ref{a37})
\begin{equation}
a^{\left( \mathrm{PF}\right) }=a_{0}^{\left( \mathrm{PF}\right)
}+a_{1}^{\left( \mathrm{PF}\right) }+a_{2}^{\left( \mathrm{PF}\right) },
\label{a39}
\end{equation}%
where
\begin{eqnarray}
a_{2}^{\left( \mathrm{PF}\right) } &=&\frac{f}{2}\eta ^{\ast \mu }\eta ^{\nu
}\partial _{[ \mu }\eta _{\nu ]},  \label{a40} \\
a_{1}^{\left( \mathrm{PF}\right) } &=&fh^{\ast \mu \rho }\left( \left(
\partial _{\rho }\eta ^{\nu }\right) h_{\mu \nu }-\eta ^{\nu }\partial _{[
\mu }h_{\nu ]\rho }\right) ,  \label{a41}
\end{eqnarray}%
and $a_{0}^{\left( \mathrm{PF}\right) }$ is the cubic vertex of the
Einstein-Hilbert Lagrangian multiplied by a real constant $f$ plus a
cosmological term\footnote{%
The terms $a_{2}^{\left( \mathrm{PF}\right) }$ and $a_{1}^{\left( \mathrm{PF}%
\right) }$ given in (\ref{a40}) and (\ref{a41}) differ from those present in~%
\cite{multi} (in the absence of collection indices) by a $\gamma $-exact and
respectively a $\delta $-exact contribution. However, the difference between
our $a_{2}^{\left( \mathrm{PF}\right) }+$ $a_{1}^{\left( \mathrm{PF}\right)
} $ and that from~\cite{multi} is a $s$-exact modulo $d$ quantity. The
associated $a_{0}^{\left( \mathrm{PF}\right) }$ is nevertheless the same in
both formulations. As a consequence, $a^{\left( \mathrm{PF}\right) }$ and
the first-order deformation from~\cite{multi} belong to the same
cohomological class from $H^{0}\left( s|d\right) $.}%
\begin{equation}
a_{0}^{\left( \mathrm{PF}\right) }=fa_{0}^{\left( \mathrm{EH-cubic}\right)
}-2\Lambda h,  \label{a0pf}
\end{equation}%
with $\Lambda $ the cosmological constant. Due to the fact that $a^{\left(
\mathrm{int}\right) }$ and $a^{\left( \mathrm{vect}\right) }$ contain
different sorts of fields, it follows that they are subject to two separate
equations
\begin{eqnarray}
sa^{\left( \mathrm{vect}\right) } &=&\partial _{\mu }m^{\left( \mathrm{vect}%
\right) \mu },  \label{a43} \\
sa^{\left( \mathrm{int}\right) } &=&\partial _{\mu }m^{\left( \mathrm{int}%
\right) \mu },  \label{a42}
\end{eqnarray}%
for some local $m^{\mu }$'s. It is known (for instance,
see~\cite{knaep}) that the general solution to (\ref{a43}) reduces
to its component of
antighost number zero and reads as%
\begin{equation}
a^{\left( \mathrm{vect}\right) }=a_{0}^{\left( \mathrm{vect}\right)
}=\sum\limits_{j>0}q_{j}\delta _{2j+1}^{D}\varepsilon ^{\mu _{1}\mu _{2}\mu
_{3}\ldots \mu _{2j}\mu _{2j+1}}V_{\mu _{1}}F_{\mu _{2}\mu _{3}}\ldots
F_{\mu _{2j}\mu _{2j+1}},  \label{g1}
\end{equation}%
with $q_{j}$ some real constants. Selecting from (\ref{g1}) only the terms
with maximum two spacetime derivatives, we conclude that we must ask $%
q_{j}=0 $ for all $j>2$, so%
\begin{equation}
a^{\left( \mathrm{vect}\right) }=a_{0}^{\left( \mathrm{vect}\right)
}=q_{1}\delta _{3}^{D}\varepsilon ^{\mu \nu \lambda }V_{\mu }F_{\nu \lambda
}+q_{2}\delta _{5}^{D}\varepsilon ^{\mu \nu \lambda \alpha \beta }V_{\mu
}F_{\nu \lambda }F_{\alpha \beta }.  \label{rr2}
\end{equation}%
The notation $\delta _{m}^{D}$ signifies the Kronecker symbol. In the sequel
we analyze the general solution to equation (\ref{a42}).

Due to (\ref{a37}) we should consider that the general solution to (\ref%
{a42}) stops at antighost number two, $a^{\left( \mathrm{int}\right)
}=a_{0}^{\left( \mathrm{int}\right) }+a_{1}^{\left(
\mathrm{int}\right) }+a_{2}^{\left( \mathrm{int}\right) }$. Equation
(\ref{a42}) is equivalent to the fact that the components of
$a^{\left( \mathrm{int}\right) }$ are subject to equations
(\ref{a29}) and (\ref{a27})--(\ref{a28}) with $I=2$ and $a$ replaced
by $a^{\left( \mathrm{int}\right) }$. It can be shown that there
exist no such solutions ending at antighost number two. For the sake
of simplicity, we omit the proof of this result, which is mainly
based on showing that there is no nontrivial $a_{2}^{\left(
\mathrm{int}\right) }$ yielding a consistent $a_{0}^{\left(
\mathrm{int}\right) }$. In view of this finding, we approach the
next situation, where the solution to (\ref{a42})
stops at antighost number one%
\begin{equation}
a^{\left( \mathrm{int}\right) }=a_{0}^{\left( \mathrm{int}\right)
}+a_{1}^{\left( \mathrm{int}\right) },  \label{a52a}
\end{equation}%
such that the components on the right-hand side of (\ref{a52a}) are subject
to the equations
\begin{eqnarray}
\gamma a_{1}^{\left( \mathrm{int}\right) } &=&0,  \label{a52} \\
\delta a_{1}^{\left( \mathrm{int}\right) }+\gamma a_{0}^{\left( \mathrm{int}%
\right) } &=&\partial _{\mu }m_{0}^{\left( \mathrm{int}\right) \mu }.
\label{a52c}
\end{eqnarray}%
In agreement with (\ref{a31}) for $I=1$ and the discussion from the end of
subsection \ref{stand}, the general solution to (\ref{a52}) is (up to
trivial, $\gamma $-exact contributions)
\begin{equation}
a_{1}^{\left( \mathrm{int}\right) }=\alpha _{1}\eta +\alpha _{1\mu }\eta
^{\mu }+\alpha _{1\mu \nu }\partial ^{\lbrack \mu }\eta ^{\nu ]},
\label{a52start}
\end{equation}%
where $\alpha _{1}$, $\alpha _{1\mu }$, and $\alpha _{1\mu \nu }$ are
nontrivial invariant polynomials from $H_{1}\left( \delta |d\right) $ (but
not necessarily from $H_{1}^{\mathrm{inv}}\left( \delta |d\right) $) in
order to produce a consistent $a_{0}^{\left( \mathrm{int}\right) }$. Because
they are nontrivial invariant polynomials of antighost number one, we can
always assume that they are linear in the undifferentiated antifields $%
V^{\ast \mu }$ and $h^{\ast \mu \nu }$, such that (\ref{a52start}) becomes%
\begin{eqnarray}
a_{1}^{\left( \mathrm{int}\right) } &=&V^{\ast \mu }\left( M_{\mu }\eta
+M_{\mu \nu }\eta ^{\nu }+M_{\mu \nu \rho }\partial ^{\lbrack \nu }\eta
^{\rho ]}\right)  \notag \\
&&+h^{\ast \mu \nu }\left( N_{\mu \nu }\eta +N_{\mu \nu \rho }\eta ^{\rho
}+N_{\mu \nu \rho \lambda }\partial ^{\lbrack \rho }\eta ^{\lambda ]}\right)
,  \label{a53}
\end{eqnarray}%
where all the coefficients, denoted by $M$ or $N$, must be $\gamma $-closed
quantities, and therefore they may depend on $F_{\mu \nu }$, $K_{\mu \alpha
|\nu \beta }$, and their derivatives. In addition, these tensors are subject
to the symmetry/antisymmetry properties
\begin{eqnarray}
M_{\mu \nu \rho } &=&-M_{\mu \rho \nu },\qquad N_{\mu \nu }=N_{\nu \mu },
\label{a54a} \\
N_{\mu \nu \rho } &=&N_{\nu \mu \rho },\qquad N_{\mu \nu \rho \lambda
}=N_{\nu \mu \rho \lambda }=-N_{\mu \nu \lambda \rho }.  \label{a54b}
\end{eqnarray}%
At this point we recall the hypothesis on the derivative order of the
deformed Lagrangian, which imposes that $a_{0}^{\left( \mathrm{int}\right) }$
as solution to (\ref{a52c}) contains at most two spacetime derivatives of
the fields. Then, relation (\ref{a53}), equation (\ref{a52c}), and
definitions (\ref{a7})--(\ref{a12}) yield the following results: A. none of
the $M$- or $N$-type tensors entering (\ref{a53}) are allowed to depend on $%
K_{\mu \alpha |\nu \beta }$ or its derivatives; B. $M_{\mu \nu \rho }$ and $%
N_{\mu \nu \rho \lambda }$ cannot involve either $F_{\mu \nu }$ or its
derivatives, and therefore they are nonderivative, constant tensors; C. the
tensors $M_{\mu }$, $M_{\mu \nu }$, $N_{\mu \nu }$, and $N_{\mu \nu \rho }$
may depend on $F_{\mu \nu }$ (and not on its derivatives), but only in a
linear manner. These results are synthesized by the formulas%
\begin{eqnarray}
M_{\mu } &=&C_{\mu }+C_{\mu \nu \rho }F^{\nu \rho },\qquad M_{\mu \nu
}=C_{\mu \nu }+C_{\mu \nu \rho \lambda }F^{\rho \lambda },  \label{a54c} \\
N_{\mu \nu } &=&D_{\mu \nu }+D_{\mu \nu \rho \lambda }F^{\rho \lambda
},\qquad N_{\mu \nu \rho }=D_{\mu \nu \rho }+D_{\mu \nu \rho \lambda \sigma
}F^{\lambda \sigma },  \label{a54d} \\
M_{\mu \nu \rho } &=&\bar{C}_{\mu \nu \rho },\qquad N_{\mu \nu \rho \lambda
}=\bar{D}_{\mu \nu \rho \lambda },  \label{a54e}
\end{eqnarray}%
where the quantities denoted by $C$, $\bar{C}$, $D$, or $\bar{D}$ are
nonderivative, constant tensors, subject to some symmetry/antisymmetry
properties such that (\ref{a54a}) and (\ref{a54b}) are fulfilled. Since we
work in $D>2$ spacetime dimensions, the only choice that complies with the
above mentioned properties and leads to consistent cross-couplings between
the Pauli-Fierz field and the vector field is\footnote{%
Strictly speaking, there is a nonvanishing solution $C_{\mu \nu \rho
}=z\delta _{3}^{D}\varepsilon _{\mu \nu \rho }$, which adds to $%
a_{1}^{\left( \mathrm{int}\right) }$ the term $z\delta _{3}^{D}\varepsilon
_{\mu \nu \rho }V^{\ast \mu }F^{\nu \rho }\eta $. Even if consistent, this
term would lead to selfinteractions in the Maxwell sector. However, $%
a_{1}^{\left( \mathrm{int}\right) }$ is restricted by hypothesis to provide
only cross-couplings between the Pauli-Fierz field and the electromagnetic
field, so this term must be removed from this context by setting $z=0$.
Apparently, there are two more possibilities, $C_{\mu \nu \rho \lambda
}=z^{\prime }\delta _{4}^{D}\varepsilon _{\mu \nu \rho \lambda }$ and $%
D_{\mu \nu \rho \lambda \sigma }=z^{\prime \prime }\delta _{3}^{D}\sigma
_{\mu \nu }\varepsilon _{\rho \lambda \sigma }$, which add to $a_{1}^{\left(
\mathrm{int}\right) }$ the terms $(z^{\prime \prime }\delta
_{3}^{D}\varepsilon _{\mu \nu \rho }h^{\ast }F^{\mu \nu }-z^{\prime }\delta
_{4}^{D}\varepsilon _{\mu \nu \rho \lambda }V^{\ast \mu }F^{\nu \lambda
})\eta ^{\rho }$. They are not eligible to enter $a_{1}^{\left( \mathrm{int}%
\right) }$ since the corresponding invariant polynomial, $z^{\prime \prime
}\delta _{3}^{D}\varepsilon _{\mu \nu \rho }h^{\ast }F^{\mu \nu }-z^{\prime
}\delta _{4}^{D}\varepsilon _{\mu \nu \rho \lambda }V^{\ast \mu }F^{\nu
\lambda }$, does not belong to $H^{1}\left( \delta |d\right) $, such that
they cannot lead to consistent pieces in $a_{0}^{\left( \mathrm{int}\right)
} $ unless $z^{\prime }=0=z^{\prime \prime }$.}%
\begin{eqnarray}
C_{\mu } &=&0,\qquad C_{\mu \nu \rho }=0,\qquad C_{\mu \nu }=y_{1}\sigma
_{\mu \nu },  \label{a54f} \\
C_{\mu \nu \rho \lambda } &=&\frac{p}{2}\left( \sigma _{\mu \rho }\sigma
_{\nu \lambda }-\sigma _{\mu \lambda }\sigma _{\nu \rho }\right) ,
\label{a54g} \\
D_{\mu \nu } &=&y_{2}\sigma _{\mu \nu },\qquad D_{\mu \nu \rho \lambda
}=D_{\mu \nu \rho }=\bar{D}_{\mu \nu \rho \lambda }=0,  \label{a54h} \\
D_{\mu \nu \rho \lambda \sigma } &=&0,\qquad \bar{C}_{\mu \nu \rho
}=y_{3}\delta _{3}^{D}\varepsilon _{\mu \nu \rho }.  \label{a54i}
\end{eqnarray}%
Substituting (\ref{a54f})--(\ref{a54i}) in (\ref{a54c})--(\ref{a54e}) and
the resulting expressions in (\ref{a53}), we obtain
\begin{equation}
a_{1}^{\left( \mathrm{int}\right) }=y_{1}V^{\ast \lambda }\eta _{\lambda
}+y_{2}h^{\ast }\eta +y_{3}\delta _{3}^{D}\varepsilon _{\mu \nu \rho
}V^{\ast \mu }\partial ^{\lbrack \nu }\eta ^{\rho ]}+pV^{\ast \mu }F_{\mu
\nu }\eta ^{\nu },  \label{a55}
\end{equation}%
where $h^{\ast }=h^{\ast \mu \nu }\sigma _{\mu \nu }$. Acting with $\delta $
on (\ref{a55}), we infer%
\begin{eqnarray}
\delta a_{1}^{\left( \mathrm{int}\right) } &=&\gamma \left[ -\left(
D-2\right) y_{2}V^{\lambda }\partial _{\lbrack \mu }h_{\lambda ]}^{\ \ \mu
}-y_{3}\delta _{3}^{D}\varepsilon _{\mu \nu \rho }F^{\lambda \mu }\partial
^{\lbrack \nu }h_{\ \ \lambda }^{\rho ]}\right.  \notag \\
&&\left. -\frac{p}{2}\left( F^{\alpha \mu }F_{\mu }^{\;\;\nu }h_{\alpha \nu
}+\frac{1}{4}F^{\alpha \mu }F_{\alpha \mu }h\right) \right] +\partial
_{\alpha }u^{\alpha }  \notag \\
&&+\delta \left\{ \left[ y_{1}+\left( D-2\right) y_{2}\right] V^{\ast
\lambda }\eta _{\lambda }\right\} .  \label{a56}
\end{eqnarray}%
Comparing (\ref{a56}) with (\ref{a52c}) and observing that (\ref{a55})
already contains a term of the type $V^{\ast \lambda }\eta _{\lambda }$, it
follows that $a_{1}^{\left( \mathrm{int}\right) }$ is consistent at
antighost number zero if and only if%
\begin{equation}
y_{1}+\left( D-2\right) y_{2}=0.  \label{conda56}
\end{equation}%
Replacing (\ref{conda56}) into (\ref{a55}) and (\ref{a56}), we get finally%
\begin{equation}
a_{1}^{\left( \mathrm{int}\right) }=y_{2}\left[ h^{\ast }\eta -\left(
D-2\right) V^{\ast \lambda }\eta _{\lambda }\right] +y_{3}\delta
_{3}^{D}\varepsilon _{\mu \nu \rho }V^{\ast \mu }\partial ^{\lbrack \nu
}\eta ^{\rho ]}+pV^{\ast \mu }F_{\mu \nu }\eta ^{\nu },  \label{a57}
\end{equation}%
\begin{eqnarray}
a_{0}^{\left( \mathrm{int}\right) } &=&\left( D-2\right) y_{2}V^{\lambda
}\partial _{\lbrack \mu }h_{\lambda ]}^{\ \ \mu }+y_{3}\delta
_{3}^{D}\varepsilon _{\mu \nu \rho }F^{\lambda \mu }\partial ^{\lbrack \nu
}h_{\ \ \lambda }^{\rho ]}  \notag \\
&&+\frac{p}{2}\left( F^{\alpha \mu }F_{\mu }^{\;\;\nu }h_{\alpha \nu }+\frac{%
1}{4}F^{\alpha \mu }F_{\alpha \mu }h\right) +\bar{a}_{0}^{\left( \mathrm{int}%
\right) },  \label{a58}
\end{eqnarray}%
where $\bar{a}_{0}^{\left( \mathrm{int}\right) }$ is the general solution to
the homogeneous equation
\begin{equation}
\gamma \bar{a}_{0}^{\left( \mathrm{int}\right) }=\partial _{\mu }\bar{m}%
^{\left( \mathrm{int}\right) \mu }.  \label{o1}
\end{equation}

Such solutions correspond to $\bar{a}_{1}^{\left( \mathrm{int}\right) }=0$
and thus they cannot deform either the gauge algebra or the gauge
transformations, but only the Lagrangian at order one in the coupling
constant. There are two main types of solutions to (\ref{o1}). The first one
corresponds to $\bar{m}^{\left( \mathrm{int}\right) \mu }=0$ and is given by
gauge-invariant, nonintegrated densities constructed from the original
fields and their spacetime derivatives. According to (\ref{a31}) for both
pure ghost and antighost numbers equal to zero, they are given by $\bar{a}%
_{0}^{\left( \mathrm{int}\right) \prime }=\bar{a}_{0}^{\left( \mathrm{int}%
\right) \prime }\left( \left[ F_{\mu \nu }\right] ,\left[ K_{\mu
\alpha |\nu \beta }\right] \right) $, up to the conditions that they
describe true cross-couplings between the two types of fields and
cannot be written in a divergence-like form. Unfortunately, this
type of solutions must depend simultaneously at least on the
linearized Riemann tensor and on the Abelian field strength in order
to provide cross-couplings, so they would lead to terms with at
least three derivatives in the deformed Lagrangian. By virtue of the
derivative order assumption, they must be discarded by setting
$\bar{a}_{0}^{\left( \mathrm{int}\right) \prime }=0$. The second
kind of solutions is associated with $\bar{m}^{\left(
\mathrm{int}\right) \mu }\neq
0 $ in (\ref{o1}), being understood that we maintain the requirements on $%
\bar{a}_{0}^{\left( \mathrm{int}\right) }$ to contain maximum two
derivatives of the fields and to describe cross-couplings. In order
to simplify the presentation, we omit the technical aspects
regarding the analysis of these solutions. \emph{The main result is
that, without loss
of generality, we can take }$\bar{a}_{0}^{\left( \mathrm{int}\right) }=0$%
\emph{\ in (\ref{a58}).} Very briefly, we mention that the procedure
used
for obtaining this result relies on decomposing $\bar{a}_{0}^{\left( \mathrm{%
int}\right) }$ along the number of derivatives, $\bar{a}_{0}^{\left(
\mathrm{int}\right) }=\omega _{0}+\omega _{1}+\omega _{2}$, where
$\omega _{i}$ contains exactly $i$ derivatives of the fields. As a
consequence, equation (\ref{o1}) becomes equivalent to three
independent equations, one for each component. The terms $\omega
_{0}$ and respectively $\omega _{1}$ are ruled out because they
cannot produce cross-couplings. As for $\omega _{2}$, it requires
the existence of a nonderivative, real, constant tensor $C^{\mu
\alpha |\nu \beta ;\sigma }$, which displays the generalized
symmetry properties of the Riemann tensor with respect to its first
four indices and is simultaneously antisymmetric in its last three
indices. Since there are no such tensors in any $D\geq 3$ spacetime
dimension, we must discard $\omega _{2}$, which finally leaves us
with $\bar{a}_{0}^{\left( \mathrm{int}\right) }=0$.

Replacing (\ref{a57}), (\ref{a58}), and $\bar{a}_{0}^{\left( \mathrm{int}%
\right) }=0$ in (\ref{a52a}), we obtain the concrete form of the general
solution $a^{\left( \mathrm{int}\right) }$ to (\ref{a42}). We can still
remove certain trivial, $s$-exact modulo $d$ terms from the resulting $%
a^{\left( \mathrm{int}\right) }$. Indeed, we have that%
\begin{equation}
a^{\left( \mathrm{int}\right) }=a^{\prime \left( \mathrm{int}\right) }+s%
\left[ -p\left( \eta ^{\ast }V^{\mu }\eta _{\mu }+\frac{1}{2}V^{\ast \mu
}V^{\nu }h_{\mu \nu }\right) \right] +\partial _{\mu }t^{\mu },
\label{relaint}
\end{equation}%
such that, in agreement with the discussion made in the beginning of this
section, we can work with%
\begin{eqnarray}
a^{\prime \left( \mathrm{int}\right) } &=&a^{\left( \mathrm{int}\right) }+s%
\left[ p\left( \eta ^{\ast }V^{\mu }\eta _{\mu }+\frac{1}{2}V^{\ast \mu
}V^{\nu }h_{\mu \nu }\right) \right] -\partial _{\mu }t^{\mu }  \notag \\
&\equiv &y_{2}\left[ h^{\ast }\eta +\left( D-2\right) \left( -V^{\ast
\lambda }\eta _{\lambda }+V^{\lambda }\partial _{\lbrack \mu }h_{\lambda
]}^{\ \ \mu }\right) \right]  \notag \\
&&+y_{3}\delta _{3}^{D}\varepsilon _{\mu \nu \rho }\left( V^{\ast \mu
}\partial ^{\lbrack \nu }\eta ^{\rho ]}+F^{\lambda \mu }\partial ^{\lbrack
\nu }h_{\ \ \lambda }^{\rho ]}\right) +p\left[ \eta ^{\ast }\eta _{\mu
}\partial ^{\mu }\eta \right.  \notag \\
&&-\frac{1}{2}V^{\ast \mu }\left( V^{\nu }\partial _{\lbrack \mu }\eta _{\nu
]}+2\left( \partial _{\nu }V_{\mu }\right) \eta ^{\nu }-h_{\mu \nu }\partial
^{\nu }\eta \right)  \notag \\
&&\left. +\frac{1}{8}F^{\mu \nu }\left( 2\partial _{\lbrack \mu }\left(
h_{\nu ]\rho }V^{\rho }\right) +F_{\mu \nu }h-4F_{\mu \rho }h_{\ \ \nu
}^{\rho }\right) \right]  \label{aintfinal}
\end{eqnarray}%
instead of $a^{\left( \mathrm{int}\right) }$.

In view of the results (\ref{a39}), (\ref{rr2}), and (\ref{aintfinal}) we
conclude that the most general, nontrivial first-order deformation of the
solution to the master equation corresponding to action (\ref{a1}) and to
its gauge transformations (\ref{a2}), which complies with all the working
hypotheses, is expressed by%
\begin{equation}
S_{1}=S_{1}^{(\mathrm{PF})}+S_{1}^{\left( \mathrm{int}\right) },  \label{s1}
\end{equation}%
where%
\begin{equation}
S_{1}^{(\mathrm{PF})}\equiv \int d^{D}x\,a^{\left( \mathrm{PF}\right) }=\int
d^{D}x\left( a_{2}^{\left( \mathrm{PF}\right) }+a_{1}^{\left( \mathrm{PF}%
\right) }+a_{0}^{\left( \mathrm{PF}\right) }\right) ,  \label{S1pf}
\end{equation}%
and%
\begin{eqnarray}
S_{1}^{\left( \mathrm{int}\right) } &=&\int d^{D}x\left( a^{\prime \left(
\mathrm{int}\right) }+a^{\left( \mathrm{vect}\right) }\right)  \notag \\
&\equiv &\int d^{D}x\left\{ y_{2}\left[ h^{\ast }\eta +\left( D-2\right)
\left( -V^{\ast \lambda }\eta _{\lambda }+V^{\lambda }\partial _{[ \mu
}h_{\lambda ]}^{\ \ \mu }\right) \right] \right.  \notag \\
&&+y_{3}\delta _{3}^{D}\varepsilon _{\mu \nu \rho }\left( V^{\ast \mu
}\partial ^{[ \nu }\eta ^{\rho ]}+F^{\lambda \mu }\partial ^{[ \nu }h_{\ \
\lambda }^{\rho ]}\right) +p\left[ \eta ^{\ast }\eta _{\mu }\partial ^{\mu
}\eta \right.  \notag \\
&&-\frac{1}{2}V^{\ast \mu }\left( V^{\nu }\partial _{[ \mu }\eta _{\nu
]}+2\left( \partial _{\nu }V_{\mu }\right) \eta ^{\nu }-h_{\mu \nu }\partial
^{\nu }\eta \right)  \notag \\
&&\left. +\frac{1}{8}F^{\mu \nu }\left( 2\partial _{[ \mu }\left( h_{\nu
]\rho }V^{\rho }\right) +F_{\mu \nu }h-4F_{\mu \rho }h_{\nu }^{\ \ \rho
}\right) \right]  \notag \\
&&\left. +q_{1}\delta _{3}^{D}\varepsilon ^{\mu \nu \lambda }V_{\mu }F_{\nu
\lambda }+q_{2}\delta _{5}^{D}\varepsilon ^{\mu \nu \lambda \alpha \beta
}V_{\mu }F_{\nu \lambda }F_{\alpha \beta }\right\} .  \label{a60int}
\end{eqnarray}%
Thus, the first-order deformation of the solution to the master equation for
the model under study is parameterized by seven independent, real constants,
namely $f$ and $\Lambda $ corresponding to $S_{1}^{(\mathrm{PF})}$ (see (\ref%
{a40}), (\ref{a41}), and (\ref{a0pf})) together with $p$, $y_{2}$, $%
y_{3}\delta _{3}^{D}$, $q_{1}\delta _{3}^{D}$, and $q_{2}\delta _{5}^{D}$
associated with $S_{1}^{\left( \mathrm{int}\right) }$.

\subsection{Computation of second-order deformations\label{deformarea II}}

Here, we approach the construction of the second-order deformation of the
solution to the master equation, governed by equation (\ref{a21}). Replacing
(\ref{s1}) into (\ref{a21}) we find that it becomes equivalent to the
equations
\begin{eqnarray}
\left( S_{1}^{\left( \mathrm{PF}\right) },S_{1}^{\left( \mathrm{PF}\right)
}\right) +\left( S_{1}^{\left( \mathrm{int}\right) },S_{1}^{\left( \mathrm{%
int}\right) }\right) ^{\left( \mathrm{PF}\right) }+2sS_{2}^{\left( \mathrm{PF%
}\right) } &=&0,  \label{a62} \\
2\left( S_{1}^{\left( \mathrm{PF}\right) },S_{1}^{\left( \mathrm{int}\right)
}\right) +\left( S_{1}^{\left( \mathrm{int}\right) },S_{1}^{\left( \mathrm{%
int}\right) }\right) ^{\left( \mathrm{int}\right) }+2sS_{2}^{\left( \mathrm{%
int}\right) } &=&0,  \label{a63}
\end{eqnarray}%
where $\left( S_{1}^{\left( \mathrm{int}\right) },S_{1}^{\left( \mathrm{int}%
\right) }\right) ^{\left( \mathrm{PF}\right) }$ comprises only BRST
generators from the Pauli-Fierz sector and each term from $\left(
S_{1}^{\left( \mathrm{int}\right) },S_{1}^{\left( \mathrm{int}\right)
}\right) ^{\left( \mathrm{int}\right) }$ contains at least one BRST
generator from the one-form sector. By writing down (\ref{a62}) and (\ref%
{a63}), it is understood that the second-order deformation decomposes as
\begin{equation}
S_{2}=S_{2}^{\left( \mathrm{PF}\right) }+S_{2}^{\left( \mathrm{int}\right) },
\label{a64}
\end{equation}%
where $S_{2}^{\left( \mathrm{PF}\right) }$ represents the component
from the Pauli-Fierz sector and $S_{2}^{\left( \mathrm{int}\right)
}$ signifies the complementary part.

Initially, we analyze equation (\ref{a62}). It is known from the
literature (for instance, see~\cite{multi} in the absence of
collection indices) that there exists $S_{2}^{\left(
\mathrm{PF}\right) }\left( f^{2},f\Lambda
\right) $ such that%
\begin{equation}
\left( S_{1}^{\left( \mathrm{PF}\right) },S_{1}^{\left( \mathrm{PF}\right)
}\right) +2sS_{2}^{\left( \mathrm{PF}\right) }\left( f^{2},f\Lambda \right)
=0,  \label{a64a}
\end{equation}%
where
\begin{equation}
S_{2}^{\left( \mathrm{PF}\right) }\left( f^{2},f\Lambda \right)
=f^{2}S_{2}^{(\mathrm{EH-quartic})}+f\Lambda \int d^{D}x\left( h^{\mu \nu
}h_{\mu \nu }-\frac{1}{2}h^{2}\right) ,  \label{a64b}
\end{equation}%
with $S_{2}^{(\mathrm{EH-quartic})}$ the second-order Einstein-Hilbert
deformation, including the quartic vertex of the Einstein-Hilbert
Lagrangian. On the other hand, direct computation based on (\ref{a60int})
leads to%
\begin{eqnarray}
&&\left( S_{1}^{\left( \mathrm{int}\right) },S_{1}^{\left( \mathrm{int}%
\right) }\right) ^{\left( \mathrm{PF}\right) }=-2s\int d^{D}x\left[ y_{2}^{2}%
\frac{\left( D-2\right) ^{2}}{4}\left( h^{2}-h^{\mu \nu }h_{\mu \nu }\right)
\right.  \notag \\
&&\left. +y_{2}y_{3}\left( D-2\right) \delta _{3}^{D}\varepsilon _{\mu \nu
\rho }\left( \partial ^{[ \nu }h^{\rho ]\lambda }\right) h_{\ \ \lambda
}^{\mu }+y_{3}^{2}\delta _{3}^{D}\left( \partial ^{[ \nu }h^{\rho ]\lambda
}\right) \partial _{[ \nu }h_{\rho ]\lambda }\right]  \notag \\
&\equiv &-2s\left( S_{2}^{\left( \mathrm{PF}\right) }\left( y_{2}^{2}\right)
+S_{2}^{\left( \mathrm{PF}\right) }\left( y_{2}y_{3}\right) +S_{2}^{\left(
\mathrm{PF}\right) }\left( y_{3}^{2}\right) \right) ,  \label{a64c}
\end{eqnarray}%
where we used the obvious notations%
\begin{eqnarray}
S_{2}^{\left( \mathrm{PF}\right) }\left( y_{2}^{2}\right) &=&y_{2}^{2}\frac{%
\left( D-2\right) ^{2}}{4}\int d^{D}x\left( h^{2}-h^{\mu \nu }h_{\mu \nu
}\right) ,  \label{a64c1} \\
S_{2}^{\left( \mathrm{PF}\right) }\left( y_{2}y_{3}\right)
&=&y_{2}y_{3}\left( D-2\right) \delta _{3}^{D}\varepsilon _{\mu \nu \rho
}\int d^{D}x\left( \partial ^{[ \nu }h^{\rho ]\lambda }\right) h_{\ \
\lambda }^{\mu },  \label{a64c2} \\
S_{2}^{\left( \mathrm{PF}\right) }\left( y_{3}^{2}\right) &=&y_{3}^{2}\delta
_{3}^{D}\int d^{D}x\left( \partial ^{[ \nu }h^{\rho ]\lambda }\right)
\partial _{[ \nu }h_{\rho ]\lambda }.  \label{a64c3}
\end{eqnarray}%
Taking into account relations (\ref{a64a})--(\ref{a64c}) it follows that (%
\ref{a62}) becomes equivalent with%
\begin{equation}
s\left[ S_{2}^{\left( \mathrm{PF}\right) }-\left( S_{2}^{\left( \mathrm{PF}%
\right) }\left( f^{2},f\Lambda \right) +S_{2}^{\left( \mathrm{PF}\right)
}\left( y_{2}^{2}\right) +S_{2}^{\left( \mathrm{PF}\right) }\left(
y_{2}y_{3}\right) +S_{2}^{\left( \mathrm{PF}\right) }\left( y_{3}^{2}\right)
\right) \right] =0,  \label{a64d}
\end{equation}%
which allows us to determine the component $S_{2}^{\left( \mathrm{PF}\right)
}$ from the second-order deformation (\ref{a64}), up to trivial, $s$-exact
contributions\footnote{%
Strictly speaking, we must add to (\ref{a64e}) the nontrivial solution $F$
to the homogeneous equation $sF=0$. However, this solution brings nothing
new and can always be absorbed into the full deformed solution to the master
equation $S$ (actually in $S_{1}^{(\mathrm{PF})}$) through a convenient
redefinition of the coupling constant and of the other constants that
parameterize $S_{1}^{(\mathrm{PF})}$. For instance, see Section 7 from~\cite%
{multi}.}, in the form%
\begin{equation}
S_{2}^{\left( \mathrm{PF}\right) }=S_{2}^{\left( \mathrm{PF}\right) }\left(
f^{2},f\Lambda \right) +S_{2}^{\left( \mathrm{PF}\right) }\left(
y_{2}^{2}\right) +S_{2}^{\left( \mathrm{PF}\right) }\left( y_{2}y_{3}\right)
+S_{2}^{\left( \mathrm{PF}\right) }\left( y_{3}^{2}\right) .  \label{a64e}
\end{equation}

Next, we pass to equation (\ref{a63}). If we denote by $\Delta ^{\left(
\mathrm{int}\right) }$ and $b^{\left( \mathrm{int}\right) }$ the
nonintegrated densities of $2\left( S_{1}^{\left( \mathrm{PF}\right)
},S_{1}^{\left( \mathrm{int}\right) }\right) +\left( S_{1}^{\left( \mathrm{%
int}\right) },S_{1}^{\left( \mathrm{int}\right) }\right) ^{\left( \mathrm{int%
}\right) }$ and $S_{2}^{\left( \mathrm{int}\right) }$ respectively,
\begin{eqnarray}
2\left( S_{1}^{\left( \mathrm{PF}\right) },S_{1}^{\left( \mathrm{int}\right)
}\right) +\left( S_{1}^{\left( \mathrm{int}\right) },S_{1}^{\left( \mathrm{%
int}\right) }\right) ^{\left( \mathrm{int}\right) } &\equiv &\int
d^{D}x\,\Delta ^{\left( \mathrm{int}\right) },  \label{notdelta} \\
S_{2}^{\left( \mathrm{int}\right) } &\equiv &\int d^{D}x\,b^{\left( \mathrm{%
int}\right) },  \label{notdeltas2}
\end{eqnarray}%
then the local form of equation (\ref{a63}) reads as
\begin{equation}
\Delta ^{\left( \mathrm{int}\right) }=-2sb^{\left( \mathrm{int}\right)
}+\partial _{\mu }n^{\mu },  \label{a65}
\end{equation}%
where
\begin{equation}
\mathrm{gh}\left( \Delta ^{\left( \mathrm{int}\right) }\right) =1,\qquad
\mathrm{gh}\left( b^{\left( \mathrm{int}\right) }\right) =0,\qquad \mathrm{gh%
}\left( n^{\mu }\right) =1,  \label{a66}
\end{equation}%
for some local currents $n^{\mu }$. By direct computation, from (\ref{S1pf})
and (\ref{a60int}) we deduce that $\Delta ^{\left( \mathrm{int}\right) }$
decomposes as%
\begin{equation}
\Delta ^{\left( \mathrm{int}\right) }=\sum\limits_{I=0}^{2}\Delta
_{I}^{\left( \mathrm{int}\right) },\qquad \mathrm{agh}\left( \Delta
_{I}^{\left( \mathrm{int}\right) }\right) =I,  \label{a67}
\end{equation}%
where
\begin{equation}
\Delta _{2}^{\left( \mathrm{int}\right) }=\gamma \left[ p\eta ^{\ast }\left(
p\left( \partial ^{\mu }\eta \right) \eta ^{\nu }h_{\mu \nu }-\left(
p+f\right) V^{\mu }\eta ^{\nu }\partial _{\lbrack \mu }\eta _{\nu ]}\right) %
\right] +\partial _{\mu }w_{2}^{\mu },  \label{a671}
\end{equation}%
\begin{eqnarray}
&&\Delta _{1}^{\left( \mathrm{int}\right) }=\delta \left[ p\eta ^{\ast
}\left( p\left( \partial ^{\mu }\eta \right) \eta ^{\nu }h_{\mu \nu }-\left(
p+f\right) V^{\mu }\eta ^{\nu }\partial _{\lbrack \mu }\eta _{\nu ]}\right) %
\right]  \notag \\
&&+\gamma \left\{ p^{2}V^{\ast \mu }\left[ \left( \partial _{\nu }V_{\mu
}\right) h_{\ \ \rho }^{\nu }\eta ^{\rho }+\frac{1}{2}\left( \partial
_{\lbrack \mu }h_{\nu ]\rho }\right) V^{\nu }\eta ^{\rho }\right. \right.
\notag \\
&&\left. -\frac{1}{4}V^{\nu }h_{[\mu }^{\ \ \rho }\left( \partial _{\nu
]}\eta _{\rho }\right) -\frac{1}{4}V^{\nu }\left( \partial _{\rho }\eta
_{\lbrack \mu }\right) h_{\nu ]}^{\ \ \rho }-\frac{3}{4}h_{\mu }^{\ \ \nu
}h_{\nu }^{\ \ \rho }\partial _{\rho }\eta \right]  \notag \\
&&+\frac{1}{2}p\left( p+f\right) V^{\ast \mu }V^{\nu }\left[ \left( \partial
_{\lbrack \mu }h_{\rho ]\nu }+\partial _{\lbrack \nu }h_{\rho ]\mu }\right)
\eta ^{\rho }-h_{\mu }^{\ \ \rho }\partial _{\nu }\eta _{\rho }\right.
\notag \\
&&\left. -h_{\nu }^{\ \ \rho }\partial _{\mu }\eta _{\rho }\right]
-y_{3}\delta _{3}^{D}\varepsilon _{\mu \nu \rho }V^{\ast \mu }\left[ fh_{\ \
\lambda }^{\nu }\partial ^{\lbrack \rho }\eta ^{\lambda ]}+\left(
2p+f\right) \eta _{\lambda }\partial ^{\lbrack \nu }h^{\rho ]\lambda }\right]
\notag \\
&&+py_{2}V^{\ast \mu }\left[ \left( D-2\right) h_{\mu \nu }\eta ^{\nu
}-V_{\mu }\eta \right] -y_{2}h^{\ast \mu \nu }\left[ f\left( h_{\mu \nu
}\eta +2V_{\mu }\eta _{\nu }\right) \right.  \notag \\
&&\left. \left. -2\left( p+f\right) \sigma _{\mu \nu }V^{\rho }\eta _{\rho }
\right] \right\} -p\left( p+f\right) V_{\mu }^{\ast }F^{\mu \nu }\eta ^{\rho
}\partial _{\lbrack \rho }\eta _{\nu ]}  \notag \\
&&+\left( 2p+f\right) V^{\ast \mu }\left[ y_{3}\delta _{3}^{D}\varepsilon
_{\mu \nu \rho }\left( \partial ^{\lbrack \nu }\eta ^{\lambda ]}\right)
\partial ^{\lbrack \rho }\eta ^{\tau ]}\sigma _{\lambda \tau }\right.  \notag
\\
&&\left. +y_{2}\left( D-2\right) \left( \partial _{\lbrack \mu }\eta _{\nu
]}\right) \eta ^{\nu }\right] +\partial _{\mu }w_{1}^{\mu },  \label{a68}
\end{eqnarray}%
\begin{eqnarray}
&&\Delta _{0}^{\left( \mathrm{int}\right) }=\delta \left\{ p^{2}V^{\ast \mu }%
\left[ \left( \partial _{\nu }V_{\mu }\right) h_{\ \ \rho }^{\nu }\eta
^{\rho }+\frac{1}{2}\left( \partial _{\lbrack \mu }h_{\nu ]\rho }\right)
V^{\nu }\eta ^{\rho }-\frac{1}{4}V^{\nu }h_{[\mu }^{\ \ \rho }\left(
\partial _{\nu ]}\eta _{\rho }\right) \right. \right.  \notag \\
&&\left. -\frac{1}{4}V^{\nu }\left( \partial _{\rho }\eta _{\lbrack \mu
}\right) h_{\nu ]}^{\ \ \rho }-\frac{3}{4}h_{\mu }^{\ \ \nu }h_{\nu }^{\ \
\rho }\partial _{\rho }\eta \right] +\frac{16}{D-2}y_{3}q_{1}\delta
_{3}^{D}h^{\ast }\eta  \notag \\
&&\left. +\frac{1}{2}p\left( p+f\right) V^{\ast \mu }V^{\nu }\left[ \left(
\partial _{\lbrack \mu }h_{\rho ]\nu }+\partial _{\lbrack \nu }h_{\rho ]\mu
}\right) \eta ^{\rho }-h_{\mu }^{\ \ \rho }\partial _{\nu }\eta _{\rho
}-h_{\nu }^{\ \ \rho }\partial _{\mu }\eta _{\rho }\right] \right\}  \notag
\\
&&+\gamma \left\{ \frac{p^{2}}{8}\left[ V_{\rho }\left( \left( \partial
^{\lbrack \mu }h^{\nu ]\rho }\right) \left( \partial _{\lbrack \mu }h_{\nu
]\lambda }\right) V^{\lambda }-2\left( \partial ^{\lbrack \mu }h^{\nu ]\rho
}\right) h_{\lambda \lbrack \mu }\left( \partial _{\nu ]}V^{\lambda }\right)
\right) \right. \right.  \notag \\
&&+h_{\rho }^{\ \ [\mu }\left( \partial ^{\nu ]}V^{\rho }\right) h_{\lambda
\lbrack \mu }\left( \partial _{\nu ]}V^{\lambda }\right) +F^{\mu \nu }h_{\ \
\lambda }^{\rho }\left( h_{\ \ [\mu }^{\lambda }\left( \partial _{\nu
]}V_{\rho }\right) -\left( \partial _{\lbrack \mu }h_{\ \ \nu ]}^{\lambda
}\right) V_{\rho }\right)  \notag \\
&&\left. +F^{\mu \nu }h_{\ \ [\mu }^{\rho }\left( \partial _{\nu ]}h_{\rho
}^{\ \ \lambda }\right) V_{\lambda }\right] +p^{2}F^{\mu \nu }\left[ F_{\mu
\rho }h_{\nu }^{\ \ \lambda }h_{\lambda }^{\ \ \rho }+\frac{1}{16}F_{\mu \nu
}\left( h^{2}-2h^{\rho \lambda }h_{\rho \lambda }\right) \right.  \notag \\
&&-h_{\nu }^{\ \ \rho }\left( \left( \partial _{\lbrack \mu }h_{\rho ]}^{\ \
\lambda }\right) V_{\lambda }-h_{[\mu }^{\ \ \lambda }\left( \partial _{\rho
]}V_{\lambda }\right) \right) +\frac{1}{2}\left( F^{\rho \lambda }h_{\mu
\rho }h_{\nu \lambda }-F_{\mu \rho }h_{\ \ \nu }^{\rho }h\right)  \notag \\
&&\left. +\frac{1}{4}\left( \left( \partial _{\lbrack \mu }h_{\nu ]}^{\ \
\rho }\right) V_{\rho }-h_{[\mu }^{\ \ \rho }\left( \partial _{\nu ]}V_{\rho
}\right) \right) h\right] +\frac{1}{4}p\left( p+f\right) \left( F^{\mu \nu
}F_{\nu \rho }\right.  \notag \\
&&\left. +\frac{1}{4}\delta _{\rho }^{\mu }F^{\nu \lambda }F_{\nu \lambda
}\right) h_{\mu \sigma }h^{\sigma \rho }+q_{1}p\delta _{3}^{D}\varepsilon
^{\mu \nu \lambda }\left( hV_{\mu }F_{\nu \lambda }-2h_{\lambda }^{\ \
\alpha }V_{\mu }F_{\nu \alpha }\right.  \notag \\
&&\left. +h_{\mu }^{\ \ \alpha }V_{\alpha }F_{\nu \lambda }\right)
+q_{2}p\delta _{5}^{D}\varepsilon ^{\mu \nu \lambda \alpha \beta }\left(
hV_{\mu }F_{\nu \lambda }F_{\alpha \beta }-4h_{\beta }^{\ \ \rho }V_{\mu
}F_{\nu \lambda }F_{\alpha \rho }\right.  \notag \\
&&\left. \left. +2h_{\mu }^{\ \ \rho }V_{\rho }F_{\nu \lambda }F_{\alpha
\beta }\right) -16y_{3}q_{1}\delta _{3}^{D}V_{\nu }\partial ^{\lbrack \nu
}h_{\ \ \rho }^{\rho ]}-\left( D-2\right) \left( D-1\right) y_{2}^{2}V_{\mu
}V^{\mu }\right\}  \notag \\
&&-4q_{1}y_{2}\delta _{3}^{D}\left( D-2\right) \varepsilon _{\mu \nu \rho
}F^{\mu \nu }\eta ^{\rho }-6q_{2}y_{2}\delta _{5}^{D}\varepsilon _{\mu \nu
\rho \alpha \beta }F^{\mu \nu }F^{\rho \alpha }\eta ^{\beta }  \notag \\
&&+\frac{1}{2}p\left( p+f\right) \left( F^{\mu \nu }F_{\nu \rho }+\frac{1}{4}%
\delta _{\rho }^{\mu }F^{\nu \lambda }F_{\nu \lambda }\right) \left( h^{\rho
\sigma }\partial _{\lbrack \mu }\eta _{\sigma ]}-2\partial _{\lbrack \mu
}h_{\ \ \sigma ]}^{\rho }\eta ^{\sigma }\right)  \notag \\
&&+y_{2}\left[ fA_{0}^{\left( \mathrm{int}\right) }\left( \partial \partial
\Phi ^{\alpha _{0}}\Phi ^{\beta _{0}}\eta _{\alpha _{1}}\right)
+pB_{0}^{\left( \mathrm{int}\right) }\left( \partial \partial \Phi ^{\alpha
_{0}}\Phi ^{\beta _{0}}\eta _{\alpha _{1}}\right) -4D\Lambda \eta \right]
\notag \\
&&+y_{3}\delta _{3}^{D}\left[ fC_{0}^{\left( \mathrm{int}\right) }\left(
\partial \partial \partial \Phi ^{\alpha _{0}}\Phi ^{\beta _{0}}\eta
_{\alpha _{1}}\right) +pD_{0}^{\left( \mathrm{int}\right) }\left( \partial
\partial \partial \Phi ^{\alpha _{0}}\Phi ^{\beta _{0}}\eta _{\alpha
_{1}}\right) \right] +\partial _{\mu }w_{0}^{\mu }.  \label{a69}
\end{eqnarray}%
In (\ref{a69}) $A_{0}^{\left( \mathrm{int}\right) }$, $B_{0}^{\left( \mathrm{%
int}\right) }$, $C_{0}^{\left( \mathrm{int}\right) }$, and $D_{0}^{\left(
\mathrm{int}\right) }$ are linear in their arguments; for instance the
notation $A_{0}^{\left( \mathrm{int}\right) }\left( \partial \partial \Phi
^{\alpha _{0}}\Phi ^{\beta _{0}}\eta _{\alpha _{1}}\right) $ means that each
term from $A_{0}^{\left( \mathrm{int}\right) }$ contains two spacetime
derivatives and is simultaneously quadratic in the fields $\Phi ^{\alpha
_{0}}$ from (\ref{a4a}) and linear in the ghosts $\eta _{\alpha _{1}}$ from (%
\ref{a4b}).

Replacing decomposition (\ref{a67}) into equation (\ref{a65}) and using (\ref%
{a3}), one can assume, without loss of generality, that $b^{\left( \mathrm{%
int}\right) }$ and $n^{\mu }$ stop at antighost number three: $b^{\left(
\mathrm{int}\right) }=\sum\limits_{I=0}^{3}b_{I}^{\left( \mathrm{int}\right)
}$, $n^{\mu }=\sum\limits_{I=0}^{3}n_{I}^{\mu }$. However, it can be shown
in a direct manner (based on the result $H_{3}^{\mathrm{inv}}\left( \delta
|d\right) =0$) that one can take $b_{3}^{\left( \mathrm{int}\right) }=0$, so
we can work with%
\begin{eqnarray}
b^{\left( \mathrm{int}\right) } &=&\sum\limits_{I=0}^{2}b_{I}^{\left(
\mathrm{int}\right) },\qquad \mathrm{agh}\left( b_{I}^{\left( \mathrm{int}%
\right) }\right) =I,  \label{a70n} \\
n^{\mu } &=&\sum\limits_{I=0}^{2}n_{I}^{\mu },\qquad \mathrm{agh}\left(
n_{I}^{\mu }\right) =I.  \label{a71n}
\end{eqnarray}%
The above expansions inserted into equation (\ref{a65}) produce the
equivalent equations%
\begin{eqnarray}
\Delta _{2}^{\left( \mathrm{int}\right) } &=&-2\gamma b_{2}^{\left( \mathrm{%
int}\right) }+\partial _{\mu }n_{2}^{\mu },  \label{new1} \\
\Delta _{1}^{\left( \mathrm{int}\right) } &=&-2\left( \delta b_{2}^{\left(
\mathrm{int}\right) }+\gamma b_{1}^{\left( \mathrm{int}\right) }\right)
+\partial _{\mu }n_{1}^{\mu },  \label{new2} \\
\Delta _{0}^{\left( \mathrm{int}\right) } &=&-2\left( \delta b_{1}^{\left(
\mathrm{int}\right) }+\gamma b_{0}^{\left( \mathrm{int}\right) }\right)
+\partial _{\mu }n_{0}^{\mu }.  \label{new3}
\end{eqnarray}%
At this stage it is useful to make the notations%
\begin{equation}
b_{2}^{\left( \mathrm{int}\right) }=-\frac{1}{2}p\eta ^{\ast }\left[ p\left(
\partial ^{\mu }\eta \right) \eta ^{\nu }h_{\mu \nu }-\left( p+f\right)
V^{\mu }\eta ^{\nu }\partial _{\lbrack \mu }\eta _{\nu ]}\right] +\bar{b}%
_{2}^{\left( \mathrm{int}\right) },  \label{wa2}
\end{equation}%
\begin{eqnarray}
&&b_{1}^{\left( \mathrm{int}\right) }=-\frac{1}{2}p^{2}V^{\ast \mu }\left[
\left( \partial _{\nu }V_{\mu }\right) h_{\ \ \rho }^{\nu }\eta ^{\rho }+%
\frac{1}{2}\left( \partial _{\lbrack \mu }h_{\nu ]\rho }\right) V^{\nu }\eta
^{\rho }\right.  \notag \\
&&\left. -\frac{1}{4}V^{\nu }h_{[\mu }^{\ \ \rho }\left( \partial _{\nu
]}\eta _{\rho }\right) -\frac{1}{4}V^{\nu }\left( \partial _{\rho }\eta
_{\lbrack \mu }\right) h_{\nu ]}^{\ \ \rho }-\frac{3}{4}h_{\mu }^{\ \ \nu
}h_{\nu }^{\ \ \rho }\partial _{\rho }\eta \right]  \notag \\
&&-\frac{1}{4}p\left( p+f\right) V^{\ast \mu }V^{\nu }\left[ \left( \partial
_{\lbrack \mu }h_{\rho ]\nu }+\partial _{\lbrack \nu }h_{\rho ]\mu }\right)
\eta ^{\rho }-h_{\mu }^{\ \ \rho }\partial _{\nu }\eta _{\rho }\right.
\notag \\
&&\left. -h_{\nu }^{\ \ \rho }\partial _{\mu }\eta _{\rho }\right] +\frac{1}{%
2}y_{3}\delta _{3}^{D}\varepsilon _{\mu \nu \rho }V^{\ast \mu }\left[ fh_{\
\ \lambda }^{\nu }\partial ^{\lbrack \rho }\eta ^{\lambda ]}+\left(
2p+f\right) \eta _{\lambda }\partial ^{\lbrack \nu }h^{\rho ]\lambda }\right]
\notag \\
&&-\frac{1}{2}py_{2}V^{\ast \mu }\left[ \left( D-2\right) h_{\mu \nu }\eta
^{\nu }-V_{\mu }\eta \right] +\frac{1}{2}y_{2}h^{\ast \mu \nu }\left[
f\left( h_{\mu \nu }\eta +2V_{\mu }\eta _{\nu }\right) \right.  \notag \\
&&\left. -2\left( p+f\right) \sigma _{\mu \nu }V^{\rho }\eta _{\rho }\right]
-\frac{8}{D-2}y_{3}q_{1}\delta _{3}^{D}h^{\ast }\eta +\bar{b}_{1}^{\left(
\mathrm{int}\right) },  \label{notb1}
\end{eqnarray}%
\begin{eqnarray}
&&b_{0}^{\left( \mathrm{int}\right) }=-\frac{p^{2}}{16}\left[ V_{\rho
}\left( \left( \partial ^{\lbrack \mu }h^{\nu ]\rho }\right) \left( \partial
_{\lbrack \mu }h_{\nu ]\lambda }\right) V^{\lambda }-2\left( \partial
^{\lbrack \mu }h^{\nu ]\rho }\right) h_{\lambda \lbrack \mu }\left( \partial
_{\nu ]}V^{\lambda }\right) \right) \right.  \notag \\
&&+h_{\rho }^{\ \ [\mu }\left( \partial ^{\nu ]}V^{\rho }\right) h_{\lambda
\lbrack \mu }\left( \partial _{\nu ]}V^{\lambda }\right) +F^{\mu \nu }h_{\ \
\lambda }^{\rho }\left( h_{\ \ [\mu }^{\lambda }\left( \partial _{\nu
]}V_{\rho }\right) -\left( \partial _{\lbrack \mu }h_{\ \ \nu ]}^{\lambda
}\right) V_{\rho }\right)  \notag \\
&&\left. +F^{\mu \nu }h_{\ \ [\mu }^{\rho }\left( \partial _{\nu ]}h_{\rho
}^{\ \ \lambda }\right) V_{\lambda }\right] -\frac{1}{2}p^{2}F^{\mu \nu }%
\left[ F_{\mu \rho }h_{\nu }^{\ \ \lambda }h_{\lambda }^{\ \ \rho }+\frac{1}{%
16}F_{\mu \nu }\left( h^{2}-2h^{\rho \lambda }h_{\rho \lambda }\right)
\right.  \notag \\
&&-h_{\nu }^{\ \ \rho }\left( \left( \partial _{\lbrack \mu }h_{\rho ]}^{\ \
\lambda }\right) V_{\lambda }-h_{[\mu }^{\ \ \lambda }\left( \partial _{\rho
]}V_{\lambda }\right) \right) +\frac{1}{2}\left( F^{\rho \lambda }h_{\mu
\rho }h_{\nu \lambda }-F_{\mu \rho }h_{\ \ \nu }^{\rho }h\right)  \notag \\
&&\left. +\frac{1}{4}\left( \left( \partial _{\lbrack \mu }h_{\nu ]}^{\ \
\rho }\right) V_{\rho }-h_{[\mu }^{\ \ \rho }\left( \partial _{\nu ]}V_{\rho
}\right) \right) h\right] -\frac{1}{8}p\left( p+f\right) \left( F^{\mu \nu
}F_{\nu \rho }\right.  \notag \\
&&\left. +\frac{1}{4}\delta _{\rho }^{\mu }F^{\nu \lambda }F_{\nu \lambda
}\right) h_{\mu \sigma }h^{\sigma \rho }-\frac{1}{2}q_{1}p\delta
_{3}^{D}\varepsilon ^{\mu \nu \lambda }\left( hV_{\mu }F_{\nu \lambda
}-2h_{\lambda }^{\ \ \alpha }V_{\mu }F_{\nu \alpha }\right.  \notag \\
&&\left. +h_{\mu }^{\ \ \alpha }V_{\alpha }F_{\nu \lambda }\right) -\frac{1}{%
2}q_{2}p\delta _{5}^{D}\varepsilon ^{\mu \nu \lambda \alpha \beta }\left(
hV_{\mu }F_{\nu \lambda }F_{\alpha \beta }-4h_{\beta }^{\ \ \rho }V_{\mu
}F_{\nu \lambda }F_{\alpha \rho }\right.  \notag \\
&&\left. +2h_{\mu }^{\ \ \rho }V_{\rho }F_{\nu \lambda }F_{\alpha \beta
}\right) +8y_{3}q_{1}\delta _{3}^{D}V_{\nu }\partial ^{\lbrack \nu }h_{\ \
\rho }^{\rho ]}+\frac{1}{2}\left( D-2\right) \left( D-1\right)
y_{2}^{2}V_{\mu }V^{\mu }  \notag \\
&&+\bar{b}_{0}^{\left( \mathrm{int}\right) }.  \label{notb0}
\end{eqnarray}%
Using the above notations and recalling the expressions (\ref{a671})--(\ref%
{a69}) of $\Delta _{I}^{\left( \mathrm{int}\right) }$, equations (\ref{new1}%
)--(\ref{new3}) (equivalent to (\ref{a65})) become
\begin{eqnarray}
\gamma \bar{b}_{2}^{\left( \mathrm{int}\right) } &=&\partial _{\mu }\rho
_{2}^{\mu },  \label{wa1n} \\
\delta \bar{b}_{2}^{\left( \mathrm{int}\right) }+\gamma \bar{b}_{1}^{\left(
\mathrm{int}\right) } &=&\partial _{\mu }\rho _{1}^{\mu }+\frac{1}{2}\chi
_{1},  \label{wa1o} \\
\delta \bar{b}_{1}^{\left( \mathrm{int}\right) }+\gamma \bar{b}_{0}^{\left(
\mathrm{int}\right) } &=&\partial _{\mu }\rho _{0}^{\mu }+\frac{1}{2}\chi
_{0},  \label{wa1p}
\end{eqnarray}%
where
\begin{equation}
\rho _{I}^{\mu }=\frac{1}{2}\left( w_{I}^{\mu }-n_{I}^{\mu }\right) ,\qquad
I=\overline{0,2},  \label{rho}
\end{equation}%
\begin{eqnarray}
\chi _{1} &=&V^{\ast \mu }\left\{ -p\left( p+f\right) F_{\mu \nu }\eta
_{\rho }\partial ^{\lbrack \rho }\eta ^{\nu ]}\right.  \notag \\
&&+\left( 2p+f\right) \left[ y_{3}\delta _{3}^{D}\varepsilon _{\mu \nu \rho
}\left( \partial ^{\lbrack \nu }\eta ^{\lambda ]}\right) \partial ^{\lbrack
\rho }\eta ^{\tau ]}\sigma _{\lambda \tau }\right.  \notag \\
&&\left. \left. +y_{2}\left( D-2\right) \left( \partial _{\lbrack \mu }\eta
_{\nu ]}\right) \eta ^{\nu }\right] \right\} ,  \label{chi1}
\end{eqnarray}%
\begin{eqnarray}
&&\chi _{0}=\delta \left\{ y_{3}\delta _{3}^{D}\varepsilon _{\mu \nu \rho
}V^{\ast \mu }\left[ fh_{\ \ \lambda }^{\nu }\partial ^{\lbrack \rho }\eta
^{\lambda ]}+\left( 2p+f\right) \eta _{\lambda }\partial ^{\lbrack \nu
}h^{\rho ]\lambda }\right] \right.  \notag \\
&&-py_{2}V^{\ast \mu }\left[ \left( D-2\right) h_{\mu \nu }\eta ^{\nu
}-V_{\mu }\eta \right] +y_{2}h^{\ast \mu \nu }\left[ f\left( h_{\mu \nu
}\eta +2V_{\mu }\eta _{\nu }\right) \right.  \notag \\
&&\left. \left. -2\left( p+f\right) \sigma _{\mu \nu }V^{\rho }\eta _{\rho }
\right] \right\} -4q_{1}y_{2}\delta _{3}^{D}\left( D-2\right) \varepsilon
_{\mu \nu \rho }F^{\mu \nu }\eta ^{\rho }  \notag \\
&&-6q_{2}y_{2}\delta _{5}^{D}\varepsilon _{\mu \nu \rho \alpha \beta }F^{\mu
\nu }F^{\rho \alpha }\eta ^{\beta }+\frac{1}{2}p\left( p+f\right) \left(
F^{\mu \nu }F_{\nu \rho }\right.  \notag \\
&&\left. +\frac{1}{4}\delta _{\rho }^{\mu }F^{\nu \lambda }F_{\nu \lambda
}\right) \left( h^{\rho \sigma }\partial _{\lbrack \mu }\eta _{\sigma
]}-2\partial _{\lbrack \mu }h_{\ \ \sigma ]}^{\rho }\eta ^{\sigma }\right)
\notag \\
&&+y_{2}\left[ fA_{0}^{\left( \mathrm{int}\right) }\left( \partial \partial
\Phi ^{\alpha _{0}}\Phi ^{\beta _{0}}\eta _{\alpha _{1}}\right)
+pB_{0}^{\left( \mathrm{int}\right) }\left( \partial \partial \Phi ^{\alpha
_{0}}\Phi ^{\beta _{0}}\eta _{\alpha _{1}}\right) -4D\Lambda \eta \right]
\notag \\
&&+y_{3}\delta _{3}^{D}\left[ fC_{0}^{\left( \mathrm{int}\right) }\left(
\partial \partial \partial \Phi ^{\alpha _{0}}\Phi ^{\beta _{0}}\eta
_{\alpha _{1}}\right) +pD_{0}^{\left( \mathrm{int}\right) }\left( \partial
\partial \partial \Phi ^{\alpha _{0}}\Phi ^{\beta _{0}}\eta _{\alpha
_{1}}\right) \right] .  \label{chi0}
\end{eqnarray}%
One can replace (\ref{wa1n}) with
\begin{equation}
\gamma \bar{b}_{2}^{\left( \mathrm{int}\right) }=0,  \label{wa1nn}
\end{equation}%
such that (\ref{a65}) is in fact equivalent to (\ref{wa1nn}) and (\ref{wa1o}%
)--(\ref{wa1p}). So far, we have shown that the second-order deformation of
the solution to the master equation, (\ref{a64}), is completely known once
we manage to solve equations (\ref{wa1nn}) and (\ref{wa1o})--(\ref{wa1p}).
This is our next concern.

From (\ref{wa1o}) we obtain a necessary condition for the existence of $\bar{%
b}_{2}^{\left( \mathrm{int}\right) }$ and $\bar{b}_{1}^{\left( \mathrm{int}%
\right) }$, namely%
\begin{equation}
\chi _{1}=\delta \varphi _{2}+\gamma \omega _{1}+\partial _{\mu }l_{1}^{\mu
},  \label{a77}
\end{equation}%
where $\mathrm{agh}\left( \varphi _{2}\right) =2=\mathrm{pgh}\left( \varphi
_{2}\right) $, $\mathrm{agh}\left( \omega _{1}\right) =1=\mathrm{pgh}\left(
\omega _{1}\right) $, $\mathrm{agh}\left( l_{1}^{\mu }\right) =1$, $\mathrm{%
pgh}\left( l_{1}^{\mu }\right) =2$. It is essential to remark that all the
functions $\varphi _{2}$, $\omega _{1}$, and $l_{1}^{\mu }$ must be local
since otherwise we cannot obtain local second-order deformations from (\ref%
{wa1o}). Assuming (\ref{a77}) holds, we act with $\delta $ on it and use its
nilpotency and its anticommutation with $\gamma $, which yields
\begin{equation}
\delta \chi _{1}=\gamma \left( -\delta \omega _{1}\right) +\partial _{\mu
}\left( \delta l_{1}^{\mu }\right) .  \label{a78}
\end{equation}%
Without entering technical details, we mention that the validity of (\ref%
{a78}) can be checked by means of standard cohomological arguments. In fact,
after direct manipulations of (\ref{chi1}), it can be shown that (\ref{a78})
(and thus also (\ref{a77})) requires that the following conditions are
simultaneously satisfied:
\begin{eqnarray}
F^{\mu \nu }F_{\nu \rho }+\frac{1}{4}\delta _{\rho }^{\mu }F_{\nu \lambda
}F^{\nu \lambda } &=&\delta \Omega _{\rho }^{\mu },  \label{a82} \\
F^{\theta \mu } &=&\delta \bar{\Omega}^{\theta \mu },  \label{a82a} \\
\partial _{\lbrack \mu }h_{\lambda ]}^{\ \ \theta } &=&\delta \Omega _{\mu
\lambda }^{\theta },  \label{a82b} \\
\left( \partial _{\lbrack \theta }h_{\nu ]}^{\ \ \theta }\right) \partial
^{\lbrack \mu }h_{\ \ \mu }^{\nu ]}-\left( \partial ^{\lbrack \nu }h^{\theta
]\mu }\right) \partial _{\nu }h_{\theta \mu } &=&\delta \Omega .
\label{a82c}
\end{eqnarray}%
All the quantities denoted by $\Omega $ or $\bar{\Omega}$ must be local;
their locality is essential in obtaining local deformations, which is one of
the main working hypotheses of our paper. One can explicitly reveal locality
obstructions to each of these conditions. For instance, assuming that
equation (\ref{a82}) is satisfied for some local $\Omega _{\rho }^{\mu }$
and taking its divergence, it follows that the relation
\begin{equation}
\partial _{\mu }\left( F^{\mu \nu }F_{\nu \rho }+\frac{1}{4}\delta _{\rho
}^{\mu }F_{\nu \lambda }F^{\nu \lambda }\right) =\delta \left( \partial
_{\mu }\Omega _{\rho }^{\mu }\right)  \label{a83}
\end{equation}%
should also take place. On the other hand, it is easy to see that%
\begin{equation}
\partial _{\mu }\left( F^{\mu \nu }F_{\nu \rho }+\frac{1}{4}\delta _{\rho
}^{\mu }F_{\nu \lambda }F^{\nu \lambda }\right) =\delta \left( -V^{\ast \nu
}F_{\nu \rho }\right) .  \label{a84}
\end{equation}%
Since $-V^{\ast \nu }F_{\nu \rho }$ obviously is not a divergence of a local
function, equation (\ref{a83}) cannot hold for some local $\Omega _{\rho
}^{\mu }$, so neither does (\ref{a82}). Acting in a similar manner with
respect to equation (\ref{a82a}), we infer $\partial _{\mu }F^{\theta \mu
}=\delta V^{\ast \theta }\neq \delta \left( \partial _{\mu }\bar{\Omega}%
^{\theta \mu }\right) $, such that (\ref{a82a}) cannot be satisfied for some
local $\bar{\Omega}^{\theta \mu }$. Related to (\ref{a82b}), if we apply $%
\partial ^{\mu }$ on it and then take its trace, we obtain $\partial ^{\mu
}\partial _{\lbrack \mu }h_{\lambda ]}^{\ \ \lambda }=\delta \left( \frac{%
h^{\ast }}{D-2}\right) \neq \delta \left( \partial ^{\mu }\Omega _{\mu
\lambda }^{\lambda }\right) $, and hence (\ref{a82b}) is not valid for some
local $\Omega _{\mu \lambda }^{\theta }$. Concerning equation (\ref{a82c}),
it can be shown directly that its left-hand side reads as $\delta \left(
-h_{\mu \nu }h^{\ast \mu \nu }\right) +\partial _{\mu }u^{\mu }$, with $%
\partial _{\mu }u^{\mu }\neq 0$ and $u^{\mu }\neq \delta u_{1}^{\mu }$ for
some local $u_{1}^{\mu }$, so (\ref{a82c}) also fails to be true. Combining
these last results, it follows that (\ref{a78}) (and hence also (\ref{a77}))
cannot hold locally unless $\chi _{1}=0$, which yields%
\begin{eqnarray}
p\left( p+f\right) &=&0,  \label{a85a} \\
\left( 2p+f\right) y_{3}\delta _{3}^{D} &=&0,  \label{a85b} \\
\left( 2p+f\right) y_{2} &=&0.  \label{a85c}
\end{eqnarray}%
There are three relevant solutions to the above equations\footnote{\label%
{relevant}By `relevant solution' we mean that the resulting
deformations lead to a maximum number of consistent couplings and
gauge
symmetries. For instance, another possible solution to (\ref{a86a})--(\ref%
{a86c}) is $p=0$, $f\neq 0$, $y_{2}=0$, $y_{3}\delta _{3}^{D}=0$. This case
is not relevant since it would mean to allow the Einstein-Hilbert
selfinteractions of the graviton, but forbid: (i) the standard couplings
graviton-photon and (ii) the diffeomorphism sector of the vector field gauge
symmetries prescribed by General Relativity.}
\begin{eqnarray}
\mathrm{Case}\ \mathrm{I} &:&p=-f\neq 0,\qquad y_{2}=0=y_{3}\delta
_{3}^{D},\qquad D>2,  \label{a86a} \\
\mathrm{Case}\ \mathrm{II} &:&p=f=0,\qquad D=3,  \label{a86b} \\
\mathrm{Case}\ \mathrm{III} &:&p=f=0,\qquad D>3,  \label{a86c}
\end{eqnarray}%
which require an individual treatment.

\subsubsection{Case I --- General Relativity\label{caseI}}

According to (\ref{a86a}), the first-order deformation (\ref{s1}) is
parameterized in this situation by four real constants, namely, $f$, $%
\Lambda $, $q_{1}\delta _{3}^{D}$, and $q_{2}\delta _{5}^{D}$. For the sake
of simplicity we set $f=1$, so $p=-1$, such that the $S_{1}$ (see (\ref{S1pf}%
) with the components (\ref{a40}), (\ref{a41}), and (\ref{a0pf}) plus (\ref%
{a60int})) takes the concrete form%
\begin{eqnarray}
&&S_{1}^{\left( \mathrm{I}\right) }=S_{1}^{(\mathrm{PF})}+S_{1}^{\left(
\mathrm{int}\right) }  \notag \\
&\equiv &\int d^{D}x\left\{ \frac{1}{2}\eta ^{\ast \mu }\eta ^{\nu }\partial
_{\lbrack \mu }\eta _{\nu ]}+h^{\ast \mu \rho }\left[ \left( \partial _{\rho
}\eta ^{\nu }\right) h_{\mu \nu }-\eta ^{\nu }\partial _{\lbrack \mu }h_{\nu
]\rho }\right] \right.  \notag \\
&&\left. +a_{0}^{\left( \mathrm{EH-cubic}\right) }-2\Lambda h\right\} +\int
d^{D}x\left\{ -\eta ^{\ast }\eta _{\mu }\partial ^{\mu }\eta \right.  \notag
\\
&&+\frac{1}{2}V^{\ast \mu }\left[ V^{\nu }\partial _{\lbrack \mu }\eta _{\nu
]}+2\left( \partial _{\nu }V_{\mu }\right) \eta ^{\nu }-h_{\mu \nu }\partial
^{\nu }\eta \right]  \notag \\
&&-\frac{1}{8}F^{\mu \nu }\left[ 2\partial _{\lbrack \mu }\left( h_{\nu
]\rho }V^{\rho }\right) +F_{\mu \nu }h-4F_{\mu \rho }h_{\nu }^{\ \rho }%
\right]  \notag \\
&&\left. +q_{1}\delta _{3}^{D}\varepsilon ^{\mu \nu \lambda }V_{\mu }F_{\nu
\lambda }+q_{2}\delta _{5}^{D}\varepsilon ^{\mu \nu \lambda \alpha \beta
}V_{\mu }F_{\nu \lambda }F_{\alpha \beta }\right\} .  \label{S1I}
\end{eqnarray}%
Replacing (\ref{a86a}) into (\ref{chi1}) and (\ref{chi0}), we find that
\begin{equation}
\chi _{1}=0,\qquad \chi _{0}=0,  \label{chicaseI}
\end{equation}%
such that equations (\ref{wa1nn}) and (\ref{wa1o})--(\ref{wa1p}) become%
\begin{eqnarray}
\gamma \bar{b}_{2}^{\left( \mathrm{int}\right) } &=&0,  \label{wa1nncaseI} \\
\delta \bar{b}_{2}^{\left( \mathrm{int}\right) }+\gamma \bar{b}_{1}^{\left(
\mathrm{int}\right) } &=&\partial _{\mu }\rho _{1}^{\mu },  \label{wa1ocaseI}
\\
\delta \bar{b}_{1}^{\left( \mathrm{int}\right) }+\gamma \bar{b}_{0}^{\left(
\mathrm{int}\right) } &=&\partial _{\mu }\rho _{0}^{\mu }.  \label{wa1pcaseI}
\end{eqnarray}%
These equations have already been considered in Section \ref{deformareaI} at
the construction of the first-order deformation, so their solutions can be
absorbed into $S_{1}^{\left( \mathrm{int}\right) }$ from (\ref{S1I}) by a
suitable redefinition of the constants $p$, $q_{1}$, and $q_{2}$. In
conclusion, we can work with%
\begin{equation}
\bar{b}_{2}^{\left( \mathrm{int}\right) }=0,\qquad \bar{b}_{1}^{\left(
\mathrm{int}\right) }=0,\qquad \bar{b}_{0}^{\left( \mathrm{int}\right) }=0.
\label{b2b1b0caseI}
\end{equation}%
Inserting the previous results together with (\ref{a86a}) for $f=1$ in (\ref%
{wa2}), (\ref{notb1}), and (\ref{notb0}) and then the resulting expressions
in (\ref{a70n}), we complete the interacting component $S_{2}^{\left(
\mathrm{int}\right) }$ from the second-order deformation of the solution to
the master equation, in agreement with notation (\ref{notdeltas2}).
Particularizing (\ref{a64b}) and (\ref{a64c1})--(\ref{a64c3}) to the case (%
\ref{a86a}) for $f=1$, we also infer $S_{2}^{\left( \mathrm{PF}\right) }$
with the help of relation (\ref{a64b}). Putting together these expressions
of $S_{2}^{\left( \mathrm{int}\right) }$ and $S_{2}^{\left( \mathrm{PF}%
\right) }$ via formula (\ref{a64}), we can state that the full second-order
deformation to the master equation in case I reads as%
\begin{eqnarray}
&&S_{2}^{\left( \mathrm{I}\right) }=S_{2}^{(\mathrm{PF})}+S_{2}^{\left(
\mathrm{int}\right) }  \notag \\
&\equiv &\left[ S_{2}^{(\mathrm{EH-quartic})}+\Lambda \int d^{D}x\left(
h^{\mu \nu }h_{\mu \nu }-\frac{1}{2}h^{2}\right) \right] -\frac{1}{2}\int
d^{D}x\left\{ \eta ^{\ast }\left( \partial ^{\mu }\eta \right) \eta ^{\nu
}h_{\mu \nu }\right.  \notag \\
&&+V^{\ast \mu }\left[ \left( \partial _{\nu }V_{\mu }\right) h_{\ \ \rho
}^{\nu }\eta ^{\rho }+\frac{1}{2}\left( \partial _{\lbrack \mu }h_{\nu ]\rho
}\right) V^{\nu }\eta ^{\rho }-\frac{1}{4}V^{\nu }h_{[\mu }^{\ \ \rho
}\left( \partial _{\nu ]}\eta _{\rho }\right) \right.  \notag \\
&&\left. -\frac{1}{4}V^{\nu }\left( \partial _{\rho }\eta _{\lbrack \mu
}\right) h_{\nu ]}^{\ \ \rho }-\frac{3}{4}h_{\mu }^{\ \ \nu }h_{\nu }^{\ \
\rho }\partial _{\rho }\eta \right] +\frac{1}{8}\left[ F^{\mu \nu }h_{\ \
\lambda }^{\rho }\left( h_{\ \ [\mu }^{\lambda }\left( \partial _{\nu
]}V_{\rho }\right) \right. \right.  \notag \\
&&\left. -\left( \partial _{\lbrack \mu }h_{\ \ \nu ]}^{\lambda }\right)
V_{\rho }\right) +F^{\mu \nu }h_{\ \ [\mu }^{\rho }\left( \partial _{\nu
]}h_{\rho }^{\ \ \lambda }\right) V_{\lambda }+V_{\rho }\left( \left(
\partial ^{\lbrack \mu }h^{\nu ]\rho }\right) \left( \partial _{\lbrack \mu
}h_{\nu ]\lambda }\right) V^{\lambda }\right.  \notag \\
&&\left. \left. -2\left( \partial ^{\lbrack \mu }h^{\nu ]\rho }\right)
h_{\lambda \lbrack \mu }\left( \partial _{\nu ]}V^{\lambda }\right) \right)
+h_{\rho }^{\ \ [\mu }\left( \partial ^{\nu ]}V^{\rho }\right) h_{\lambda
\lbrack \mu }\left( \partial _{\nu ]}V^{\lambda }\right) \right]  \notag \\
&&+F^{\mu \nu }\left[ F_{\mu \rho }h_{\nu }^{\ \ \lambda }h_{\lambda }^{\ \
\rho }+\frac{1}{2}\left( F^{\rho \lambda }h_{\mu \rho }h_{\nu \lambda
}-F_{\mu \rho }h_{\ \ \nu }^{\rho }h\right) \right.  \notag \\
&&+\frac{1}{16}F_{\mu \nu }\left( h^{2}-2h^{\rho \lambda }h_{\rho \lambda
}\right) -h_{\nu }^{\ \ \rho }\left( \left( \partial _{\lbrack \mu }h_{\rho
]}^{\ \ \lambda }\right) V_{\lambda }-h_{[\mu }^{\ \ \lambda }\left(
\partial _{\rho ]}V_{\lambda }\right) \right)  \notag \\
&&\left. +\frac{1}{4}\left( \left( \partial _{\lbrack \mu }h_{\nu ]}^{\ \
\rho }\right) V_{\rho }-h_{[\mu }^{\ \ \rho }\left( \partial _{\nu ]}V_{\rho
}\right) \right) h\right] -q_{1}\delta _{3}^{D}\varepsilon ^{\mu \nu \lambda
}\left( hV_{\mu }F_{\nu \lambda }\right.  \notag \\
&&\left. -2h_{\lambda }^{\ \ \alpha }V_{\mu }F_{\nu \alpha }+h_{\mu }^{\ \
\alpha }V_{\alpha }F_{\nu \lambda }\right) -q_{2}\delta _{5}^{D}\varepsilon
^{\mu \nu \lambda \alpha \beta }\left( hV_{\mu }F_{\nu \lambda }F_{\alpha
\beta }\right.  \notag \\
&&\left. \left. -4h_{\beta }^{\ \ \rho }V_{\mu }F_{\nu \lambda }F_{\alpha
\rho }+2h_{\mu }^{\ \ \rho }V_{\rho }F_{\nu \lambda }F_{\alpha \beta
}\right) \right\} .  \label{S2I}
\end{eqnarray}%
The deformation procedure goes on indefinitely in the sense that it produces
an infinite number of nontrivial higher-order components of the deformed
solution to the master equation%
\begin{equation}
S_{n}^{\left( \mathrm{I}\right) }\neq 0,\qquad \mathrm{for\ all}\qquad n>0.
\label{SkcaseI}
\end{equation}%
Nevertheless, we will see in Section \ref{analysiscaseI} that the first two
deformations derived so far for case I are enough in order to describe the
overall deformed theory at all orders in the coupling constant, which turns
out to describe nothing but the standard graviton-vector interactions from
General Relativity.

\subsubsection{Case II --- new solutions in $D=3$\label{caseII}}

In this situation we substitute (\ref{a86b}) into (\ref{chi1}) and (\ref%
{chi0}) and obtain that\footnote{%
Note that in $D=3$ we have $q_{2}\delta _{5}^{D}=0$.}
\begin{equation}
\chi _{1}=0,\qquad \chi _{0}=-4y_{2}\left( q_{1}\varepsilon _{\mu \nu \rho
}F^{\mu \nu }\eta ^{\rho }+3\Lambda \eta \right) .  \label{chicaseII}
\end{equation}%
Thus, from (\ref{wa1p}) we obtain a necessary condition for the existence of
$\bar{b}_{1}^{\left( \mathrm{int}\right) }$ and $\bar{b}_{0}^{\left( \mathrm{%
int}\right) }$, namely%
\begin{equation}
\chi _{0}=\delta \varphi _{1}+\gamma \omega _{0}+\partial _{\mu }l_{0}^{\mu
},  \label{condchi0}
\end{equation}%
where $\mathrm{agh}\left( \varphi _{1}\right) =1=\mathrm{pgh}\left( \varphi
_{1}\right) $, $\mathrm{agh}\left( \omega _{0}\right) =0=\mathrm{pgh}\left(
\omega _{0}\right) $, $\mathrm{agh}\left( l_{0}^{\mu }\right) =0$, $\mathrm{%
pgh}\left( l_{0}^{\mu }\right) =1$. We insist that all the quantities $%
\varphi _{1}$, $\omega _{0}$, and $l_{0}^{\mu }$ from (\ref{condchi0}) must
be local in order to render a local second-order deformation via (\ref{wa1p}%
). This is the second place where we analyze the possible obstructions in
finding local deformations. It is clear from (\ref{chicaseII}) that $\chi
_{0}$ is a nontrivial element from $H^{1}\left( \gamma \right) $ of
antighost number zero, $\gamma \chi _{0}=0$, since it is written as $\chi
_{0}=\alpha _{0M}\left( F_{\mu \nu }\right) e^{1M}$, where $\alpha _{0M}$
are invariant polynomials not depending on the antifields and $e^{1M}$ are
the elements of a basis in the space of polynomials with pure ghost number
one in $\eta _{\mu }$ and $\eta $. The latter term from the right-hand side
of (\ref{chicaseII}) is derivative-free while the non-vanishing actions of $%
\delta $ and $\gamma $ contain at least one derivative, so it cannot be
written as in (\ref{condchi0}) and, as a consequence, we must require $%
y_{2}\Lambda =0$. (From the latter definition in (\ref{a11}) we have that $%
\gamma (\partial ^{\mu }V_{\mu })=\square \eta $, so we can indeed write $%
\eta =\gamma (\square ^{-1}\partial ^{\mu }V_{\mu })$. But $\square
^{-1}\partial ^{\mu }V_{\mu }$ is not local, so this solution must be
discarded.) Regarding the former term, proportional with $\varepsilon _{\mu
\nu \rho }F^{\mu \nu }\eta ^{\rho }$, since $\mathrm{agh}\left( \varphi
_{1}\right) =1$, it follows that $\varphi _{1}$ is linear in the antifields $%
\Phi _{\alpha _{0}}^{\ast }=(h^{\ast \mu \nu },V^{\ast \mu })$. On behalf of
definitions (\ref{a7}), it would produce in (\ref{condchi0}) terms with two
spacetime derivatives. But $\varepsilon _{\mu \nu \rho }F^{\mu \nu }\eta
^{\rho }$ contains only pieces with at most one derivative, so the locality
assumption requires $\varphi _{1}=0$ in (\ref{condchi0}), such that this
becomes
\begin{equation}
-4y_{2}q_{1}\varepsilon _{\mu \nu \rho }F^{\mu \nu }\eta ^{\rho }=\gamma
\omega _{0}+\partial _{\mu }l_{0}^{\mu }.  \label{conchi0a}
\end{equation}%
From definitions (\ref{a11}) it is clear now that (\ref{conchi0a}) cannot
hold for some local $\omega _{0}$ and $l_{0}^{\mu }$. By virtue of the above
discussion we must impose $\chi _{0}=0$, which is equivalent with the
supplementary conditions%
\begin{equation}
y_{2}q_{1}=0,\qquad y_{2}\Lambda =0,  \label{a85b1}
\end{equation}%
displaying two relevant solutions%
\begin{eqnarray}
y_{2} &=&0,  \label{a86b1} \\
q_{1} &=&0=\Lambda .  \label{a86b2}
\end{eqnarray}%
Thus, the second case admits two subcases, deserving separate analyses.

\textbf{Subcase II.1} results from (\ref{a86b}) and (\ref{a86b1}), so it
corresponds to the choice%
\begin{equation}
D=3,\qquad p=f=q_{2}\delta _{5}^{D}=y_{2}=0.  \label{caseII.1}
\end{equation}%
We observe that the deformations lie in three spacetime dimensions and are
parameterized by three constants, namely $\Lambda $, $y_{3}$, and $q_{1}$.
Under these circumstances, the first-order deformation $S_{1}$ (see (\ref%
{S1pf}) with the components (\ref{a40}), (\ref{a41}), and (\ref{a0pf}) plus (%
\ref{a60int}), all particularized to (\ref{caseII.1})) is expressed by%
\begin{eqnarray}
&&S_{1}^{\left( \mathrm{II.1}\right) }=S_{1}^{(\mathrm{PF})}+S_{1}^{\left(
\mathrm{int}\right) }\equiv -2\Lambda \int d^{3}x\,h  \notag \\
&&+\int d^{3}x\,\varepsilon _{\mu \nu \rho }\left[ y_{3}\left( V^{\ast \mu
}\partial ^{\lbrack \nu }\eta ^{\rho ]}+F^{\lambda \mu }\partial ^{\lbrack
\nu }h_{\ \ \lambda }^{\rho ]}\right) +q_{1}V^{\mu }F^{\nu \rho }\right] .
\label{S1II.1}
\end{eqnarray}%
Substituting relations (\ref{caseII.1}) into (\ref{chi1}) and (\ref{chi0}),
we find that
\begin{equation}
\chi _{1}=0,\qquad \chi _{0}=0,  \label{chicaseII.1}
\end{equation}%
so the discussion from subsection \ref{caseI} applies here as well and we
can take%
\begin{equation}
\bar{b}_{2}^{\left( \mathrm{int}\right) }=0,\qquad \bar{b}_{1}^{\left(
\mathrm{int}\right) }=0,\qquad \bar{b}_{0}^{\left( \mathrm{int}\right) }=0
\label{bcaseII.1}
\end{equation}%
in (\ref{wa2}), (\ref{notb1}), and (\ref{notb0}). Consequently, with the
help of formulas (\ref{a64}), (\ref{a64e}) (with the components (\ref{a64c1}%
)--(\ref{a64c3})), (\ref{notdeltas2}), and (\ref{a70n}) (with the components
(\ref{wa2})--(\ref{notb0})) written in the presence of conditions (\ref%
{caseII.1}) and (\ref{bcaseII.1}) we determine the second-order deformation
in the form%
\begin{eqnarray}
&&S_{2}^{\left( \mathrm{II.1}\right) }=S_{2}^{(\mathrm{PF})}+S_{2}^{\left(
\mathrm{int}\right) }\equiv y_{3}^{2}\int d^{3}x\left( \partial ^{\lbrack
\nu }h^{\rho ]\lambda }\right) \partial _{\lbrack \nu }h_{\rho ]\lambda }
\notag \\
&&+8y_{3}q_{1}\int d^{3}x\left( -h^{\ast }\eta +V_{\nu }\partial ^{\lbrack
\nu }h_{\ \ \rho }^{\rho ]}\right) .  \label{S2caseII.1}
\end{eqnarray}

Next, we approach the consistency of $S_{2}^{\left( \mathrm{II.1}\right) }$,
i.e. we solve the equation introducing the third-order deformation of the
solution to the master equation%
\begin{equation}
\left( S_{1}^{\left( \mathrm{II.1}\right) },S_{2}^{\left( \mathrm{II.1}%
\right) }\right) +sS_{3}^{\left( \mathrm{II.1}\right) }=0.  \label{eqS3}
\end{equation}%
By direct computation we obtain%
\begin{equation}
\left( S_{1}^{\left( \mathrm{II.1}\right) },S_{2}^{\left( \mathrm{II.1}%
\right) }\right) =s\left( 4y_{3}^{2}q_{1}\int d^{3}x\,\varepsilon _{\mu \nu
\rho }h_{\ \ \lambda }^{\mu }\partial ^{\lbrack \nu }h_{\ \ \lambda }^{\rho
]}\right) +48\Lambda y_{3}q_{1}\int d^{3}x\,\eta .  \label{S1S2caseII.1}
\end{equation}%
Substituting the last result into (\ref{eqS3}) we arrive at%
\begin{equation}
s\left( S_{3}^{\left( \mathrm{II.1}\right) }+4y_{3}^{2}q_{1}\int
d^{3}x\,\varepsilon _{\mu \nu \rho }h_{\ \ \lambda }^{\mu }\partial
^{\lbrack \nu }h_{\ \ \lambda }^{\rho ]}\right) +48\Lambda y_{3}q_{1}\int
d^{3}x\,\eta =0.  \label{II.1S3}
\end{equation}%
The last equation possesses local solutions if and only if the integrand of
the last term from the left-hand side of (\ref{II.1S3}) is written in a $s$%
-exact modulo $d$ form from local functions. We discussed a similar term in
the beginning of Section \ref{caseII} (see the second term on the right-hand
side of (\ref{chicaseII}) and equation (\ref{condchi0})) and concluded that
it cannot be written in a $s$-exact modulo $d$ form from local functions
until its coefficient vanishes. Then, we can state that (\ref{II.1S3}) holds
if and only if
\begin{equation}
\Lambda y_{3}q_{1}=0.  \label{a85b11}
\end{equation}%
The relevant solutions to the above equation are\footnote{%
The solution $y_{3}=0$ and $\Lambda q_{1}\neq 0$ yields no
couplings: the original gauge transformations (\ref{a2}) are
maintained and two
gauge-invariant terms are added to the starting Lagrangian (\ref{a1}): $%
-2k\Lambda h$ and $kq_{1}\delta _{3}^{D}\varepsilon ^{\mu \nu \rho }V_{\mu
}F_{\nu \rho }$.}%
\begin{eqnarray}
y_{3} &\neq &0,\qquad \Lambda \neq 0,\qquad q_{1}=0,  \label{a86b11} \\
y_{3} &\neq &0,\qquad q_{1}\neq 0,\qquad \Lambda =0.  \label{a86b12}
\end{eqnarray}%
Thus, the first subcase from case II splits again into two complementary
situations.

In \textbf{subcase II.1.1}, where (\ref{caseII.1}) and (\ref{a86b11}) hold
simultaneously,
\begin{equation}
D=3,\qquad p=f=q_{2}\delta _{5}^{D}=y_{2}=q_{1}=0,\qquad y_{3}\neq 0,\qquad
\Lambda \neq 0,  \label{subcaseII.1.1}
\end{equation}%
we have that the deformed solution to the master equation is parameterized
by two constants, $y_{3}$ and $\Lambda $. Its first two components result
from (\ref{S1II.1}) and (\ref{S2caseII.1}) where we set $q_{1}=0$ and read
as
\begin{eqnarray}
S_{1}^{\left( \mathrm{II.1.1}\right) } &=&\int d^{3}x\left[ -2\Lambda
h+y_{3}\varepsilon _{\mu \nu \rho }\left( V^{\ast \mu }\partial ^{\lbrack
\nu }\eta ^{\rho ]}+F^{\lambda \mu }\partial ^{\lbrack \nu }h_{\ \ \lambda
}^{\rho ]}\right) \right] ,  \label{S1caseII.1.1} \\
S_{2}^{\left( \mathrm{II.1.1}\right) } &=&y_{3}^{2}\int d^{3}x\left(
\partial ^{\lbrack \nu }h^{\rho ]\lambda }\right) \partial _{\lbrack \nu
}h_{\rho ]\lambda }.  \label{S2caseII.1.1}
\end{eqnarray}%
Consequently, $\left( S_{1}^{\left( \mathrm{II.1.1}\right) },S_{2}^{\left(
\mathrm{II.1.1}\right) }\right) =0$, so (\ref{eqS3}) becomes%
\begin{equation}
sS_{3}^{\left( \mathrm{II.1.1}\right) }=0,  \label{consistS2caseII.1.1}
\end{equation}%
whose solution can be taken to be trivial%
\begin{equation}
S_{3}^{\left( \mathrm{II.1.1}\right) }=0  \label{S3II.1.1}
\end{equation}%
(the solution to the homogeneous equation (\ref{consistS2caseII.1.1}) can be
absorbed into (\ref{S1caseII.1.1}) by a suitable redefinition of the
involved constants). Inserting (\ref{S3II.1.1}) into the next deformation
equation%
\begin{equation}
\frac{1}{2}\left( S_{2}^{\left( \mathrm{II.1.1}\right) },S_{2}^{\left(
\mathrm{II.1.1}\right) }\right) +\left( S_{1}^{\left( \mathrm{II.1.1}\right)
},S_{3}^{\left( \mathrm{II.1.1}\right) }\right) +sS_{4}^{\left( \mathrm{%
II.1.1}\right) }=0  \label{eqS4caseII.1.1}
\end{equation}%
and observing that $\left( S_{2}^{\left( \mathrm{II.1.1}\right)
},S_{2}^{\left( \mathrm{II.1.1}\right) }\right) =0$, we can again take%
\begin{equation}
S_{4}^{\left( \mathrm{II.1.1}\right) }=0.  \label{S4II.1.1}
\end{equation}%
It is easy to see that in fact we can set%
\begin{equation}
S_{n}^{\left( \mathrm{II.1.1}\right) }=0,\qquad \mathrm{for\ all}\qquad n>2.
\label{SkcaseII.1.1}
\end{equation}%
We can therefore conclude that in subcase II.1.1, described by conditions (%
\ref{subcaseII.1.1}), the deformation procedure stops nontrivially at a
finite step ($n=2$) and the deformed solution to the master equation,
consistent to all orders in the deformation parameter, takes the form%
\begin{eqnarray}
&&S^{\left( \mathrm{II.1.1}\right) }=\bar{S}+kS_{1}^{\left( \mathrm{II.1.1}%
\right) }+k^{2}S_{2}^{\left( \mathrm{II.1.1}\right) }\equiv \int d^{3}x\left[
\mathcal{L}_{0}^{\left( \mathrm{PF}\right) }-\frac{1}{4}F_{\mu \nu }F^{\mu
\nu }\right.  \notag \\
&&+h^{\ast \mu \nu }\partial _{(\mu }\eta _{\nu )}+V^{\ast \mu }\partial
_{\mu }\eta -2k\Lambda h  \notag \\
&&\left. +ky_{3}\varepsilon _{\mu \nu \rho }\left( V^{\ast \mu }\partial
^{\lbrack \nu }\eta ^{\rho ]}+F^{\lambda \mu }\partial ^{\lbrack \nu }h_{\ \
\lambda }^{\rho ]}\right) +k^{2}y_{3}^{2}\left( \partial ^{\lbrack \nu
}h^{\rho ]\lambda }\right) \partial _{\lbrack \nu }h_{\rho ]\lambda }\right]
,  \label{SfincaseII.1.1}
\end{eqnarray}%
where $\mathcal{L}_{0}^{\left( \mathrm{PF}\right) }$ is the Pauli-Fierz
Lagrangian.

We choose not to expose in detail the remaining possibilities, whose
investigation is merely technical, but simply state their main conclusions.
Thus, in \textbf{subcase II.1.2}, where (\ref{caseII.1}) and (\ref{a86b12})
are assumed to take place concurrently, the deformed solution to the master
equation is parameterized by $y_{3}$ and $q_{1}$ and starts like in (\ref%
{S1II.1}) and (\ref{S2caseII.1}) where we set $\Lambda =0$. There appear no
obstructions in solving the higher-order deformation equations, of order
three and four, while that of order five requires the supplementary
condition $y_{3}^{3}q_{1}^{2}=0$. Its relevant solution is $q_{1}=0$ since
in the opposite situation, $y_{3}=0$, there are no cross-couplings at all
between the graviton and the vector field: the original gauge
transformations are not affected and the Lagrangian is modified by an
Abelian Chern-Simons term $kq_{1}\varepsilon _{\mu \nu \rho }V^{\mu }F^{\nu
\rho }$. Based on $q_{1}=0$, it can be shown that all the deformations of
order three or higher can be made to vanish, such that \emph{the resulting
deformed solution to the master equation precisely reduces to a particular
solution of subcase II.1.1}: it is expressed by (\ref{SfincaseII.1.1}) for $%
\Lambda =0$. Regarding \textbf{subcase II.2}, it is pictured by conditions (%
\ref{a86b}) and (\ref{a86b2}), so the deformations `live' again in a
three-dimensional spacetime, being parameterized by $y_{2}$ and $y_{3}$. The
first-order deformation reduces to (\ref{a60int}) where we set $D=3$ and $%
p=0=q_{1}$. There are no obstructions in finding the deformation of order
two in the coupling constant, but the existence of the third-order
deformation imposes the additional condition $y_{2}=0$, which further
implies that all the deformations of order three or higher are trivial.
Therefore, \emph{the fully deformed solution to the master equation is
nothing but the same particular solution from subcase II.1.1}, being equal
to (\ref{SfincaseII.1.1}) for $\Lambda =0$.

Combining all the results exposed so far, we can state that the most general
solution of the deformation procedure in case II is provided by a
three-dimensional, consistent solution to the master equation that stops at
the second order in the deformation parameter, is parameterized by $y_{3}$
and $\Lambda $, and reads as in (\ref{SfincaseII.1.1}). We will argue in
Section \ref{speccoupl} that this solution describes a new mechanism for
coupling a spin-two field to a massless vector field in $D=3$, which is
completely different from the standard one, based on General Relativity
prescriptions.

\subsubsection{Case III --- nothing new\label{caseIII}}

Case III is subject to conditions (\ref{a86c}), so it is valid only in $D>3$
spacetime dimensions\footnote{\label{footnotecaseIII}Note that $D>3$ implies
automatically $y_{3}\delta _{3}^{D}=0=q_{1}\delta _{3}^{D}$.}, being
parameterized in the first instance by $y_{2}$, $q_{2}\delta _{5}^{D}$, and $%
\Lambda $. In agreement with (\ref{a86c}), formulas (\ref{chi1}) and (\ref%
{chi0}) will be%
\begin{equation}
\chi _{1}=0,\qquad \chi _{0}=-2y_{2}\left( 3q_{2}\delta _{5}^{D}\varepsilon
_{\mu \nu \rho \alpha \beta }F^{\mu \nu }F^{\rho \alpha }\eta ^{\beta
}+2D\Lambda \eta \right) ,  \label{chicaseIII}
\end{equation}%
such that (\ref{wa1p}) yields the same necessary condition for the existence
of $\bar{b}_{1}^{\left( \mathrm{int}\right) }$ and $\bar{b}_{0}^{\left(
\mathrm{int}\right) }$ like in case II%
\begin{equation}
\chi _{0}=\delta \varphi _{1}+\gamma \omega _{0}+\partial _{\mu }l_{0}^{\mu
},  \label{condchi0caseIII}
\end{equation}%
where $\mathrm{agh}\left( \varphi _{1}\right) =1=\mathrm{pgh}\left( \varphi
_{1}\right) $, $\mathrm{agh}\left( \omega _{0}\right) =0=\mathrm{pgh}\left(
\omega _{0}\right) $, $\mathrm{agh}\left( l_{0}^{\mu }\right) =0$, $\mathrm{%
pgh}\left( l_{0}^{\mu }\right) =1$. The locality of the second-order
deformation requires that all $\varphi _{1}$, $\omega _{0}$, and $l_{0}^{\mu
}$ are local functions. From (\ref{chicaseIII}) and definitions (\ref{a7})
and (\ref{a11}) it is obvious that (\ref{condchi0caseIII}) cannot be
satisfied for some local $\varphi _{1}$, $\omega _{0}$, and $l_{0}^{\mu }$
until we set $\chi _{0}=0$, which further demands%
\begin{equation}
y_{2}q_{2}\delta _{5}^{D}=0,\qquad y_{2}\Lambda =0.  \label{a85c1}
\end{equation}%
There are obviously two complementary solutions to these equations%
\begin{eqnarray}
q_{2}\delta _{5}^{D} &=&0,\qquad \Lambda =0,  \label{a86c1} \\
y_{2} &=&0.  \label{caseIII.2}
\end{eqnarray}

Once more, we try to simplify the presentation by avoiding the technical
details involved and mentioning only the key points. \textbf{Subcase III.1},
described by (\ref{a86c}) and (\ref{a86c1}), is parameterized by a single
constant, $y_{2}$, such that the first-order deformation reduces to the
component of (\ref{a60int}) proportional with this parameter%
\begin{equation}
S_{1}^{\left( \mathrm{III.1}\right) }=y_{2}\int d^{D}x\left[ h^{\ast }\eta
+\left( D-2\right) \left( -V^{\ast \lambda }\eta _{\lambda }+V^{\lambda
}\partial _{\lbrack \mu }h_{\lambda ]}^{\ \ \mu }\right) \right] .
\label{S1III}
\end{equation}%
The second-order deformation, $S_{2}^{\left( \mathrm{III.1}\right) }$, is
then easily obtained from the observation that $\chi _{1}=0=\chi _{0}$, so
equations (\ref{wa1nn}) and (\ref{wa1o})--(\ref{wa1p}) reduce, like in case
I, to (\ref{wa1nncaseI})--(\ref{wa1pcaseI}), whose solution can be taken to
vanish, like in (\ref{b2b1b0caseI}). Consequently, the nonintegrated density
of $S_{2}^{\left( \mathrm{III.1}\right) }$ contains only terms of antighost
number zero and reduces to the integrand of (\ref{a64c1}) plus the terms
proportional with $y_{2}^{2}$ from (\ref{notb0}). It is easy to show that
the existence of a local third-order deformation requires $y_{2}=0$, so
\emph{subcase III.1 leads to no nontrivial deformations}, $S_{n}^{\left(
\mathrm{III.1}\right) }=0$, for all $n\geq 1$. \textbf{Subcase III.2},
pictured by (\ref{a86c}) and (\ref{caseIII.2}), is parameterized by $%
q_{2}\delta _{5}^{D}$ and $\Lambda $ (see also footnote \ref{footnotecaseIII}%
). Only the first-order deformation is found non-trivial, being equal to
\begin{equation}
S_{1}^{\left( \mathrm{III.2}\right) }=\int d^{D}x\left( -2\Lambda
h+q_{2}\delta _{5}^{D}\varepsilon ^{\mu \nu \lambda \alpha \beta
}V_{\mu }F_{\nu \lambda }F_{\alpha \beta }\right) .  \label{S1III.2}
\end{equation}%
Analyzing (\ref{S1III.2}), we can state that \emph{subcase III.2 is
not interesting} since the deformation procedure does not modify the
original gauge transformations (\ref{a2}), but mainly adds to the
original action (\ref{a1}) two gauge-invariant terms: a cosmological
one and a generalized Abelian Chern-Simons action. In conclusion,
case III brings no new information on the possible couplings
vbetween a spin-two field and a massless one-form.

\subsection{Analysis of the deformed theory\label{analysis}}

The main aim of this section is to give an appropriate
interpretation of the Lagrangian formulation of the deformed
theories obtained previously from the deformation of the solution to
the master equation. We will analyze the first two cases separately
since we have seen that the third one gives nothing interesting. It
is useful to recall the relationship between some quantities
appearing in the deformed solution of the master equation, $S$, and
the associated gauge theory: the component of antighost number zero
from the former is nothing but the Lagrangian action of the coupled
model, the piece of antighost number one provides the gauge
transformations of the deformed theory, and the terms of antighost
number two contain the structure functions defined by the
commutators among the deformed gauge transformations. More
precisely, the gauge transformations of the coupled theory result
from the terms of antighost number one present in $S$ (generically
written as $\Phi _{\alpha _{0}}^{\ast }Z_{\;\;\alpha _{1}}^{\alpha
_{0}}\eta ^{\alpha _{1}}$) by replacing the ghosts with the gauge
parameters $\epsilon ^{\alpha _{1}}$, $\delta _{\epsilon }\Phi
^{\alpha _{0}}=Z_{\;\;\alpha _{1}}^{\alpha _{0}}\epsilon ^{\alpha
_{1}}$.
The functions%
\begin{equation}
Z_{\;\;\alpha _{1}}^{\alpha _{0}}=\bar{Z}_{\;\;\alpha _{1}}^{\alpha
_{0}}+kZ_{1\;\;\alpha _{1}}^{\alpha _{0}}+k^{2}Z_{2\;\;\alpha _{1}}^{\alpha
_{0}}+\cdots  \label{zdef}
\end{equation}%
define the gauge generators of the coupled model, where the components $\bar{%
Z}_{\;\;\alpha _{1}}^{\alpha _{0}}$ are responsible for the original gauge
transformations.

\subsubsection{Case I: standard couplings from General Relativity\label%
{analysiscaseI}}

We discussed in detail in Section \ref{caseI} a first case of obtaining
consistent interactions between a Pauli-Fierz field and a vector field. This
is defined by conditions (\ref{a86a}), in which situation the deformed
solution to the master equation starts like%
\begin{eqnarray}
S^{(\mathrm{I})} &=&\bar{S}+kS_{1}^{(\mathrm{I})}+k^{2}S_{2}^{(\mathrm{I}%
)}+\cdots  \notag \\
&=&\bar{S}+k\left( S_{1}^{(\mathrm{PF})}+S_{1}^{(\mathrm{int})}\right)
+k^{2}\left( S_{2}^{(\mathrm{PF})}+S_{2}^{(\mathrm{int})}\right) +\cdots ,
\label{SdefI}
\end{eqnarray}%
where $\bar{S}$, $S_{1}^{(\mathrm{I})}$, and $S_{2}^{(\mathrm{I})}$ read as
in (\ref{a16}), (\ref{S1I}), and (\ref{S2I}) respectively.

In order to identify the main ingredients of the coupled model in the first
case we use the result proved in Section 5 of~\cite{noijhepdirac}, according
to which the local BRST cohomologies of the Pauli-Fierz model and of the
linearized version of vielbein formulation of spin-two field theory are
isomorphic. Because the local BRST cohomology (in ghost numbers zero and
one) controls the deformation procedure, it results that this isomorphism
allows one to pass in a consistent manner from the Pauli-Fierz model to the
linearized version of the vielbein formulation and conversely during the
deformation procedure. Nevertheless, the linearized vielbein formulation
possesses more fields (the antisymmetric part of the linearized vielbein)
and more gauge parameters (Lorentz parameters) than the Pauli-Fierz model.
The switch from the former version to the latter is realized via the above
mentioned isomorphism by imposing some partial gauge-fixing conditions,
chosen to annihilate the antisymmetric components of the vielbein. An
appropriate interpretation of the Lagrangian description of the interacting
theory in case I requires the generalized expression of these partial
gauge-fixing conditions~\cite{siegelfields}%
\begin{equation}
\sigma _{\mu [ a}e_{b]}^{\mu }=0  \label{wzz1}
\end{equation}%
and the development of the vielbein $e_{a}^{\mu }$ and of its inverse $%
e_{\mu }^{a}$ up to the second order in the coupling constant in terms of
the Pauli-Fierz field
\begin{eqnarray}
e_{a}^{\mu } &=&\overset{(0)}{e}_{a}^{\mu }+k\overset{(1)}{e}_{a}^{\mu
}+k^{2}\overset{(2)}{e}_{a}^{\mu }+\cdots =\delta _{a}^{\mu }-\frac{k}{2}%
h_{a}^{\mu }+\frac{3k^{2}}{8}h_{a}^{\nu }h_{\nu }^{\mu }+\cdots ,
\label{id1} \\
e_{\mu }^{a} &=&\overset{(0)}{e}_{\mu }^{a}+k\overset{(1)}{e}_{\mu
}^{a}+k^{2}\overset{(2)}{e}_{\mu }^{a}+\cdots =\delta _{\mu }^{a}+\frac{k}{2}%
h_{\mu }^{a}-\frac{k^{2}}{8}h_{\nu }^{a}h_{\mu }^{\nu }+\cdots .
\label{uv2a}
\end{eqnarray}%
The expansion of the inverse of the metric tensor $g^{\mu \nu }$ and of the
square root from the minus determinant of the metric tensor $\sqrt{-g}=\sqrt{%
-\det g_{\mu \nu }}$ in terms of the Pauli-Fierz field,
\begin{eqnarray}
g^{\mu \nu } &=&\overset{(0)}{g^{\mu \nu }}+k\overset{(1)}{g^{\mu \nu }}%
+k^{2}\overset{(2)}{g^{\mu \nu }}+\cdots =\sigma ^{\mu \nu }-kh^{\mu \nu
}+k^{2}h_{\rho }^{\mu }h^{\rho \nu }+\cdots ,  \label{a94} \\
\sqrt{-g} &=&\overset{(0)}{\sqrt{-g}}+k\overset{(1)}{\sqrt{-g}}+k^{2}\overset%
{(2)}{\sqrt{-g}}+\cdots  \notag \\
&=&1+\frac{k}{2}h+\frac{k^{2}}{8}\left( h^{2}-2h_{\mu \nu }h^{\mu \nu
}\right) +\cdots ,  \label{a96}
\end{eqnarray}%
will also be necessary in what follows. We note that the metric tensor is%
\begin{equation}
g_{\mu \nu }=\sigma _{\mu \nu }+kh_{\mu \nu }.  \label{metric}
\end{equation}

The interacting Lagrangian at order one in the coupling constant, $\mathcal{L%
}_{1}^{\left( \mathrm{int}\right) }$, is the nonintegrated density of the
piece of antighost number zero from the first-order deformation in the
interacting sector, $S_{1}^{(\mathrm{int})}$. Using (\ref{S1I}) and
expansions (\ref{id1})--(\ref{a96}), we can write
\begin{eqnarray}
\mathcal{L}_{1}^{\left( \mathrm{int}\right) } &=&-\frac{1}{4}F^{\mu \nu
}\partial _{\lbrack \mu }\left( h_{\nu ]\rho }V^{\rho }\right) -\frac{1}{8}%
F^{\mu \nu }F_{\mu \nu }h+\frac{1}{2}F^{\mu \nu }F_{\mu \rho }h_{\nu }^{\rho
}  \notag \\
&&+q_{1}\delta _{3}^{D}\varepsilon ^{\mu \nu \lambda }V_{\mu }F_{\nu \lambda
}+q_{2}\delta _{5}^{D}\varepsilon ^{\mu \nu \lambda \alpha \beta }V_{\mu
}F_{\nu \lambda }F_{\alpha \beta }  \notag \\
&\equiv &-\frac{1}{4}\left[ \left( \overset{(0)}{\sqrt{-g}}\overset{\left(
0\right) }{g^{\mu \nu }}\overset{\left( 1\right) }{g^{\rho \lambda }}+%
\overset{(0)}{\sqrt{-g}}\overset{\left( 1\right) }{g^{\mu \nu }}\overset{%
\left( 0\right) }{g^{\rho \lambda }}+\overset{(1)}{\sqrt{-g}}\overset{\left(
0\right) }{g^{\mu \nu }}\overset{\left( 0\right) }{g^{\rho \lambda }}\right)
\overset{\left( 0\right) }{\bar{F}_{\mu \rho }}\overset{\left( 0\right) }{%
\bar{F}_{\nu \lambda }}\right.  \notag \\
&&\left. +\overset{(0)}{\sqrt{-g}}\overset{\left( 0\right) }{g^{\mu \nu }}%
\overset{\left( 0\right) }{g^{\rho \lambda }}\left( \overset{\left( 1\right)
}{\bar{F}_{\mu \rho }}\overset{\left( 0\right) }{\bar{F}_{\nu \lambda }}+%
\overset{\left( 0\right) }{\bar{F}_{\mu \rho }}\overset{\left( 1\right) }{%
\bar{F}_{\nu \lambda }}\right) \right]  \notag \\
&&+q_{1}\delta _{3}^{D}\overset{\left( 0\right) }{\sqrt{-g}}\overset{\left(
0\right) }{e}_{a_{1}}^{\mu _{1}}\overset{\left( 0\right) }{e}_{a_{2}}^{\mu
_{2}}\overset{\left( 0\right) }{e}_{a_{3}}^{\mu _{3}}\varepsilon
^{a_{1}a_{2}a_{3}}\overset{\left( 0\right) }{\bar{V}}_{\mu _{1}}\overset{%
\left( 0\right) }{\bar{F}}_{\mu _{2}\mu _{3}}  \notag \\
&&+q_{2}\delta _{5}^{D}\overset{\left( 0\right) }{\sqrt{-g}}\overset{\left(
0\right) }{e}_{a_{1}}^{\mu _{1}}\cdots \overset{\left( 0\right) }{e}%
_{a_{5}}^{\mu _{5}}\varepsilon ^{a_{1}a_{2}a_{3}a_{4}a_{5}}\overset{\left(
0\right) }{\bar{V}}_{\mu _{1}}\overset{\left( 0\right) }{\bar{F}}_{\mu
_{2}\mu _{3}}\overset{\left( 0\right) }{\bar{F}}_{\mu _{4}\mu _{5}},
\label{a93}
\end{eqnarray}%
where
\begin{equation}
\overset{\left( 0\right) }{\bar{V}}_{\mu }=\overset{\left( 0\right) }{e}%
_{\mu }^{a}V_{a},\overset{\left( 0\right) }{\quad \bar{F}_{\mu \nu }}%
=\partial _{\lbrack \mu }\left( \overset{\left( 0\right) }{e}_{\nu
]}^{a}V_{a}\right) ,\quad \overset{\left( 1\right) }{\bar{F}_{\mu \nu }}%
=\partial _{\lbrack \mu }\left( \overset{\left( 1\right) }{e}_{\nu
]}^{a}V_{a}\right) .  \label{lw1}
\end{equation}%
Along the same line, the interacting Lagrangian at order two, $\mathcal{L}%
_{2}^{\left( \mathrm{int}\right) }$, results from $S_{2}^{(\mathrm{int})}$
at antighost number zero. Taking into account formula (\ref{S2I}) and
expansions (\ref{id1})--(\ref{a96}), we have that
\begin{eqnarray}
\mathcal{L}_{2}^{\left( \mathrm{int}\right) } &\equiv &-\frac{1}{4}\left[
\overset{(0)}{\sqrt{-g}}\overset{\left( 0\right) }{g^{\mu \nu }}\overset{%
\left( 0\right) }{g^{\rho \lambda }}\left( \overset{\left( 0\right) }{\bar{F}%
_{\mu \rho }}\overset{\left( 2\right) }{\bar{F}_{\nu \lambda }}+\overset{%
\left( 2\right) }{\bar{F}_{\mu \rho }}\overset{\left( 0\right) }{\bar{F}%
_{\nu \lambda }}\right) \right.  \notag \\
&&+\left( \overset{(0)}{\sqrt{-g}}\overset{\left( 0\right) }{g^{\mu \nu }}%
\overset{\left( 1\right) }{g^{\rho \lambda }}+\overset{(0)}{\sqrt{-g}}%
\overset{\left( 1\right) }{g^{\mu \nu }}\overset{\left( 0\right) }{g^{\rho
\lambda }}+\overset{(1)}{\sqrt{-g}}\overset{\left( 0\right) }{g^{\mu \nu }}%
\overset{\left( 0\right) }{g^{\rho \lambda }}\right) \left( \overset{\left(
0\right) }{\bar{F}_{\mu \rho }}\overset{\left( 1\right) }{\bar{F}_{\nu
\lambda }}+\overset{\left( 1\right) }{\bar{F}_{\mu \rho }}\overset{\left(
0\right) }{\bar{F}_{\nu \lambda }}\right)  \notag \\
&&+\left( \overset{(0)}{\sqrt{-g}}\overset{\left( 0\right) }{g^{\mu \nu }}%
\overset{\left( 2\right) }{g^{\rho \lambda }}+\overset{(0)}{\sqrt{-g}}%
\overset{\left( 2\right) }{g^{\mu \nu }}\overset{\left( 0\right) }{g^{\rho
\lambda }}+\overset{(0)}{\sqrt{-g}}\overset{\left( 1\right) }{g^{\mu \nu }}%
\overset{\left( 1\right) }{g^{\rho \lambda }}\right.  \notag \\
&&\left. \left. +\overset{(1)}{\sqrt{-g}}\overset{\left( 0\right) }{g^{\mu
\nu }}\overset{\left( 1\right) }{g^{\rho \lambda }}+\overset{(1)}{\sqrt{-g}}%
\overset{\left( 1\right) }{g^{\mu \nu }}\overset{\left( 0\right) }{g^{\rho
\lambda }}+\overset{(2)}{\sqrt{-g}}\overset{\left( 0\right) }{g^{\mu \nu }}%
\overset{\left( 0\right) }{g^{\rho \lambda }}\right) \overset{\left(
0\right) }{\bar{F}_{\mu \rho }}\overset{\left( 0\right) }{\bar{F}_{\nu
\lambda }}\right]  \notag \\
&&+q_{1}\delta _{3}^{D}\varepsilon ^{a_{1}a_{2}a_{3}}\left[ \overset{\left(
1\right) }{\sqrt{-g}}\overset{\left( 0\right) }{e}_{a_{1}}^{\mu _{1}}\overset%
{\left( 0\right) }{e}_{a_{2}}^{\mu _{2}}\overset{\left( 0\right) }{e}%
_{a_{3}}^{\mu _{3}}\overset{\left( 0\right) }{\bar{V}}_{\mu _{1}}\overset{%
\left( 0\right) }{\bar{F}}_{\mu _{2}\mu _{3}}\right.  \notag \\
&&+\overset{\left( 0\right) }{\sqrt{-g}}\left( \overset{\left( 1\right) }{e}%
_{a_{1}}^{\mu _{1}}\overset{\left( 0\right) }{e}_{a_{2}}^{\mu _{2}}\overset{%
\left( 0\right) }{e}_{a_{3}}^{\mu _{3}}\overset{\left( 0\right) }{\bar{V}}%
_{\mu _{1}}\overset{\left( 0\right) }{\bar{F}}_{\mu _{2}\mu _{3}}\right.
\notag \\
&&+2\overset{\left( 0\right) }{e}_{a_{1}}^{\mu _{1}}\overset{\left( 0\right)
}{e}_{a_{2}}^{\mu _{2}}\overset{\left( 1\right) }{e}_{a_{3}}^{\mu _{3}}%
\overset{\left( 0\right) }{\bar{V}}_{\mu _{1}}\overset{\left( 0\right) }{%
\bar{F}}_{\mu _{2}\mu _{3}}+\overset{\left( 0\right) }{e}_{a_{1}}^{\mu _{1}}%
\overset{\left( 0\right) }{e}_{a_{2}}^{\mu _{2}}\overset{\left( 0\right) }{e}%
_{a_{3}}^{\mu _{3}}\overset{\left( 1\right) }{\bar{V}}_{\mu _{1}}\overset{%
\left( 0\right) }{\bar{F}}_{\mu _{2}\mu _{3}}  \notag \\
&&\left. \left. +\overset{\left( 0\right) }{e}_{a_{1}}^{\mu _{1}}\overset{%
\left( 0\right) }{e}_{a_{2}}^{\mu _{2}}\overset{\left( 0\right) }{e}%
_{a_{3}}^{\mu _{3}}\overset{\left( 0\right) }{\bar{V}}_{\mu _{1}}\overset{%
\left( 1\right) }{\bar{F}}_{\mu _{2}\mu _{3}}\right) \right]  \notag \\
&&+q_{2}\delta _{5}^{D}\varepsilon ^{a_{1}a_{2}a_{3}a_{4}a_{5}}\left[
\overset{\left( 1\right) }{\sqrt{-g}}\overset{\left( 0\right) }{e}%
_{a_{1}}^{\mu _{1}}\cdots \overset{\left( 0\right) }{e}_{a_{5}}^{\mu _{5}}%
\overset{\left( 0\right) }{\bar{V}}_{\mu _{1}}\overset{\left( 0\right) }{%
\bar{F}}_{\mu _{2}\mu _{3}}\overset{\left( 0\right) }{\bar{F}}_{\mu _{4}\mu
_{5}}\right.  \notag \\
&&\overset{\left( 0\right) }{+\sqrt{-g}}\left( \overset{\left( 1\right) }{e}%
_{a_{1}}^{\mu _{1}}\overset{\left( 0\right) }{e}_{a_{2}}^{\mu _{2}}\cdots
\overset{\left( 0\right) }{e}_{a_{5}}^{\mu _{5}}\overset{\left( 0\right) }{%
\bar{V}}_{\mu _{1}}\overset{\left( 0\right) }{\bar{F}}_{\mu _{2}\mu _{3}}%
\overset{\left( 0\right) }{\bar{F}}_{\mu _{4}\mu _{5}}\right.  \notag \\
&&+4\overset{\left( 0\right) }{e}_{a_{1}}^{\mu _{1}}\cdots \overset{\left(
0\right) }{e}_{a_{4}}^{\mu _{4}}\overset{\left( 1\right) }{e}_{a_{5}}^{\mu
_{5}}\overset{\left( 0\right) }{\bar{V}}_{\mu _{1}}\overset{\left( 0\right) }%
{\bar{F}}_{\mu _{2}\mu _{3}}\overset{\left( 0\right) }{\bar{F}}_{\mu _{4}\mu
_{5}}  \notag \\
&&+\overset{\left( 0\right) }{e}_{a_{1}}^{\mu _{1}}\cdots \overset{\left(
0\right) }{e}_{a_{5}}^{\mu _{5}}\overset{\left( 1\right) }{\bar{V}}_{\mu
_{1}}\overset{\left( 0\right) }{\bar{F}}_{\mu _{2}\mu _{3}}\overset{\left(
0\right) }{\bar{F}}_{\mu _{4}\mu _{5}}  \notag \\
&&\left. \left. +2\overset{\left( 0\right) }{e}_{a_{1}}^{\mu _{1}}\cdots
\overset{\left( 0\right) }{e}_{a_{5}}^{\mu _{5}}\overset{\left( 0\right) }{%
\bar{V}}_{\mu _{1}}\overset{\left( 0\right) }{\bar{F}}_{\mu _{2}\mu _{3}}%
\overset{\left( 1\right) }{\bar{F}}_{\mu _{4}\mu _{5}}\right) \right] ,
\label{a98}
\end{eqnarray}%
with
\begin{equation}
\overset{\left( 1\right) }{\bar{V}}_{\mu }=\overset{\left( 1\right) }{e}%
_{\mu }^{a}V_{a},\qquad \overset{\left( 2\right) }{\bar{F}_{\mu \nu }}%
=\partial _{\lbrack \mu }\left( \overset{\left( 2\right) }{e}_{\nu
]}^{a}V_{a}\right) .  \label{lw2}
\end{equation}%
From the expressions of $\mathcal{L}_{1}^{(\mathrm{int})}$ and $\mathcal{L}%
_{2}^{(\mathrm{int})}$, we observe that the first three terms from the full
interacting Lagrangian in case I
\begin{equation}
\mathcal{L}_{\mathrm{I}}^{(\mathrm{int})}=\mathcal{L}_{0}^{(\mathrm{vect})}+k%
\mathcal{L}_{1}^{(\mathrm{int})}+k^{2}\mathcal{L}_{2}^{(\mathrm{int})}+\cdots
\label{Lint}
\end{equation}%
coincide with the first orders of the Lagrangian describing the standard
vector field-graviton cross-couplings from General Relativity
\begin{eqnarray}
\mathcal{L}^{(\mathrm{vector-graviton})} &=&-\frac{1}{4}\sqrt{-g}g^{\mu \nu
}g^{\rho \lambda }\bar{F}_{\mu \rho }\bar{F}_{\nu \lambda }+k\left(
q_{1}\delta _{3}^{D}\varepsilon ^{\mu _{1}\mu _{2}\mu _{3}}\bar{V}_{\mu _{1}}%
\bar{F}_{\mu _{2}\mu _{3}}\right.  \notag \\
&&\left. +q_{2}\delta _{5}^{D}\varepsilon ^{\mu _{1}\mu _{2}\mu _{3}\mu
_{4}\mu _{5}}\bar{V}_{\mu _{1}}\bar{F}_{\mu _{2}\mu _{3}}\bar{F}_{\mu
_{4}\mu _{5}}\right) ,  \label{a99}
\end{eqnarray}%
where the fully deformed field strength $\bar{F}_{\mu \nu }$ and the
Levi-Civita symbol with curved indices $\varepsilon ^{\mu _{1}\ldots \mu
_{D}}$ are given by
\begin{eqnarray}
\bar{F}_{\mu \nu } &=&\partial _{\lbrack \mu }\left( e_{\nu
]}^{a}V_{a}\right) \equiv \overset{\left( 0\right) }{\bar{F}_{\mu \nu }}+k%
\overset{\left( 1\right) }{\bar{F}_{\mu \nu }}+k^{2}\overset{\left( 2\right)
}{\bar{F}_{\mu \nu }}+\cdots  \notag \\
&=&\partial _{\lbrack \mu }\left( \overset{\left( 0\right) }{e}_{\nu
]}^{a}V_{a}\right) +k\partial _{\lbrack \mu }\left( \overset{\left( 1\right)
}{e}_{\nu ]}^{a}V_{a}\right) +k^{2}\partial _{\lbrack \mu }\left( \overset{%
\left( 2\right) }{e}_{\nu ]}^{a}V_{a}\right) +\cdots ,  \label{fs}
\end{eqnarray}%
\begin{equation}
\varepsilon ^{\mu _{1}\ldots \mu _{D}}=\sqrt{-g}e_{a_{1}}^{\mu _{1}}\cdots
e_{a_{D}}^{\mu _{D}}\varepsilon ^{a_{1}\ldots a_{D}}.  \label{gdgd}
\end{equation}%
The self-interactions of the Pauli-Fierz field at orders one and two in the
coupling constant, $\mathcal{L}_{1,2}^{\left( \mathrm{PF}\right) }$, result
from the terms of antighost number zero present in $S_{1}^{(\mathrm{PF})}$
and $S_{2}^{(\mathrm{PF})}$ (see (\ref{S1I}) and (\ref{S2I})), so the full
Lagrangian describing the self-interactions of the graviton in case I starts
like
\begin{equation}
\tilde{\mathcal{L}}_{\mathrm{I}}^{\left( \mathrm{PF}\right) }=\mathcal{L}%
_{0}^{\left( \mathrm{PF}\right) }+k\mathcal{L}_{1}^{\left( \mathrm{PF}%
\right) }+k^{2}\mathcal{L}_{2}^{\left( \mathrm{PF}\right) }+\ldots ,
\label{LPF}
\end{equation}%
where $\mathcal{L}_{0}^{\left( \mathrm{PF}\right) }$ is the Pauli-Fierz
Lagrangian. Using (\ref{a94})--(\ref{metric}), one finds that the first
three terms from $\tilde{\mathcal{L}}_{\mathrm{I}}^{\left( \mathrm{PF}%
\right) }$ are nothing but the first orders of the Einstein-Hilbert
Lagrangian with a cosmological term~\cite{multi}%
\begin{equation}
\mathcal{L}^{\left( \mathrm{EH}\right) }=\frac{2}{k^{2}}\sqrt{-g}\left(
R-2k^{2}\Lambda \right) ,  \label{LHE}
\end{equation}%
where $R$ is the full scalar curvature.

As explained in the beginning of this section, the terms present in (\ref%
{SdefI}) (see (\ref{a16}), (\ref{S1I}), and (\ref{S2I})) that are linear in
the antifields $V^{\ast \mu }$ provide the deformed gauge transformations of
the vector field
\begin{eqnarray}
&&\delta _{\epsilon }^{\mathrm{(I)}}V_{\alpha }=\left( \delta _{\alpha
}^{\mu }-\frac{k}{2}h_{\alpha }^{\mu }+\frac{3k^{2}}{8}h_{\nu }^{\mu
}h_{\alpha }^{\nu }+\cdots \right) \partial _{\mu }\epsilon +\left[ \frac{k}{%
2}\partial _{[ \alpha }\epsilon _{\beta ]}\right.  \notag \\
&&\left. +k^{2}\left( -\frac{1}{4}\left( \partial _{[ \alpha }h_{\beta
]\gamma }\right) \epsilon ^{\gamma }+\frac{1}{8}h_{\gamma [ \alpha }\partial
_{\beta ]}\epsilon ^{\gamma }+\frac{1}{8}\left( \partial _{\gamma }\epsilon
_{[ \alpha }\right) h_{\beta ]}^{\gamma }\right) V^{\beta }+\cdots \right]
\notag \\
&&+\left( \partial _{\mu }V_{\alpha }\right) \left( k\delta _{\beta }^{\mu }-%
\frac{k^{2}}{2}h_{\beta }^{\mu }+\frac{3k^{3}}{8}h_{\nu }^{\mu }h_{\beta
}^{\nu }+\cdots \right) \epsilon ^{\beta }.  \label{a100}
\end{eqnarray}%
In the last formula the indices of the one-form, even if written in
Latin letters, are flat. In standard, Latin notation the above gauge
transformations can be written as%
\begin{equation*}
\delta _{\epsilon }^{\mathrm{(I)}}V_{a}=\overset{\left( 0\right) }{\delta }%
_{\epsilon }V_{a}+k\overset{\left( 1\right) }{\delta }_{\epsilon }V_{a}+k^{2}%
\overset{\left( 2\right) }{\delta }_{\epsilon }V_{a}+\cdots ,
\end{equation*}%
where the first orders of the gauge transformations read as
\begin{eqnarray}
\overset{\left( 0\right) }{\delta }_{\epsilon }V_{a} &=&\overset{\left(
0\right) }{e}_{a}^{\mu }\partial _{\mu }\epsilon ,  \label{a101} \\
\overset{\left( 1\right) }{\delta }_{\epsilon }V_{a} &=&\overset{\left(
1\right) }{e}_{a}^{\mu }\partial _{\mu }\epsilon +\overset{\left( 0\right) }{%
\epsilon }_{ab}V^{b}+\left( \partial _{\mu }V_{a}\right) \overset{(0)}{\bar{%
\epsilon}}^{\mu },  \label{a1011} \\
\overset{\left( 2\right) }{\delta }_{\epsilon }V_{a} &=&\overset{\left(
2\right) }{e}_{a}^{\mu }\partial _{\mu }\epsilon +\overset{\left( 1\right) }{%
\epsilon }_{ab}V^{b}+\left( \partial _{\mu }V_{a}\right) \overset{(1)}{\bar{%
\epsilon}}^{\mu }  \label{a102}
\end{eqnarray}%
and the various orders of the gauge parameters are expressed by
\begin{eqnarray}
\overset{(0)}{\bar{\epsilon}}^{\mu } &=&\epsilon ^{\mu }\equiv \epsilon
^{a}\delta _{a}^{\mu },\qquad \overset{(1)}{\bar{\epsilon}}^{\mu }=-\frac{1}{%
2}\epsilon ^{a}h_{a}^{\mu },  \label{a1031} \\
\overset{(0)}{\epsilon }_{ab} &=&\frac{1}{2}\partial _{[ a}\epsilon _{b]},
\label{a1032} \\
\overset{(1)}{\epsilon }_{ab} &=&-\frac{1}{4}\epsilon ^{c}\partial _{[
a}h_{b]c}+\frac{1}{8}h_{[a}^{c}\partial _{b]}\epsilon _{c}+\frac{1}{8}\left(
\partial _{c}\epsilon _{[ a}\right) h_{b]}^{c}.  \label{a1033}
\end{eqnarray}%
Based on the above notations, we can re-write the gauge transformations of
the vector field with a flat index as%
\begin{eqnarray}
\delta _{\epsilon }^{\mathrm{(I)}}V_{a} &=&\left( \overset{\left( 0\right) }{%
e}_{a}^{\mu }+k\overset{\left( 1\right) }{e}_{a}^{\mu }+\cdots \right)
\partial _{\mu }\epsilon +k\left( \overset{\left( 0\right) }{\epsilon }%
_{ab}+k\overset{\left( 1\right) }{\epsilon }_{ab}+\cdots \right) V^{b}
\notag \\
&&+k\left( \partial _{\mu }V_{a}\right) \left( \overset{(0)}{\bar{\epsilon}}%
^{\mu }+k\overset{(1)}{\bar{\epsilon}}^{\mu }+\cdots \right) .  \label{a104}
\end{eqnarray}%
The gauge parameters $\overset{(0)}{\epsilon }_{ab}$ and $\overset{(1)}{%
\epsilon }_{ab}$ are precisely the first two terms from the Lorentz
parameters expressed in terms of the flat parameters $\epsilon ^{a}$ via the
partial gauge-fixing (\ref{wzz1}). Indeed, (\ref{wzz1}) leads to
\begin{equation}
\delta _{\epsilon }\left( \sigma _{\mu [ a}e_{b]}^{\mu }\right) =0.
\label{id5}
\end{equation}%
Using
\begin{equation}
\delta _{\epsilon }e_{a}^{\mu }=\bar{\epsilon}^{\rho }\partial _{\rho
}e_{a}^{\mu }-e_{a}^{\rho }\partial _{\rho }\bar{\epsilon}^{\mu }+\epsilon
_{a}^{\;\;b}e_{b}^{\mu }  \label{id6}
\end{equation}%
and inserting (\ref{id1}) together with the expansions
\begin{eqnarray}
\bar{\epsilon}^{\mu } &=&\overset{(0)}{\bar{\epsilon}}^{\mu }+k\overset{(1)}{%
\bar{\epsilon}}^{\mu }+\cdots =\left( \delta _{a}^{\mu }-\frac{k}{2}%
h_{a}^{\mu }+\cdots \right) \epsilon ^{a},  \label{uv15} \\
\epsilon _{ab} &=&\overset{(0)}{\epsilon }_{ab}+k\overset{(1)}{\epsilon }%
_{ab}+\cdots  \label{uv12}
\end{eqnarray}%
in (\ref{id5}), we arrive precisely at (\ref{a1032}) and (\ref{a1033}). At
this point it is easy to see that the first orders of the gauge
transformations (\ref{a104}) coincide with those arising from the
perturbative expansion of the formula
\begin{equation}
\delta _{\epsilon }^{\mathrm{(I)}}V_{a}=e_{a}^{\mu }\partial _{\mu }\epsilon
+k\epsilon _{ab}V^{b}+k\left( \partial _{\mu }V_{a}\right) \bar{\epsilon}%
^{\mu }.  \label{fulltrans}
\end{equation}%
Concerning the vector field with a curved index $\bar{V}_{\mu }$%
\begin{equation}
\bar{V}_{\mu }=e_{\mu }^{a}V_{a},  \label{x1}
\end{equation}%
its gauge transformations will be correctly described at the level of the
first orders in the coupling constant by the well-known gauge transformations%
\begin{equation}
\delta _{\epsilon }^{\mathrm{(I)}}\bar{V}_{\mu }=\partial _{\mu }\epsilon
+k\left( \partial _{\mu }\bar{\epsilon}^{\nu }\right) \bar{V}_{\nu }+k\left(
\partial _{\nu }\bar{V}_{\mu }\right) \bar{\epsilon}^{\nu }  \label{x2}
\end{equation}%
of the vector field (in interaction with the Einstein-Hilbert graviton) from
General Relativity. Finally, from the terms present in (\ref{SdefI}) linear
in the Pauli-Fierz antifields $h^{\ast \mu \nu }$ (see (\ref{a16}), (\ref%
{S1I}), and (\ref{S2I})) one infers that the deformed gauge transformations
of the metric tensor (\ref{metric}) reproduce the first orders of
diffeomorphisms%
\begin{equation}
\delta _{\epsilon }^{\mathrm{(I)}}g_{\mu \nu }=k\epsilon _{(\mu ;\nu )},
\label{diff}
\end{equation}%
where $\epsilon _{\mu ;\nu }$ is the (full) covariant derivative of $%
\epsilon _{\mu }$.

So far, we argued that \emph{in the first case the consistent interactions
between a graviton and a vector field are described in all }$D>2$\emph{\
dimensions} by the first orders of the Lagrangian and gauge transformations
prescribed \emph{by the standard rules of General Relativity} (see (\ref{a99}%
), (\ref{LHE}), (\ref{x2}), and (\ref{diff})). Our result follows as
a consequence of applying a cohomological procedure based on
the\textquotedblleft free\textquotedblright\ BRST symmetry in the
presence of a few natural assumptions: locality, smoothness in the
coupling constant, Poincar\'{e} invariance, Lorentz covariance, and
preservation of the number of derivatives on each field. General
covariance was not imposed a priori, but was gained in a natural way
from the cohomological setting developed here under the previously
mentioned
hypotheses. It can be shown that formulas (\ref{a99}), (\ref{LHE}), (\ref{x2}%
), and (\ref{diff}) apply in fact to all orders in the coupling constant, so
we can state that \emph{the fully interacting Lagrangian action in case I}
reads as%
\begin{eqnarray}
&&S^{\mathrm{L(I)}}\left[ g_{\mu \nu },\bar{V}_{\mu }\right] =\int d^{D}x%
\left[ \frac{2}{k^{2}}\sqrt{-g}\left( R-2k^{2}\Lambda \right) -\frac{1}{4}%
\sqrt{-g}g^{\mu \nu }g^{\rho \lambda }\bar{F}_{\mu \rho }\bar{F}_{\nu
\lambda }\right.  \notag \\
&&\left. +k\left( q_{1}\delta _{3}^{D}\varepsilon ^{\mu _{1}\mu _{2}\mu _{3}}%
\bar{V}_{\mu _{1}}\bar{F}_{\mu _{2}\mu _{3}}+q_{2}\delta _{5}^{D}\varepsilon
^{\mu _{1}\mu _{2}\mu _{3}\mu _{4}\mu _{5}}\bar{V}_{\mu _{1}}\bar{F}_{\mu
_{2}\mu _{3}}\bar{F}_{\mu _{4}\mu _{5}}\right) \right]  \label{LagactcaseI}
\end{eqnarray}%
and is invariant under \emph{the deformed gauge transformations}
\begin{equation}
\delta _{\epsilon }^{\mathrm{(I)}}g_{\mu \nu }=k\epsilon _{(\mu ;\nu
)},\qquad \delta _{\epsilon }^{\mathrm{(I)}}\bar{V}_{\mu }=\partial _{\mu
}\epsilon +k\left( \partial _{\mu }\bar{\epsilon}^{\nu }\right) \bar{V}_{\nu
}+k\left( \partial _{\nu }\bar{V}_{\mu }\right) \bar{\epsilon}^{\nu }.
\label{gaugesymcaseI}
\end{equation}%
The validity of (\ref{LagactcaseI}) and (\ref{gaugesymcaseI}) to all orders
in the coupling constant can be done by developing the same technique used
in Section 7 of~\cite{multi}.

\subsubsection{Case II: new couplings in $D=3$\label{speccoupl}}

As discussed in Section \ref{caseII}, the second case of interest allowing
for nontrivial, consistent couplings between a Pauli-Fierz field and a
vector field is pictured by the deformed solution to the master equation
given in (\ref{SfincaseII.1.1}). We can re-write the deformation in a more
convenient way by adding to (\ref{SfincaseII.1.1}) some $s$-exact terms,
since we know that this does not affect the physical content of the coupled
model (see (\ref{a24})). Because the most general couplings in case II are
obtained in subcase II.1.1, described by conditions (\ref{subcaseII.1.1}),
we will denote the deformed solution (\ref{SfincaseII.1.1}) to which we add
the previously mentioned $s$-exact terms and where we set $y_{3}=1$ by $%
S^{\left( \mathrm{II}\right) }$%
\begin{eqnarray}
&&S^{\left( \mathrm{II}\right) }\equiv \left. S^{\left( \mathrm{II.1.1}%
\right) }\right\vert _{y_{3}=1}-s\left[ 2k^{2}\int d^{3}x\left( h^{\ast \mu
\nu }h_{\mu \nu }+\eta ^{\ast \mu }\eta _{\mu }\right) \right]  \notag \\
&=&\int d^{3}x\left[ \mathcal{L}_{0}^{\left( \mathrm{PF}\right) }-\frac{1}{4}%
F_{\mu \nu }F^{\mu \nu }-2k\Lambda h\right.  \notag \\
&&-kF^{\mu \nu }\varepsilon _{\mu \nu \rho }\partial ^{\lbrack \theta }h_{\
\ \theta }^{\rho ]}+2k^{2}\left( \partial ^{\lbrack \mu }h_{\ \ \mu }^{\rho
]}\right) \partial _{\lbrack \nu }h_{\rho ]}^{\ \ \nu }  \notag \\
&&\left. +h^{\ast \mu \nu }\partial _{(\mu }\eta _{\nu )}+V^{\ast \mu
}\left( \partial _{\mu }\eta +k\varepsilon _{\mu \nu \rho }\partial
^{\lbrack \nu }\eta ^{\rho ]}\right) \right] .  \label{SfincaseII}
\end{eqnarray}%
Essentially, it is \emph{not} trivial and \emph{is consistent} to all orders
in the coupling constant, namely%
\begin{equation}
\left( S^{\left( \mathrm{II}\right) },S^{\left( \mathrm{II}\right) }\right)
=0.  \label{consSII}
\end{equation}

From the terms of antighost number zero we deduce the Lagrangian action of
the coupled model%
\begin{eqnarray}
S^{\mathrm{L(II)}}[h_{\mu \nu },V_{\mu }] &=&\int d^{3}x\left[ \mathcal{L}%
_{0}^{\left( \mathrm{PF}\right) }-\frac{1}{4}F_{\mu \nu }F^{\mu \nu
}-2k\Lambda h\right.  \notag \\
&&\left. -kF^{\mu \nu }\varepsilon _{\mu \nu \rho }\partial ^{[ \theta }h_{\
\ \theta }^{\rho ]}+2k^{2}\left( \partial ^{[ \mu }h_{\ \ \mu }^{\rho
]}\right) \partial _{[ \nu }h_{\rho ]}^{\ \ \nu }\right] ,
\label{LagactcaseII}
\end{eqnarray}%
where $\mathcal{L}_{0}^{\left( \mathrm{PF}\right) }$ is the Pauli-Fierz
Lagrangian and $\Lambda $ is the cosmological constant. The component of
antighost number one provides the gauge symmetries of (\ref{LagactcaseII})
(see the discussion from the preamble of Section \ref{analysis})%
\begin{equation}
\delta _{\epsilon }^{\mathrm{(II)}}h_{\mu \nu }=\partial _{(\mu }\epsilon
_{\nu )},\qquad \delta _{\epsilon }^{\mathrm{(II)}}V_{\mu }=\partial _{\mu
}\epsilon +k\varepsilon _{\mu \nu \rho }\partial ^{[ \nu }\epsilon ^{\rho ]}.
\label{gaugecaseII}
\end{equation}%
The absence of terms of antighost number strictly greater than one shows
that the above gauge transformations are independent (irreducible) and their
algebra remains Abelian, like the original one. Action (\ref{LagactcaseII})
can be set in a more suggestive form by introducing a deformed field strength%
\begin{equation}
F_{\mu \nu }^{\prime }=F_{\mu \nu }+2k\varepsilon _{\mu \nu \rho }\partial
^{[ \theta }h_{\ \ \theta }^{\rho ]},  \label{deffieldstr}
\end{equation}%
in terms of which we can write%
\begin{equation}
S^{\mathrm{L(II)}}[h_{\mu \nu },V_{\mu }]=\int d^{3}x\left( \mathcal{L}%
_{0}^{\left( \mathrm{PF}\right) }-2k\Lambda h-\frac{1}{4}F_{\mu \nu
}^{\prime }F^{\prime \mu \nu }\right) .  \label{LagactcaseIIfin}
\end{equation}%
Under this form, action (\ref{LagactcaseIIfin}) is manifestly invariant
under the gauge transformations (\ref{gaugecaseII}): its first two terms are
known to be invariant under linearized diffeomorphisms and the third is
gauge-invariant under (\ref{gaugecaseII}) since the deformed field strength
is so%
\begin{equation}
\delta _{\epsilon }^{\mathrm{(II)}}F_{\mu \nu }^{\prime }=0.
\label{invfieldstr}
\end{equation}

This result is new and will be generalized in Section \ref{comm} to the case
of couplings between a graviton and an arbitrary $p$-form.\ In conclusion,
this case yields another possibility to establish nontrivial couplings
between the Pauli-Fierz field and a vector field. It is \emph{complementary
to} case I (\emph{General Relativity}) and is valid only in $D=3$. The
resulting Lagrangian action and gauge transformations are \emph{not} series
in the coupling constant. The Lagrangian contains pieces of maximum order
two in the coupling constant, which are mixing-component terms (there is no
interaction vertex at least cubic in the fields) and emphasize the
deformation of the standard Abelian field strength of the vector field like
in (\ref{deffieldstr}). Concerning the new gauge transformations, only those
of the massless vector field are modified at order one in the coupling
constant by adding to the original $U\left( 1\right) $ gauge symmetry a term
linear in the antisymmetric first-order derivatives of the Pauli-Fierz gauge
parameters. As a consequence, the gauge algebra, defined by the commutators
among the deformed gauge transformations, remains Abelian, just like for the
free theory. We cannot stress enough that these two cases (I and II) cannot
coexist, even in $D=3$, due to the consistency conditions (\ref{a85a})--(\ref%
{a85c}).

\section{No cross-couplings in multi-graviton theories intermediated by a
vector field\label{manyspintwoem}}

As it has been proved in~\cite{multi}, there are no direct
cross-couplings that can be introduced among a finite collection of
gravitons and also no cross-couplings among different gravitons
intermediated by a scalar field. Similar conclusions have been drawn
in~\cite{noijhepdirac,noiRS} related to the couplings between a
finite collection of spin-two fields and a Dirac or a massive
Rarita-Schwinger field. In this section, under the same hypotheses
like before, namely, locality, smoothness in the coupling constant,
Poincar\'{e} invariance, Lorentz covariance, and preservation of the
number of derivatives on each field, we investigate the existence of
cross-couplings among different gravitons intermediated by a
massless vector field. The Greek field indices are (Lorentz) flat:
they are lowered and raised with a flat metric of `mostly plus'
signature, $\sigma _{\mu \nu }=\left( -+\ldots +\right) $.

\subsection{First- and second-order deformations. Consistency conditions}

\subsubsection{Generalities}

We start now from a finite sum of Pauli-Fierz actions and a single Maxwell
action in $D>2$
\begin{eqnarray}
S_{0}^{\mathrm{L}}\left[ h_{\mu \nu }^{A},V_{\mu }\right] &=&\int d^{D}x%
\left[ -\frac{1}{2}\left( \partial _{\mu }h_{\nu \rho }^{A}\right) \partial
^{\mu }h_{A}^{\nu \rho }+\left( \partial _{\mu }h_{A}^{\mu \rho }\right)
\partial ^{\nu }h_{\nu \rho }^{A}\right.  \notag \\
&&\left. -\left( \partial _{\mu }h^{A}\right) \partial _{\nu }h_{A}^{\nu \mu
}+\frac{1}{2}\left( \partial _{\mu }h^{A}\right) \partial ^{\mu }h_{A}-\frac{%
1}{4}F_{\mu \nu }F^{\mu \nu }\right] ,  \label{i1}
\end{eqnarray}%
where $h_{A}$ is the trace of the Pauli-Fierz field $h_{A}^{\mu \nu }$ ($%
h_{A}=\sigma _{\mu \nu }h_{A}^{\mu \nu }$), with $A=\overline{1,n}$ and $n>1$%
. The collection indices $A$, $B$, etc., are raised and lowered with a
quadratic form $k_{AB}$ that determines a positively-defined metric in the
internal space. It can always be normalized to $\delta _{AB}$ by a simple
linear field redefinition, so from now on we take $k_{AB}=\delta _{AB}$ and
re-write (\ref{i1}) as%
\begin{equation}
S_{0}^{\mathrm{L}}\left[ h_{\mu \nu }^{A},V_{\mu }\right] =\int d^{D}x\left[
\sum_{A=1}^{n}\mathcal{L}_{0}^{\left( \mathrm{PF}\right) }\left( h_{\mu \nu
}^{A},\partial _{\lambda }h_{\mu \nu }^{A}\right) +\mathcal{L}_{0}^{\left(
\mathrm{vect}\right) }\right] ,  \label{i1a}
\end{equation}%
where $\mathcal{L}_{0}^{\left( \mathrm{PF}\right) }\left( h_{\mu \nu
}^{A},\partial _{\lambda }h_{\mu \nu }^{A}\right) $ is the Pauli-Fierz
Lagrangian for the graviton $A$. Action (\ref{i1}) is invariant under the
gauge transformations
\begin{equation}
\delta _{\epsilon }h_{\mu \nu }^{A}=\partial _{(\mu }\epsilon _{\nu
)}^{A},\qquad \delta _{\epsilon }V_{\mu }=\partial _{\mu }\epsilon .
\label{i2}
\end{equation}%
The BRST complex comprises the fields, ghosts, and antifields
\begin{eqnarray}
\hat{\Phi}^{\alpha _{0}} &=&(h_{\mu \nu }^{A},V_{\mu }),\qquad \hat{\eta}%
_{\alpha _{1}}=(\eta _{\mu }^{A},\eta ),  \label{i3} \\
\hat{\Phi}_{\alpha _{0}}^{\ast } &=&(h_{A}^{\ast \mu \nu },V^{\ast \mu
}),\qquad \hat{\eta}^{\ast \alpha _{1}}=(\eta _{A}^{\ast \mu },\eta ^{\ast
}),  \label{i4}
\end{eqnarray}%
whose degrees are the same like in the case of a single Pauli-Fierz field.
The BRST differential decomposes exactly like in (\ref{a3}) and its
components act on the BRST generators via the relations
\begin{eqnarray}
\delta h_{A}^{\ast \mu \nu } &=&2H_{A}^{\mu \nu },\qquad \delta V^{\ast \mu
}=-\partial _{\nu }F^{\nu \mu },  \label{i5} \\
\delta \eta _{A}^{\ast \mu } &=&-2\partial _{\nu }h_{A}^{\ast \nu \mu
},\qquad \delta \eta ^{\ast }=-\partial _{\mu }V^{\ast \mu },  \label{i6} \\
\delta \hat{\Phi}^{\alpha _{0}} &=&0,\qquad \delta \hat{\eta}_{\alpha
_{1}}=0,  \label{i7} \\
\gamma \hat{\Phi}_{\alpha _{0}}^{\ast } &=&0,\qquad \gamma \hat{\eta}^{\ast
\alpha _{1}}=0,  \label{i8} \\
\gamma h_{\mu \nu }^{A} &=&\partial _{(\mu }\eta _{\nu )}^{A},\qquad \gamma
V_{\mu }=\partial _{\mu }\eta ,  \label{i9} \\
\gamma \eta _{\mu }^{A} &=&0,\qquad \gamma \eta =0,  \label{i10}
\end{eqnarray}%
where $H_{A}^{\mu \nu }=K_{A}^{\mu \nu }-\frac{1}{2}\sigma ^{\mu \nu }K_{A}$
is the linearized Einstein tensor of the Pauli-Fierz field $h_{A}^{\mu \nu }$%
. The solution to the master equation for this free model takes the simple
form%
\begin{equation}
\bar{S}^{\prime }=S_{0}^{\mathrm{L}}\left[ h_{\mu \nu }^{A},V_{\mu }\right]
+\int d^{D}x\left( h_{A}^{\ast \mu \nu }\partial _{(\mu }\eta _{\nu
)}^{A}+V^{\ast \mu }\partial _{\mu }\eta \right) .  \label{i11}
\end{equation}

\subsubsection{First-order deformation}

The first-order deformation of the solution to the master equation
decomposes like in the case of a single graviton in a sum of three
independent components%
\begin{equation}
\hat{a}=\hat{a}^{\left( \mathrm{PF}\right) }+\hat{a}^{\left( \mathrm{int}%
\right) }+\hat{a}^{\left( \mathrm{vect}\right) }.  \label{i12}
\end{equation}%
The first-order deformation in the Pauli-Fierz sector, $\hat{a}^{\left(
\mathrm{PF}\right) }$, can be shown to expand as%
\begin{equation}
\hat{a}^{\left( \mathrm{PF}\right) }=\hat{a}_{2}^{\left( \mathrm{PF}\right)
}+\hat{a}_{1}^{\left( \mathrm{PF}\right) }+\hat{a}_{0}^{\left( \mathrm{PF}%
\right) },  \label{i13}
\end{equation}%
where
\begin{equation}
\hat{a}_{2}^{\left( \mathrm{PF}\right) }=\frac{1}{2}f_{BC}^{A}\eta
_{A}^{\ast \mu }\eta ^{B\nu }\partial _{[ \mu }\eta _{\nu ]}^{C},
\label{i14}
\end{equation}%
with $f_{BC}^{A}$ some real constants. The requirement that $\hat{a}%
_{2}^{\left( \mathrm{PF}\right) }$ produces a consistent $\hat{a}%
_{1}^{\left( \mathrm{PF}\right) }$ as solution to the equation $\delta \hat{a%
}_{2}^{\left( \mathrm{PF}\right) }+\gamma \hat{a}_{1}^{\left( \mathrm{PF}%
\right) }=\partial _{\mu }\hat{m}_{1}^{\left( \mathrm{PF}\right) \mu }$
restricts the coefficients $f_{BC}^{A}$ to be symmetric with respect to
their lower indices (commutativity of the algebra defined by $f_{BC}^{A}$)~%
\cite{multi}\footnote{%
The term (\ref{i14}) differs from that corresponding to~\cite{multi} through
a $\gamma $-exact term, which does not affect (\ref{i15}).}
\begin{equation}
f_{BC}^{A}=f_{CB}^{A}.  \label{i15}
\end{equation}%
Based on (\ref{i15}), it follows that
\begin{equation}
\hat{a}_{1}^{\left( \mathrm{PF}\right) }=f_{BC}^{A}h_{A}^{\ast \mu \rho
}\left( \left( \partial _{\rho }\eta ^{B\nu }\right) h_{\mu \nu }^{C}-\eta
^{B\nu }\partial _{[ \mu }h_{\nu ]\rho }^{C}\right) .  \label{i16}
\end{equation}%
Asking that $\hat{a}_{1}^{\left( \mathrm{PF}\right) }$ provides a consistent
$\hat{a}_{0}^{\left( \mathrm{PF}\right) }$ as solution to the equation $%
\delta \hat{a}_{1}^{\left( \mathrm{PF}\right) }+\gamma \hat{a}_{0}^{\left(
\mathrm{PF}\right) }=\partial _{\mu }\hat{m}_{0}^{\left( \mathrm{PF}\right)
\mu }$ further constrains the coefficients with lowered indices, $%
f_{ABC}=k_{AD}f_{BC}^{D}\equiv \delta _{AD}f_{BC}^{D}$, to be fully
symmetric~\cite{multi}\footnote{%
The piece (\ref{i16}) differs from that corresponding to~\cite{multi}
through a $\delta $-exact term, which does not change (\ref{i17}).}
\begin{equation}
f_{ABC}=\frac{1}{3}f_{\left( ABC\right) }.  \label{i17}
\end{equation}%
From (\ref{i17}) we obtain that $\hat{a}_{0}^{\left( \mathrm{PF}\right) }$
coincides with that from~\cite{multi} (where it is denoted by $a_{0}$ and
the coefficients $f_{ABC}$ by $a_{abc}$)%
\begin{equation}
\hat{a}_{0}^{\left( \mathrm{PF}\right) }=f_{ABC}\hat{a}_{0}^{\left( \mathrm{%
cubic}\right) ABC}-2\Lambda _{A}h^{A},  \label{i16a}
\end{equation}%
where $\hat{a}_{0}^{\left( \mathrm{cubic}\right) ABC}$ contains only
vertices that are cubic in the Pauli-Fierz fields and reduce to the cubic
Einstein-Hilbert vertex in the absence of collection indices. $\Lambda _{A}$
play the role of cosmological constants. Employing exactly the same line
like in \ref{deformareaI}, we find that the first-order deformation giving
the cross-couplings between the gravitons and the vector fields ends at
antighost number one
\begin{equation}
\hat{a}^{\left( \mathrm{int}\right) }=\hat{a}_{1}^{\left( \mathrm{int}%
\right) }+\hat{a}_{0}^{\left( \mathrm{int}\right) },  \label{alphaint}
\end{equation}%
where
\begin{eqnarray}
\hat{a}_{1}^{\left( \mathrm{int}\right) } &=&y_{2A}\left[ h^{\ast A}\eta
-\left( D-2\right) V^{\ast \lambda }\eta _{\lambda }^{A}\right]  \notag \\
&&+y_{3}^{A}\delta _{3}^{D}\varepsilon _{\mu \nu \rho }V^{\ast \mu }\partial
^{[ \nu }\eta _{A}^{\rho ]}+p_{A}V^{\ast \mu }F_{\mu \nu }\eta ^{A\nu },
\label{i18}
\end{eqnarray}%
\begin{eqnarray}
\hat{a}_{0}^{\left( \mathrm{int}\right) } &=&\left( D-2\right)
y_{2A}V^{\lambda }\partial _{[ \mu }h_{\lambda ]}^{A\ \mu }+y_{3}^{A}\delta
_{3}^{D}\varepsilon _{\mu \nu \rho }F^{\lambda \mu }\partial ^{[ \nu }h_{A\
\lambda }^{\rho ]}  \notag \\
&&+\frac{p_{A}}{2}\left( F^{\alpha \mu }F_{\mu }^{\;\;\nu }h_{\alpha \nu
}^{A}+\frac{1}{4}F^{\alpha \mu }F_{\alpha \mu }h^{A}\right)  \label{i19}
\end{eqnarray}%
and $y_{2A}$, $y_{3}^{A}$ together with $p_{A}$ are some arbitrary, real
constants. Like in Section \ref{deformareaI}, we eliminate some $s$-exact
modulo $d$ terms from $\hat{a}^{\left( \mathrm{int}\right) }$ and work with%
\begin{equation}
\hat{a}^{\prime \left( \mathrm{int}\right) }=\hat{a}^{\left( \mathrm{int}%
\right) }+s\left[ p_{A}\left( \eta ^{\ast }V^{\mu }\eta _{\mu }^{A}+\frac{1}{%
2}V^{\ast \mu }V^{\nu }h_{\mu \nu }^{A}\right) \right] -\partial _{\mu }\hat{%
t}^{\mu }.  \label{i19a}
\end{equation}%
The component $\hat{a}^{\left( \mathrm{vect}\right) }$ coincides with that
from Section \ref{deformareaI} (see (\ref{rr2}))
\begin{equation}
\hat{a}^{\left( \mathrm{vect}\right) }=a^{\left( \mathrm{vect}\right)
}=q_{1}\delta _{3}^{D}\varepsilon ^{\mu \nu \lambda }V_{\mu }F_{\nu \lambda
}+q_{2}\delta _{5}^{D}\varepsilon ^{\mu \nu \lambda \alpha \beta }V_{\mu
}F_{\nu \lambda }F_{\alpha \beta }.  \label{ig1}
\end{equation}

Putting together (\ref{i13}) and (\ref{alphaint})--(\ref{ig1}) with the help
of (\ref{i12}), we can write the first-order deformation of the solution to
the master for a single vector field and a collection of Pauli-Fierz fields
like%
\begin{equation}
\hat{S}_{1}=\hat{S}_{1}^{(\mathrm{PF})}+\hat{S}_{1}^{\left( \mathrm{int}%
\right) },  \label{S1coll}
\end{equation}%
where%
\begin{eqnarray}
\hat{S}_{1}^{(\mathrm{PF})} &\equiv &\int d^{D}x\left( \hat{a}_{2}^{\left(
\mathrm{PF}\right) }+\hat{a}_{1}^{\left( \mathrm{PF}\right) }+\hat{a}%
_{0}^{\left( \mathrm{PF}\right) }\right)  \notag \\
&=&\int d^{D}x\left\{ \frac{1}{2}f_{BC}^{A}\eta _{A}^{\ast \mu }\eta ^{B\nu
}\partial _{[ \mu }\eta _{\nu ]}^{C}+f_{BC}^{A}h_{A}^{\ast \mu \rho }\left[
\left( \partial _{\rho }\eta ^{B\nu }\right) h_{\mu \nu }^{C}\right. \right.
\notag \\
&&\left. \left. -\eta ^{B\nu }\partial _{[ \mu }h_{\nu ]\rho }^{C} \right]
+f_{ABC}\hat{a}_{0}^{\left( \mathrm{cubic}\right) ABC}-2\Lambda
_{A}h^{A}\right\} ,  \label{S1PFcoll}
\end{eqnarray}%
\begin{eqnarray}
\hat{S}_{1}^{\left( \mathrm{int}\right) } &\equiv &\int d^{D}x\left( \hat{a}%
^{\prime \left( \mathrm{int}\right) }+\hat{a}^{\left( \mathrm{vect}\right)
}\right)  \notag \\
&=&\int d^{D}x\left\{ y_{2A}\left[ h^{\ast A}\eta +\left( D-2\right) \left(
-V^{\ast \lambda }\eta _{\lambda }^{A}+V^{\lambda }\partial _{[ \mu
}h_{\lambda ]}^{A\mu }\right) \right] \right.  \notag \\
&&+y_{3}^{A}\delta _{3}^{D}\varepsilon _{\mu \nu \rho }\left( V^{\ast \mu
}\partial ^{[ \nu }\eta _{A}^{\rho ]}+F^{\lambda \mu }\partial ^{[ \nu
}h_{A\ \lambda }^{\rho ]}\right) +p_{A}\left[ \eta ^{\ast }\eta _{\mu
}^{A}\partial ^{\mu }\eta \right.  \notag \\
&&-\frac{1}{2}V^{\ast \mu }\left( V^{\nu }\partial _{[ \mu }\eta _{\nu
]}^{A}+2\left( \partial _{\nu }V_{\mu }\right) \eta ^{A\nu }-h_{\mu \nu
}^{A}\partial ^{\nu }\eta \right)  \notag \\
&&\left. +\frac{1}{8}F^{\mu \nu }\left( 2\partial _{[ \mu }\left( h_{\nu
]\rho }^{A}V^{\rho }\right) +F_{\mu \nu }h^{A}-4F_{\mu \rho }h_{\nu }^{A\rho
}\right) \right]  \notag \\
&&\left. +q_{1}\delta _{3}^{D}\varepsilon ^{\mu \nu \lambda }V_{\mu }F_{\nu
\lambda }+q_{2}\delta _{5}^{D}\varepsilon ^{\mu \nu \lambda \alpha \beta
}V_{\mu }F_{\nu \lambda }F_{\alpha \beta }\right\} .  \label{S1intcoll}
\end{eqnarray}%
It is parameterized by seven types of real, constant coefficients, namely $%
f_{BC}^{A}$, $\Lambda _{A}$, $y_{2A}$, $y_{3}^{A}\delta _{3}^{D}$, $p_{A}$, $%
q_{1}\delta _{3}^{D}$, and $q_{2}\delta _{5}^{D}$, with $f_{BC}^{A}$ fully
symmetric (see (\ref{i17})).

\subsubsection{Consistency of the first-order deformation}

Next, we investigate the consistency of the first-order deformation,
expressed by equation (\ref{a21}), with $S_{1,2}$ replaced by $\hat{S}_{1,2}$%
\begin{equation}
\left( \hat{S}_{1},\hat{S}_{1}\right) +2s\hat{S}_{2}=0.  \label{a21coll}
\end{equation}%
We decompose the second-order deformation as
\begin{equation}
\hat{S}_{2}=\hat{S}_{2}^{\left( \mathrm{PF}\right) }+\hat{S}_{2}^{\left(
\mathrm{int}\right) },  \label{S2dec}
\end{equation}%
where $\hat{S}_{2}^{\left( \mathrm{PF}\right) }$ is responsible only for the
self-interactions of the Pauli-Fierz fields and $\hat{S}_{2}^{\left( \mathrm{%
int}\right) }$ for the cross-couplings between the gravitons and the vector
field. Using (\ref{S1coll}), we find that (\ref{a21coll}) becomes equivalent
with two independent equations
\begin{eqnarray}
\left( \hat{S}_{1}^{\left( \mathrm{PF}\right) },\hat{S}_{1}^{\left( \mathrm{%
PF}\right) }\right) +\left( \hat{S}_{1}^{\left( \mathrm{int}\right) },\hat{S}%
_{1}^{\left( \mathrm{int}\right) }\right) ^{\left( \mathrm{PF}\right) }+2s%
\hat{S}_{2}^{\left( \mathrm{PF}\right) } &=&0,  \label{i24} \\
2\left( \hat{S}_{1}^{\left( \mathrm{PF}\right) },\hat{S}_{1}^{\left( \mathrm{%
int}\right) }\right) +\left( \hat{S}_{1}^{\left( \mathrm{int}\right) },\hat{S%
}_{1}^{\left( \mathrm{int}\right) }\right) ^{\left( \mathrm{int}\right) }+2s%
\hat{S}_{2}^{\left( \mathrm{int}\right) } &=&0,  \label{i25}
\end{eqnarray}%
where $\left( \hat{S}_{1}^{\left( \mathrm{int}\right) },\hat{S}_{1}^{\left(
\mathrm{int}\right) }\right) ^{\left( \mathrm{PF}\right) }$ contains only
Pauli-Fierz BRST generators and each term of $\left( \hat{S}_{1}^{\left(
\mathrm{int}\right) },\hat{S}_{1}^{\left( \mathrm{int}\right) }\right)
^{\left( \mathrm{int}\right) }$ includes at least one BRST generator from
the Maxwell sector.

Initially, we analyze the existence of $\hat{S}_{2}^{\left( \mathrm{PF}%
\right) }$, governed by equation (\ref{i24}). By direct computation we find%
\begin{eqnarray}
&&\left( \hat{S}_{1}^{\left( \mathrm{int}\right) },\hat{S}_{1}^{\left(
\mathrm{int}\right) }\right) ^{\left( \mathrm{PF}\right) }=-2s\int d^{D}x%
\left[ y_{2A}y_{2B}\frac{\left( D-2\right) ^{2}}{4}\left( h^{A}h^{B}-h^{A\mu
\nu }h_{\mu \nu }^{B}\right) \right.  \notag \\
&&\left. +y_{2A}y_{3}^{B}\delta _{3}^{D}\left( D-2\right) \varepsilon _{\mu
\nu \rho }h_{\ \lambda }^{A\mu }\left( \partial ^{[ \nu }h_{B}^{\rho
]\lambda }\right) +y_{3}^{A}y_{3B}\delta _{3}^{D}\left( \partial ^{[ \nu
}h_{A}^{\rho ]\lambda }\right) \partial _{[ \nu }h_{\rho ]\lambda }^{B}
\right]  \notag \\
&\equiv &-2s\left( \hat{S}_{2}^{\left( \mathrm{PF}\right) }\left(
y_{2A}y_{2B}\right) +\hat{S}_{2}^{\left( \mathrm{PF}\right) }\left(
y_{2A}y_{3}^{B}\right) +\hat{S}_{2}^{\left( \mathrm{PF}\right) }\left(
y_{3}^{A}y_{3B}\right) \right) ,  \label{S1S1intPF}
\end{eqnarray}%
where%
\begin{eqnarray}
\hat{S}_{2}^{\left( \mathrm{PF}\right) }\left( y_{2A}y_{2B}\right)
&=&y_{2A}y_{2B}\frac{\left( D-2\right) ^{2}}{4}\int d^{D}x\left(
h^{A}h^{B}-h^{A\mu \nu }h_{\mu \nu }^{B}\right) ,  \label{S2PFy2} \\
\hat{S}_{2}^{\left( \mathrm{PF}\right) }\left( y_{2A}y_{3}^{B}\right)
&=&y_{2A}y_{3}^{B}\delta _{3}^{D}\left( D-2\right) \varepsilon _{\mu \nu
\rho }\int d^{D}xh_{\ \lambda }^{A\mu }\left( \partial ^{[ \nu }h_{B}^{\rho
]\lambda }\right) ,  \label{S2PFy23} \\
\hat{S}_{2}^{\left( \mathrm{PF}\right) }\left( y_{3}^{A}y_{3}^{B}\right)
&=&y_{3}^{A}y_{3B}\delta _{3}^{D}\int d^{D}x\left( \partial ^{[ \nu
}h_{A}^{\rho ]\lambda }\right) \partial _{[ \nu }h_{\rho ]\lambda }^{B}.
\label{S2PFy3}
\end{eqnarray}%
Replacing (\ref{S1S1intPF}) into (\ref{i24}), it becomes equivalent to
\begin{eqnarray}
&&\left( \hat{S}_{1}^{\left( \mathrm{PF}\right) },\hat{S}_{1}^{\left(
\mathrm{PF}\right) }\right) +2s\left[ \hat{S}_{2}^{\left( \mathrm{PF}\right)
}-\hat{S}_{2}^{\left( \mathrm{PF}\right) }\left( y_{2A}y_{2B}\right) \right.
\notag \\
&&\left. -\hat{S}_{2}^{\left( \mathrm{PF}\right) }\left(
y_{2A}y_{3}^{B}\right) -\hat{S}_{2}^{\left( \mathrm{PF}\right) }\left(
y_{3}^{A}y_{3}^{B}\right) \right] =0,  \label{new}
\end{eqnarray}%
so the existence of $\hat{S}_{2}^{\left( \mathrm{PF}\right) }$ requires that
$\left( \hat{S}_{1}^{\left( \mathrm{PF}\right) },\hat{S}_{1}^{\left( \mathrm{%
PF}\right) }\right) $ is $s$-exact, where $\hat{S}_{1}^{\left( \mathrm{PF}%
\right) }$ reads as in (\ref{S1PFcoll}). It has been shown in~\cite{multi}
(Section 5.4) that this requirement restricts the coefficients $f_{AB}^{C}$
to satisfy the supplementary conditions%
\begin{equation}
f_{A\left[ B\right. }^{D}f_{\left. C\right] D}^{E}=0.  \label{i26}
\end{equation}%
Combining (\ref{i15}), (\ref{i17}), and (\ref{i26}), we conclude that the
coefficients $f_{AB}^{C}$ define the structure constants of a real,
commutative, symmetric, and associative (finite-dimensional) algebra. The
analysis realized in~\cite{multi} (Section 6) shows that such an algebra has
a trivial structure: it is a direct sum of one-dimensional ideals.
Therefore, $f_{AB}^{C}=0$ whenever two indices are different%
\begin{equation}
f_{AB}^{C}=0,\qquad \mathrm{if}\qquad \left( A\neq B\qquad \mathrm{or}\qquad
B\neq C\qquad \mathrm{or}\qquad C\neq A\right) .  \label{fabcdif}
\end{equation}%
For notational simplicity, we denote $f_{ABC}$ for $A=B=C$ by%
\begin{equation}
f_{AAA}\equiv f_{A}\qquad \mathrm{without\ summation\ over}\ A.  \label{faaa}
\end{equation}%
Using (\ref{fabcdif}), it follows that $\left( \hat{S}_{1}^{\left( \mathrm{PF%
}\right) },\hat{S}_{1}^{\left( \mathrm{PF}\right) }\right) $ cannot couple
different gravitons: it will be written as a sum of $s$-exact terms, each
term involving a single graviton%
\begin{eqnarray}
&&\left( \hat{S}_{1}^{\left( \mathrm{PF}\right) },\hat{S}_{1}^{\left(
\mathrm{PF}\right) }\right) =-2s\left\{ \sum_{A=1}^{n}f_{A}\left[
f_{A}S_{2}^{(\mathrm{EH-quartic})A}+\Lambda _{A}\int d^{D}x\left( h^{A\mu
\nu }h_{\mu \nu }^{A}\right. \right. \right.  \notag \\
&&\left. \left. \left. -\frac{1}{2}\left( h^{A}\right) ^{2}\right) \right]
\right\} \equiv -2s\sum_{A=1}^{n}\hat{S}_{2}^{\left( \mathrm{PF}\right)
}\left( f_{A}^{2},f_{A}\Lambda _{A}\right) .  \label{newint2}
\end{eqnarray}%
Each $S_{2}^{(\mathrm{EH-quartic})A}$ is the second-order Einstein-Hilbert
deformation in the sector of the graviton $A$. It includes the quartic
Einstein-Hilbert Lagrangian for the field $h_{\mu \nu }^{A}$ and is written
\emph{only} in terms of the BRST generators from the $A$ sector, namely $%
h_{\mu \nu }^{A}$, $\eta ^{A\mu }$, and their antifields. Also, it is
important to note that (\ref{fabcdif}) restricts $\hat{S}_{1}^{\left(
\mathrm{PF}\right) }$ to have the same property (see (\ref{S1PFcoll})) of
being written as a sum of individual components, each component involving a
single graviton sector%
\begin{eqnarray}
&&\hat{S}_{1}^{\left( \mathrm{PF}\right) }=\sum_{A=1}^{n}\left\{ f_{A}\int
d^{D}x\left[ \frac{1}{2}\eta ^{\ast A\mu }\eta ^{A\nu }\partial _{[ \mu
}\eta _{\nu ]}^{A}+h^{\ast A\mu \rho }\left[ \left( \partial _{\rho }\eta
^{A\nu }\right) h_{\mu \nu }^{A}\right. \right. \right.  \notag \\
&&\left. \left. \left. -\eta ^{A\nu }\partial _{[ \mu }h_{\nu ]\rho }^{A}
\right] +\hat{a}_{0}^{(\mathrm{EH-cubic})A}\right] \right\}
-2\sum_{A=1}^{n}\left( \Lambda _{A}\int d^{D}x\,h^{A}\right) .
\label{S1PFcollfin}
\end{eqnarray}%
Now, $\hat{a}_{0}^{(\mathrm{EH-cubic})A}$ is nothing but the cubic
Einstein-Hilbert Lagrangian involving \emph{only} the graviton field $h_{\mu
\nu }^{A}$. Substituting (\ref{newint2}) into (\ref{new}) we find the
equation%
\begin{eqnarray}
&&s\left[ \hat{S}_{2}^{\left( \mathrm{PF}\right) }-\hat{S}_{2}^{\left(
\mathrm{PF}\right) }\left( y_{2A}y_{2B}\right) -\hat{S}_{2}^{\left( \mathrm{%
PF}\right) }\left( y_{2A}y_{3}^{B}\right) \right.  \notag \\
&&\left. -\hat{S}_{2}^{\left( \mathrm{PF}\right) }\left(
y_{3}^{A}y_{3}^{B}\right) -\sum_{A=1}^{n}\hat{S}_{2}^{\left( \mathrm{PF}%
\right) }\left( f_{A}^{2},f_{A}\Lambda _{A}\right) \right] =0,
\label{newint3}
\end{eqnarray}%
whose solution reads as (up to the solution of the homogeneous equation, $s%
\hat{S}_{2}^{\prime \left( \mathrm{PF}\right) }=0$, which can be
incorporated into (\ref{S1PFcollfin}) by a suitable redefinition of the
constants involved)%
\begin{eqnarray}
\hat{S}_{2}^{\left( \mathrm{PF}\right) } &=&\hat{S}_{2}^{\left( \mathrm{PF}%
\right) }\left( y_{2A}y_{2B}\right) +\hat{S}_{2}^{\left( \mathrm{PF}\right)
}\left( y_{2A}y_{3}^{B}\right) +\hat{S}_{2}^{\left( \mathrm{PF}\right)
}\left( y_{3}^{A}y_{3}^{B}\right)  \notag \\
&&+\sum_{A=1}^{n}\hat{S}_{2}^{\left( \mathrm{PF}\right) }\left(
f_{A}^{2},f_{A}\Lambda _{A}\right) .  \label{S2PFcollfin}
\end{eqnarray}%
Inspecting (\ref{S1PFcollfin}) and (\ref{S2PFcollfin}), we observe that the
latter component contains at this stage three pieces that mix different
graviton sectors, namely those proportional with $y_{iA}y_{jB}$ for $i,j=2,3$
and $A\neq B$.

Next, we approach the solution $\hat{S}_{2}^{\left( \mathrm{int}\right) }$
to equation (\ref{i25}). We act like in Section \ref{deformarea II}. If we
make the notations%
\begin{eqnarray}
2\left( \hat{S}_{1}^{\left( \mathrm{PF}\right) },\hat{S}_{1}^{\left( \mathrm{%
int}\right) }\right) +\left( \hat{S}_{1}^{\left( \mathrm{int}\right) },\hat{S%
}_{1}^{\left( \mathrm{int}\right) }\right) ^{\left( \mathrm{int}\right) }
&\equiv &\int d^{D}x\,\hat{\Delta}%
^{\left( \mathrm{int}\right) },  \label{notdeltacoll} \\
\hat{S}_{2}^{\left( \mathrm{int}\right) } &\equiv &\int d^{D}x\,\hat{b}%
^{\left( \mathrm{int}\right) },  \label{notS2intcoll}
\end{eqnarray}%
then equation (\ref{i25}) takes the local form%
\begin{equation}
\hat{\Delta}^{\left( \mathrm{int}\right) }=-2s\hat{b}^{\left( \mathrm{int}%
\right) }+\partial _{\mu }\hat{n}^{\mu }.  \label{i28}
\end{equation}%
Developing $\hat{\Delta}^{\left( \mathrm{int}\right) }$ according to the
antighost number, we obtain that
\begin{equation}
\hat{\Delta}^{\left( \mathrm{int}\right) }=\sum\limits_{I=0}^{2}\hat{\Delta}%
_{I}^{\left( \mathrm{int}\right) },\qquad \mathrm{agh}\left( \hat{\Delta}%
_{I}^{\left( \mathrm{int}\right) }\right) =I,\qquad I=\overline{0,2},
\label{i30}
\end{equation}%
with
\begin{eqnarray}
\hat{\Delta}_{2}^{\left( \mathrm{int}\right) } &=&\gamma \left[ \eta ^{\ast
}\left( p_{A}p_{B}\left( \partial ^{\mu }\eta \right) \eta ^{A\nu }h_{\mu
\nu }^{B}\right. \right.  \notag \\
&&\left. \left. -\left( f_{AB}^{C}p_{C}+p_{A}p_{B}\right) V^{\mu }\eta
^{A\nu }\partial _{[ \mu }\eta _{\nu ]}^{B}\right) \right] +\partial _{\mu }%
\hat{w}_{2}^{\mu },  \label{i301}
\end{eqnarray}%
\begin{eqnarray}
&&\hat{\Delta}_{1}^{\left( \mathrm{int}\right) }=\delta \left[ \eta ^{\ast
}\left( p_{A}p_{B}\left( \partial ^{\mu }\eta \right) \eta ^{A\nu }h_{\mu
\nu }^{B}-\left( f_{AB}^{C}p_{C}+p_{A}p_{B}\right) V^{\mu }\eta ^{A\nu
}\partial _{[ \mu }\eta _{\nu ]}^{B}\right) \right]  \notag \\
&&+\gamma \left\{ p_{A}p_{B}V^{\ast \mu }\left[ \left( \partial _{\nu
}V_{\mu }\right) h_{\ \ \rho }^{A\nu }\eta ^{B\rho }+\frac{1}{2}\left(
\partial _{[ \mu }h_{\nu ]\rho }^{A}\right) V^{\nu }\eta ^{B\rho }\right.
\right.  \notag \\
&&\left. -\frac{1}{4}V^{\nu }h_{[\mu }^{A\,\rho }\left( \partial _{\nu
]}\eta _{\rho }^{B}\right) -\frac{1}{4}V^{\nu }\left( \partial _{\rho }\eta
_{[ \mu }^{A}\right) h_{\nu ]}^{B\,\rho }-\frac{3}{4}h_{\mu }^{A\,\nu
}h_{\nu }^{B\,\rho }\partial _{\rho }\eta \right]  \notag \\
&&+\frac{1}{2}\left( f_{AB}^{C}p_{C}+p_{A}p_{B}\right) V^{\ast \mu }V^{\nu }%
\left[ \left( \partial _{[ \mu }h_{\rho ]\nu }^{A}+\partial _{[ \nu }h_{\rho
]\mu }^{A}\right) \eta ^{B\rho }\right.  \notag \\
&&\left. -h_{\mu }^{A\,\rho }\partial _{\nu }\eta _{\rho }^{B}-h_{\nu
}^{A\,\rho }\partial _{\mu }\eta _{\rho }^{B}\right] -\delta
_{3}^{D}\varepsilon ^{\mu \nu \rho }V_{\mu }^{\ast }\left[
y_{3C}f_{AB}^{C}h_{\nu }^{A\,\lambda }\partial _{[ \rho }\eta _{\lambda
]}^{B}\right.  \notag \\
&&\left. +\left( 2y_{3B}p_{A}+y_{3C}f_{AB}^{C}\right) \eta ^{A\lambda
}\partial _{[ \nu }h_{\rho ]\lambda }^{B}\right]  \notag \\
&&+y_{2A}p_{B}V^{\ast \mu }\left[ \left( D-2\right) h_{\mu \nu }^{A}\eta
^{B\nu }-\delta ^{AB}V_{\mu }\eta \right]  \notag \\
&&\left. -h^{\ast A\mu \nu }\left[ y_{2C}f_{AB}^{C}\left( h_{\mu \nu
}^{B}\eta +2V_{\mu }\eta _{\nu }^{B}\right) -2\left(
y_{2A}p_{B}+y_{2C}f_{AB}^{C}\right) \sigma _{\mu \nu }V^{\rho }\eta _{\rho
}^{B}\right] \right\}  \notag \\
&&-\left( f_{AB}^{C}p_{C}+p_{A}p_{B}\right) V_{\mu }^{\ast }F^{\mu \nu }\eta
^{A\rho }\partial _{[ \rho }\eta _{\nu ]}^{B}+V_{\mu }^{\ast }\left[ \left(
y_{3A}p_{B}+y_{3B}p_{A}\right. \right.  \notag \\
&&\left. +y_{3C}f_{AB}^{C}\right) \delta _{3}^{D}\varepsilon ^{\mu \nu \rho
}\left( \partial _{[ \nu }\eta _{\lambda ]}^{A}\right) \partial _{[ \rho
}\eta _{\tau ]}^{B}\sigma ^{\lambda \tau }+\left(
y_{2A}p_{B}+y_{2B}p_{A}\right.  \notag \\
&&\left. \left. +y_{2C}f_{AB}^{C}\right) \left( D-2\right) \sigma ^{\mu \nu
}\left( \partial _{[ \nu }\eta _{\rho ]}^{A}\right) \eta ^{B\rho } \right]
+\partial _{\mu }\hat{w}_{1}^{\mu },  \label{i31}
\end{eqnarray}%
\begin{eqnarray}
&&\hat{\Delta}_{0}^{\left( \mathrm{int}\right) }=\delta \left\{
p_{A}p_{B}V^{\ast \mu }\left[ \left( \partial _{\nu }V_{\mu }\right) h_{\ \
\rho }^{A\nu }\eta ^{B\rho }+\frac{1}{2}\left( \partial _{[ \mu }h_{\nu
]\rho }^{A}\right) V^{\nu }\eta ^{B\rho }\right. \right.  \notag \\
&&\left. -\frac{1}{4}V^{\nu }h_{[\mu }^{A\,\rho }\left( \partial _{\nu
]}\eta _{\rho }^{B}\right) -\frac{1}{4}V^{\nu }\left( \partial _{\rho }\eta
_{[ \mu }^{A}\right) h_{\nu ]}^{B\,\rho }-\frac{3}{4}h_{\mu }^{A\,\nu
}h_{\nu }^{B\,\rho }\partial _{\rho }\eta \right]  \notag \\
&&+\frac{1}{2}\left( f_{AB}^{C}p_{C}+p_{A}p_{B}\right) V^{\ast \mu }V^{\nu }%
\left[ \left( \partial _{[ \mu }h_{\rho ]\nu }^{A}+\partial _{[ \nu }h_{\rho
]\mu }^{A}\right) \eta ^{B\rho }\right.  \notag \\
&&\left. \left. -h_{\mu }^{A\,\rho }\partial _{\nu }\eta _{\rho }^{B}-h_{\nu
}^{A\,\rho }\partial _{\mu }\eta _{\rho }^{B}\right] +\frac{16}{D-2}%
y_{3A}q_{1}\delta _{3}^{D}h^{\ast A}\eta \right\}  \notag \\
&&+\gamma \left\{ \frac{p_{A}p_{B}}{8}\left[ V_{\rho }\left( \left( \partial
^{[ \mu }h^{A\nu ]\rho }\right) \left( \partial _{[ \mu }h_{\nu ]\lambda
}^{B}\right) V^{\lambda }-2\left( \partial ^{[ \mu }h^{A\nu ]\rho }\right)
h_{\lambda [ \mu }^{B}\left( \partial _{\nu ]}V^{\lambda }\right) \right)
\right. \right.  \notag \\
&&+h_{\rho }^{A\,[\mu }\left( \partial ^{\nu ]}V^{\rho }\right) h_{\lambda [
\mu }^{B}\left( \partial _{\nu ]}V^{\lambda }\right) +F^{\mu \nu }h_{\ \
\lambda }^{A\rho }\left( h_{\ \ [\mu }^{B\lambda }\left( \partial _{\nu
]}V_{\rho }\right) -\left( \partial _{[ \mu }h_{\ \ \nu ]}^{B\lambda
}\right) V_{\rho }\right)  \notag \\
&&\left. +F^{\mu \nu }h_{\ \ [\mu }^{A\rho }\left( \partial _{\nu ]}h_{\rho
}^{B\,\lambda }\right) V_{\lambda }\right] +p_{A}p_{B}F^{\mu \nu }\left[
F_{\mu \rho }h_{\nu }^{A\,\lambda }h_{\lambda }^{B\,\rho }+\frac{1}{16}%
F_{\mu \nu }\left( h^{A}h^{B}\right. \right.  \notag \\
&&\left. -2h^{A\rho \lambda }h_{\rho \lambda }^{B}\right) -h_{\nu }^{A\,\rho
}\left( \left( \partial _{[ \mu }h_{\rho ]}^{B\ \lambda }\right) V_{\lambda
}-h_{[\mu }^{B\ \lambda }\left( \partial _{\rho ]}V_{\lambda }\right) \right)
\notag \\
&&\left. +\frac{1}{2}\left( F^{\rho \lambda }h_{\mu \rho }^{A}h_{\nu \lambda
}^{B}-F_{\mu \rho }h_{\ \ \nu }^{A\rho }h^{B}\right) +\frac{1}{4}\left(
\left( \partial _{[ \mu }h_{\nu ]}^{A\ \rho }\right) V_{\rho }-h_{[\mu }^{A\
\rho }\left( \partial _{\nu ]}V_{\rho }\right) \right) h^{B}\right]  \notag
\\
&&+\frac{1}{4}\left( f_{AB}^{C}p_{C}+p_{A}p_{B}\right) \left( F^{\mu \nu
}F_{\nu \rho }+\frac{1}{4}\delta _{\rho }^{\mu }F^{\nu \lambda }F_{\nu
\lambda }\right) h_{\mu \sigma }^{A}h^{B\sigma \rho }  \notag \\
&&+q_{1}\delta _{3}^{D}p_{A}\varepsilon ^{\mu \nu \lambda }\left(
h^{A}V_{\mu }F_{\nu \lambda }-2h_{\lambda }^{A\,\alpha }V_{\mu }F_{\nu
\alpha }+h_{\mu }^{A\,\alpha }V_{\alpha }F_{\nu \lambda }\right)  \notag \\
&&+q_{2}\delta _{5}^{D}p_{A}\varepsilon ^{\mu \nu \lambda \alpha \beta
}\left( h^{A}V_{\mu }F_{\nu \lambda }F_{\alpha \beta }-4h_{\beta }^{A\,\rho
}V_{\mu }F_{\nu \lambda }F_{\alpha \rho }+2h_{\mu }^{A\,\rho }V_{\rho
}F_{\nu \lambda }F_{\alpha \beta }\right)  \notag \\
&&\left. -16y_{3A}q_{1}\delta _{3}^{D}V^{\nu }\partial _{[ \nu }h_{\rho
]}^{A\,\rho }-\left( D-2\right) \left( D-1\right) y_{2A}y_{2}^{A}V_{\mu
}V^{\mu }\right\}  \notag \\
&&-4q_{1}\delta _{3}^{D}y_{2A}\left( D-2\right) \varepsilon _{\mu \nu \rho
}F^{\mu \nu }\eta ^{A\rho }-6q_{2}\delta _{5}^{D}y_{2A}\varepsilon _{\mu \nu
\rho \alpha \beta }F^{\mu \nu }F^{\rho \alpha }\eta ^{A\beta }  \notag \\
&&+\frac{1}{2}\left( f_{AB}^{C}p_{C}+p_{A}p_{B}\right) \left( F^{\mu \nu
}F_{\nu \rho }+\frac{1}{4}\delta _{\rho }^{\mu }F^{\nu \lambda }F_{\nu
\lambda }\right) \left( h^{A\rho \sigma }\partial _{[ \mu }\eta _{\sigma
]}^{B}\right.  \notag \\
&&\left. -2\partial _{[ \mu }h_{\ \ \sigma ]}^{A\rho }\eta ^{B\sigma
}\right) +y_{2A}\left[ -4D\Lambda ^{A}\eta +f_{BC}^{A}\hat{A}_{0}^{\left(
\mathrm{int}\right) BC}\left( \partial \partial \hat{\Phi}^{\alpha _{0}}\hat{%
\Phi}^{\beta _{0}}\hat{\eta}_{\alpha _{1}}\right) \right.  \notag \\
&&\left. +p_{B}\hat{B}_{0}^{\left( \mathrm{int}\right) AB}\left( \partial
\partial \hat{\Phi}^{\alpha _{0}}\hat{\Phi}^{\beta _{0}}\hat{\eta}_{\alpha
_{1}}\right) \right] +y_{3A}\delta _{3}^{D}\left[ f_{BC}^{A}\hat{C}%
_{0}^{\left( \mathrm{int}\right) BC}\left( \partial \partial \partial \hat{%
\Phi}^{\alpha _{0}}\hat{\Phi}^{\beta _{0}}\hat{\eta}_{\alpha _{1}}\right)
\right.  \notag \\
&&\left. +p_{B}\hat{D}_{0}^{\left( \mathrm{int}\right) AB}\left( \partial
\partial \partial \hat{\Phi}^{\alpha _{0}}\hat{\Phi}^{\beta _{0}}\hat{\eta}%
_{\alpha _{1}}\right) \right] +\partial _{\mu }\hat{w}_{0}^{\mu }.
\label{i31.1}
\end{eqnarray}%
In (\ref{i31.1}) the functions $\hat{A}_{0}^{\left( \mathrm{int}\right) BC}$%
, $\hat{B}_{0}^{\left( \mathrm{int}\right) AB}$, $\hat{C}_{0}^{\left(
\mathrm{int}\right) BC}$, and $\hat{D}_{0}^{\left( \mathrm{int}\right) AB}$
are linear in their arguments, just like in (\ref{a69}).

Acting exactly like in the case of a single graviton (see Section \ref%
{deformarea II}), we deduce that $\hat{b}^{\left( \mathrm{int}\right) }$ and
$\hat{n}^{\mu }$ from (\ref{i28}) can be taken to stop at antighost number
two and one respectively%
\begin{eqnarray}
\hat{b}^{\left( \mathrm{int}\right) } &=&\sum\limits_{I=0}^{2}\hat{b}%
_{I}^{\left( \mathrm{int}\right) },\qquad \mathrm{agh}\left( \hat{b}%
_{I}^{\left( \mathrm{int}\right) }\right) =I,\qquad I=\overline{0,2},
\label{i32} \\
\hat{n}^{\mu } &=&\sum\limits_{I=0}^{1}\hat{n}_{I}^{\mu },\qquad \mathrm{agh}%
\left( \hat{n}_{I}^{\mu }\right) =I,\qquad I=\overline{0,1}.  \label{i33}
\end{eqnarray}%
If we make the notations
\begin{eqnarray}
\hat{b}_{2}^{\left( \mathrm{int}\right) } &=&-\frac{1}{2}\eta ^{\ast }\left[
p_{A}p_{B}\left( \partial ^{\mu }\eta \right) \eta ^{A\nu }h_{\mu \nu
}^{B}-\left( f_{AB}^{C}p_{C}+p_{A}p_{B}\right) V^{\mu }\eta ^{A\nu }\partial
_{\lbrack \mu }\eta _{\nu ]}^{B}\right]  \notag \\
&&+\hat{b}_{2}^{\prime \left( \mathrm{int}\right) },  \label{izw}
\end{eqnarray}%
\begin{eqnarray}
&&\hat{b}_{1}^{\left( \mathrm{int}\right) }=-\frac{p_{A}p_{B}}{2}V^{\ast \mu
}\left[ \left( \partial _{\nu }V_{\mu }\right) h_{\ \ \rho }^{A\nu }\eta
^{B\rho }+\frac{1}{2}\left( \partial _{\lbrack \mu }h_{\nu ]\rho
}^{A}\right) V^{\nu }\eta ^{B\rho }\right.  \notag \\
&&\left. -\frac{1}{4}V^{\nu }h_{[\mu }^{A\,\rho }\left( \partial _{\nu
]}\eta _{\rho }^{B}\right) -\frac{1}{4}V^{\nu }\left( \partial _{\rho }\eta
_{\lbrack \mu }^{A}\right) h_{\nu ]}^{B\,\rho }-\frac{3}{4}h_{\mu }^{A\,\nu
}h_{\nu }^{B\,\rho }\partial _{\rho }\eta \right]  \notag \\
&&-\frac{1}{4}\left( f_{AB}^{C}p_{C}+p_{A}p_{B}\right) V^{\ast \mu }V^{\nu }%
\left[ \left( \partial _{\lbrack \mu }h_{\rho ]\nu }^{A}+\partial _{\lbrack
\nu }h_{\rho ]\mu }^{A}\right) \eta ^{B\rho }\right.  \notag \\
&&\left. -h_{\mu }^{A\,\rho }\partial _{\nu }\eta _{\rho }^{B}-h_{\nu
}^{A\,\rho }\partial _{\mu }\eta _{\rho }^{B}\right] +\frac{1}{2}\delta
_{3}^{D}\varepsilon ^{\mu \nu \rho }V_{\mu }^{\ast }\left[
y_{3C}f_{AB}^{C}h_{\nu }^{A\,\lambda }\partial _{\lbrack \rho }\eta
_{\lambda ]}^{B}\right.  \notag \\
&&\left. +\left( 2y_{3B}p_{A}+y_{3C}f_{AB}^{C}\right) \eta ^{A\lambda
}\partial _{\lbrack \nu }h_{\rho ]\lambda }^{B}\right]  \notag \\
&&-\frac{1}{2}y_{2A}p_{B}V^{\ast \mu }\left[ \left( D-2\right) h_{\mu \nu
}^{A}\eta ^{B\nu }-\delta ^{AB}V_{\mu }\eta \right]  \notag \\
&&+\frac{1}{2}h^{\ast A\mu \nu }\left[ y_{2C}f_{AB}^{C}\left( h_{\mu \nu
}^{B}\eta +2V_{\mu }\eta _{\nu }^{B}\right) -2\left( y_{2A}p_{B}\right.
\right.  \notag \\
&&\left. \left. +y_{2C}f_{AB}^{C}\right) \sigma _{\mu \nu }V^{\rho }\eta
_{\rho }^{B}\right] -\frac{8}{D-2}y_{3A}q_{1}\delta _{3}^{D}h^{\ast A}\eta +%
\hat{b}_{1}^{\prime \left( \mathrm{int}\right) },  \label{izx}
\end{eqnarray}%
\begin{eqnarray}
\hat{b}_{0}^{\left( \mathrm{int}\right) } &=&-\frac{p_{A}p_{B}}{16}\left[
V_{\rho }\left( \left( \partial ^{\lbrack \mu }h^{A\nu ]\rho }\right) \left(
\partial _{\lbrack \mu }h_{\nu ]\lambda }^{B}\right) V^{\lambda }-2\left(
\partial ^{\lbrack \mu }h^{A\nu ]\rho }\right) h_{\lambda \lbrack \mu
}^{B}\left( \partial _{\nu ]}V^{\lambda }\right) \right) \right.  \notag \\
&&+h_{\rho }^{A\ [\mu }\left( \partial ^{\nu ]}V^{\rho }\right) h_{\lambda
\lbrack \mu }^{B}\left( \partial _{\nu ]}V^{\lambda }\right) +F^{\mu \nu
}h_{\ \ \lambda }^{A\rho }\left( h_{\ \ [\mu }^{B\lambda }\left( \partial
_{\nu ]}V_{\rho }\right) \right.  \notag \\
&&\left. \left. -\left( \partial _{\lbrack \mu }h_{\ \ \nu ]}^{B\lambda
}\right) V_{\rho }\right) +F^{\mu \nu }h_{\ \ [\mu }^{A\rho }\left( \partial
_{\nu ]}h_{\rho }^{B\,\lambda }\right) V_{\lambda }\right]  \notag \\
&&-\frac{p_{A}p_{B}}{2}F^{\mu \nu }\left[ F_{\mu \rho }h_{\nu }^{A\,\lambda
}h_{\lambda }^{B\,\rho }+\frac{1}{16}F_{\mu \nu }\left( h^{A}h^{B}-2h^{A\rho
\lambda }h_{\rho \lambda }^{B}\right) \right.  \notag \\
&&-h_{\nu }^{A\,\rho }\left( \left( \partial _{\lbrack \mu }h_{\rho ]}^{B\
\lambda }\right) V_{\lambda }-h_{[\mu }^{B\ \lambda }\left( \partial _{\rho
]}V_{\lambda }\right) \right) +\frac{1}{2}\left( F^{\rho \lambda }h_{\mu
\rho }^{A}h_{\nu \lambda }^{B}\right.  \notag \\
&&\left. \left. -F_{\mu \rho }h_{\ \ \nu }^{A\rho }h^{B}\right) +\frac{1}{4}%
\left( \left( \partial _{\lbrack \mu }h_{\nu ]}^{A\ \rho }\right) V_{\rho
}-h_{[\mu }^{A\ \rho }\left( \partial _{\nu ]}V_{\rho }\right) \right) h^{B}%
\right]  \notag \\
&&-\frac{1}{8}\left( f_{AB}^{C}p_{C}+p_{A}p_{B}\right) \left( F^{\mu \nu
}F_{\nu \rho }+\frac{1}{4}\delta _{\rho }^{\mu }F^{\nu \lambda }F_{\nu
\lambda }\right) h_{\mu \sigma }^{A}h^{B\sigma \rho }  \notag \\
&&-\frac{p_{A}}{2}q_{1}\delta _{3}^{D}\varepsilon ^{\mu \nu \lambda }\left(
h^{A}V_{\mu }F_{\nu \lambda }-2h_{\lambda }^{A\,\alpha }V_{\mu }F_{\nu
\alpha }+h_{\mu }^{A\,\alpha }V_{\alpha }F_{\nu \lambda }\right)  \notag \\
&&-\frac{p_{A}}{2}q_{2}\delta _{5}^{D}\varepsilon ^{\mu \nu \lambda \alpha
\beta }\left( h^{A}V_{\mu }F_{\nu \lambda }F_{\alpha \beta }-4h_{\beta
}^{A\,\rho }V_{\mu }F_{\nu \lambda }F_{\alpha \rho }\right.  \notag \\
&&\left. +2h_{\mu }^{A\,\rho }V_{\rho }F_{\nu \lambda }F_{\alpha \beta
}\right) +8y_{3A}q_{1}\delta _{3}^{D}V^{\nu }\partial _{\lbrack \nu }h_{\rho
]}^{A\,\rho }  \notag \\
&&+\frac{1}{2}\left( D-2\right) \left( D-1\right) \left(
y_{2A}y_{2}^{A}\right) V_{\mu }V^{\mu }+\hat{b}_{0}^{\prime \left( \mathrm{%
int}\right) }  \label{izy}
\end{eqnarray}%
and take into account expansions (\ref{i32})--(\ref{i33}) and (\ref{a3}),
then equation (\ref{i28}) becomes equivalent with the tower of equations
\begin{eqnarray}
\gamma \hat{b}_{2}^{\prime \left( \mathrm{int}\right) } &=&0,  \label{i34} \\
\delta \hat{b}_{2}^{\prime \left( \mathrm{int}\right) }+\gamma \hat{b}%
_{1}^{\prime \left( \mathrm{int}\right) } &=&\partial _{\mu }\hat{\rho}%
_{1}^{\mu }+\frac{1}{2}\hat{\chi}_{1},  \label{i35} \\
\delta \hat{b}_{1}^{\prime \left( \mathrm{int}\right) }+\gamma \hat{b}%
_{0}^{\prime \left( \mathrm{int}\right) } &=&\partial _{\mu }\hat{\rho}%
_{0}^{\mu }+\frac{1}{2}\hat{\chi}_{0},  \label{i36}
\end{eqnarray}%
where $\hat{\rho}_{I}^{\mu }=\frac{1}{2}\left( \hat{w}_{I}^{\mu }-\hat{n}%
_{I}^{\mu }\right) $ and%
\begin{eqnarray}
\hat{\chi}_{1} &=&V_{\mu }^{\ast }\left[ -\left(
f_{AB}^{C}p_{C}+p_{A}p_{B}\right) F^{\mu \nu }\eta ^{A\rho }\partial
_{\lbrack \rho }\eta _{\nu ]}^{B}+\left( y_{3A}p_{B}+y_{3B}p_{A}\right.
\right.  \notag \\
&&\left. +y_{3C}f_{AB}^{C}\right) \delta _{3}^{D}\varepsilon ^{\mu \nu \rho
}\left( \partial _{\lbrack \nu }\eta _{\lambda ]}^{A}\right) \partial
_{\lbrack \rho }\eta _{\tau ]}^{B}\sigma ^{\lambda \tau }+\left(
y_{2A}p_{B}+y_{2B}p_{A}\right.  \notag \\
&&\left. \left. +y_{2C}f_{AB}^{C}\right) \left( D-2\right) \sigma ^{\mu \nu
}\left( \partial _{\lbrack \nu }\eta _{\rho ]}^{A}\right) \eta ^{B\rho }
\right] ,  \label{chihat1}
\end{eqnarray}%
\begin{eqnarray}
\hat{\chi}_{0} &=&\delta \left\{ \delta _{3}^{D}\varepsilon ^{\mu \nu \rho
}V_{\mu }^{\ast }\left[ y_{3C}f_{AB}^{C}h_{\nu }^{A\,\lambda }\partial
_{\lbrack \rho }\eta _{\lambda ]}^{B}+\left( 2y_{3B}p_{A}\right. \right.
\right.  \notag \\
&&\left. \left. +y_{3C}f_{AB}^{C}\right) \eta ^{A\lambda }\partial _{\lbrack
\nu }h_{\rho ]\lambda }^{B}\right] -y_{2A}p_{B}V^{\ast \mu }\left[ \left(
D-2\right) h_{\mu \nu }^{A}\eta ^{B\nu }\right.  \notag \\
&&\left. -\delta ^{AB}V_{\mu }\eta \right] +h^{\ast A\mu \nu }\left[
y_{2C}f_{AB}^{C}\left( h_{\mu \nu }^{B}\eta +2V_{\mu }\eta _{\nu
}^{B}\right) -2\left( y_{2A}p_{B}\right. \right.  \notag \\
&&\left. \left. \left. +y_{2C}f_{AB}^{C}\right) \sigma _{\mu \nu }V^{\rho
}\eta _{\rho }^{B}\right] \right\} -4q_{1}y_{2A}\delta _{3}^{D}\left(
D-2\right) \varepsilon _{\mu \nu \rho }F^{\mu \nu }\eta ^{A\rho }  \notag \\
&&-6q_{2}y_{2A}\delta _{5}^{D}\varepsilon _{\mu \nu \rho \alpha \beta
}F^{\mu \nu }F^{\rho \alpha }\eta ^{A\beta }+\frac{1}{2}\left(
f_{AB}^{C}p_{C}+p_{A}p_{B}\right) \left( F^{\mu \nu }F_{\nu \rho }\right.
\notag \\
&&\left. +\frac{1}{4}\delta _{\rho }^{\mu }F^{\nu \lambda }F_{\nu \lambda
}\right) \left( h^{A\rho \sigma }\partial _{\lbrack \mu }\eta _{\sigma
]}^{B}-2\partial _{\lbrack \mu }h_{\ \ \sigma ]}^{A\rho }\eta ^{B\sigma
}\right)  \notag \\
&&+y_{2A}\left[ -4D\Lambda ^{A}\eta +f_{BC}^{A}\hat{A}_{0}^{\left( \mathrm{%
int}\right) BC}\left( \partial \partial \hat{\Phi}^{\alpha _{0}}\hat{\Phi}%
^{\beta _{0}}\hat{\eta}_{\alpha _{1}}\right) \right.  \notag \\
&&\left. +p_{B}\hat{B}_{0}^{\left( \mathrm{int}\right) AB}\left( \partial
\partial \hat{\Phi}^{\alpha _{0}}\hat{\Phi}^{\beta _{0}}\hat{\eta}_{\alpha
_{1}}\right) \right]  \notag \\
&&+y_{3A}\delta _{3}^{D}\left[ f_{BC}^{A}\hat{C}_{0}^{\left( \mathrm{int}%
\right) BC}\left( \partial \partial \partial \hat{\Phi}^{\alpha _{0}}\hat{%
\Phi}^{\beta _{0}}\hat{\eta}_{\alpha _{1}}\right) \right.  \notag \\
&&\left. +p_{B}\hat{D}_{0}^{\left( \mathrm{int}\right) AB}\left( \partial
\partial \partial \hat{\Phi}^{\alpha _{0}}\hat{\Phi}^{\beta _{0}}\hat{\eta}%
_{\alpha _{1}}\right) \right] .  \label{chihat0}
\end{eqnarray}%
The component $\hat{S}_{2}^{\left( \mathrm{int}%
\right) }$, given by (\ref{notS2intcoll}), is thus completely
determined once
we compute $\hat{b}^{\left( \mathrm{int}\right) }$, which expands as in (\ref%
{i32}). The only unknown components from $\hat{b}^{\left( \mathrm{int}%
\right) }$ are $\left( \hat{b}_{I}^{\prime \left( \mathrm{int}\right)
}\right) _{I=\overline{0,2}}$ appearing in formulas (\ref{izw})--(\ref{izy}%
). They are subject to equations (\ref{i34})--(\ref{i36}). In conclusion,
the final step needed in order to construct $\hat{S}_{2}^{\left( \mathrm{int}%
\right) }$ is to solve equations (\ref{i34})--(\ref{i36}).

Related to equation (\ref{i35}), we observe that the existence of $\hat{b}%
_{2}^{\prime \left( \mathrm{int}\right) }$ and $\hat{b}_{1}^{\prime \left(
\mathrm{int}\right) }$ requires that (\ref{chihat1}) must be written as%
\begin{equation}
\hat{\chi}_{1}=\delta \hat{\varphi}_{2}+\gamma \hat{\omega}_{1}+\partial
_{\mu }\hat{l}_{1}^{\mu },  \label{i39}
\end{equation}%
where $\hat{\varphi}_{2}$, $\hat{\omega}_{1}$, and $\hat{l}_{1}^{\mu }$
exhibit the same properties like the corresponding unhatted quantities from (%
\ref{a77}). We require that the second-order deformation is local, so $\hat{%
\varphi}_{2}$, $\hat{\omega}_{1}$, and $\hat{l}_{1}^{\mu }$ must be local
functions. Assuming (\ref{i39}) is fulfilled, we apply $\delta $ on it and
find the necessary condition
\begin{equation}
\delta \hat{\chi}_{1}=\gamma \left( -\delta \hat{\omega}_{1}\right)
+\partial _{\mu }\left( \delta \hat{l}_{1}^{\mu }\right) .  \label{i40}
\end{equation}%
We do not insist on the investigation of equation (\ref{i40}), which can be
done by standard cohomological techniques, but simply state that it can be
shown to hold if the following conditions are simultaneously satisfied
\begin{eqnarray}
F^{\mu \nu }F_{\nu \rho }+\frac{1}{4}\delta _{\rho }^{\mu }F_{\nu \lambda
}F^{\nu \lambda } &=&\delta \Omega _{\rho }^{\mu },  \label{i44} \\
F^{\theta \mu } &=&\delta \bar{\Omega}^{\theta \mu },  \label{i45} \\
\partial _{\lbrack \mu }h_{\lambda ]}^{A\ \theta } &=&\delta \Omega _{\mu
\lambda }^{A\theta },  \label{i46} \\
\left( \partial _{\lbrack \theta }h_{\nu ]}^{A\ \theta }\right) \partial
^{\lbrack \mu }h_{\ \ \ \mu }^{B\nu ]}-\left( \partial ^{\lbrack \nu
}h^{A\theta ]\mu }\right) \partial _{\nu }h_{\theta \mu }^{B} &=&\delta
\Omega ^{AB}.  \label{i47a}
\end{eqnarray}%
All the quantities denoted by $\Omega $ or $\bar{\Omega}$ must be
local in order to produce local deformations. It is easy to see, by
arguments similar to those exposed in the end of the preamble of
Section \ref{deformarea II}, that none of
equations (\ref{i44})--(\ref{i47a}) is fulfilled (for local functions), so (%
\ref{i40}), and therefore (\ref{i39}), cannot hold unless
\begin{equation}
\hat{\chi}_{1}=0,  \label{chihat10}
\end{equation}%
which further implies the following equations
\begin{eqnarray}
f_{AB}^{C}p_{C}+p_{A}p_{B} &=&0,  \label{i47b} \\
\left( p_{A}y_{3B}+p_{B}y_{3A}+f_{AB}^{C}y_{3C}\right) \delta _{3}^{D} &=&0,
\label{i47c} \\
p_{A}y_{2B}+p_{B}y_{2A}+f_{AB}^{C}y_{2C} &=&0.  \label{i47d}
\end{eqnarray}%
We recall that the constants $f_{AB}^{C}$ are not arbitrary. They have been
restricted previously to define the structure constants of a real,
commutative, symmetric, and associative (finite-dimensional) algebra, so in
addition they satisfy relations (\ref{fabcdif}).

Let us analyze briefly the solutions to (\ref{i47b})--(\ref{i47d}). Taking
into account (\ref{fabcdif}) and recalling (\ref{faaa}), equations (\ref%
{i47b})--(\ref{i47d}) become equivalent to%
\begin{eqnarray}
p_{A}p_{B} &=&0,\qquad \mathrm{for\ all}\qquad A\neq B,  \label{i48a} \\
\left( p_{A}y_{3B}+p_{B}y_{3A}\right) \delta _{3}^{D} &=&0,\qquad \mathrm{%
for\ all}\qquad A\neq B,  \label{i48b} \\
p_{A}y_{2B}+p_{B}y_{2A} &=&0,\qquad \mathrm{for\ all}\qquad A\neq B,
\label{i48c} \\
p_{A}\left( f_{A}+p_{A}\right) &=&0,\qquad \mathrm{without\ summation\ over}%
\ A,  \label{i48d} \\
\left( f_{A}+2p_{A}\right) y_{3A}\delta _{3}^{D} &=&0,\qquad \mathrm{%
without\ summation\ over}\ A,  \label{i48e} \\
\left( f_{A}+2p_{A}\right) y_{2A} &=&0,\qquad \mathrm{without\ summation\
over}\ A.  \label{i48f}
\end{eqnarray}%
Unlike Section \ref{deformarea II}, where we searched \emph{only}
the solutions relevant from the point of view of deformations, here
we must discuss \emph{all }the solutions, since our aim is to see
whether they allow
or not cross-couplings among different gravitons. Inspecting (\ref{i48a})--(%
\ref{i48f}), we observe that there appear two complementary cases related to
the $p_{A}$'s : either at least one is nonvanishing, say $p_{1}$, or all the
$p_{A}$'s vanish. In \textbf{case I}%
\begin{equation}
p_{1}\neq 0,  \label{caseIIcoll}
\end{equation}%
so from (\ref{i48d}) for $A=1$ it follows that at least $f_{1}$ is
non-vanishing%
\begin{equation}
f_{1}=-p_{1}\neq 0,  \label{i50}
\end{equation}%
while (\ref{i48a}) restricts all the other $p_{B}$'s to vanish%
\begin{equation}
p_{B}=0,\qquad B=\overline{2,n}.  \label{i51}
\end{equation}%
Thus, (\ref{i48b}) and (\ref{i48c}) for $A=1$ and $B\neq 1$ imply%
\begin{equation}
p_{1}y_{3B}\delta _{3}^{D}=0,\qquad p_{1}y_{2B}=0,\qquad B=\overline{2,n},
\label{i52b}
\end{equation}%
while (\ref{i48e}) and (\ref{i48f}) for $A=1$ together with (\ref{i50}) lead
to%
\begin{equation}
p_{1}y_{31}\delta _{3}^{D}=0,\qquad p_{1}y_{21}=0.  \label{i52c}
\end{equation}%
The last two sets of equations, (\ref{i52b}) and (\ref{i52c}), display a
unique solution%
\begin{equation}
y_{3A}\delta _{3}^{D}=0=y_{2A},\qquad A=\overline{1,n}.  \label{i52d}
\end{equation}%
In \textbf{case II}%
\begin{equation}
p_{A}=0,\qquad A=\overline{1,n},  \label{caseIcoll}
\end{equation}%
equations (\ref{i48a})--(\ref{i48d}) are identically satisfied, while the
other two take the simple form%
\begin{eqnarray}
f_{A}y_{3A}\delta _{3}^{D} &=&0,\qquad \mathrm{without\ summation\ over}\ A,
\label{i49a} \\
f_{A}y_{2A} &=&0,\qquad \mathrm{without\ summation\ over}\ A.  \label{i49b}
\end{eqnarray}%
Therefore, we have a single option, namely the set $\left\{ 1,2,\ldots
,n\right\} $ is divided into two complementary subsets such that $A=\left(
\bar{A},A^{\prime }\right) $ with $\bar{A}\neq A^{\prime }$ and $(f_{\bar{A}%
}=0,\,y_{3A^{\prime }}\delta _{3}^{D}=0,\,y_{2A^{\prime }}=0)$. Re-ordering
the indices we can always write%
\begin{equation}
f_{\bar{A}}=0,\qquad \bar{A}=\overline{1,m},\qquad y_{3A^{\prime }}\delta
_{3}^{D}=0=y_{2A^{\prime }},\qquad A^{\prime }=\overline{m+1,n}.
\label{caseIbcoll}
\end{equation}%
The above solution contains two limit situations: $m=n$ and $m=0$.

\subsection{Main cases. Coupled theories}

\subsubsection{Case I: no-go results in General Relativity\label{nogocollGR}}

As we have discussed previously, the first case is governed by the solution%
\begin{equation}
p_{1}=-f_{1}\neq 0,\quad \left( p_{B}\right) _{B=\overline{2,n}}=0,\quad
\left( y_{3A}\delta _{3}^{D}\right) _{A=\overline{1,n}}=0=\left(
y_{2A}\right) _{A=\overline{1,n}},  \label{caseIcollfull}
\end{equation}%
so the deformed solution to the master equation in all $D>2$ spacetime
dimensions is maximally parameterized by $\left( f_{A}\right) _{A=\overline{%
1,n}}$, $p_{1}=-f_{1}\neq 0$, $\left( \Lambda _{A}\right) _{A=\overline{1,n}%
} $, $q_{1}\delta _{3}^{D}$, and $q_{2}\delta _{5}^{D}$. Of course, it is
possible that some of $f_{B}$ (for $B\neq 1$), $\Lambda _{A}$, $q_{1}$, or $%
q_{2}$ vanish. Inserting (\ref{caseIcollfull}) into (\ref{chihat0}) we find%
\begin{equation}
\hat{\chi}_{0}=0.  \label{chihat00}
\end{equation}%
Combining this result with (\ref{chihat10}) we observe that the tower of
equations (\ref{i34})--(\ref{i36}) takes the `homogeneous' form%
\begin{eqnarray}
\gamma \hat{b}_{2}^{\prime \left( \mathrm{int}\right) } &=&0,  \label{i53a}
\\
\delta \hat{b}_{2}^{\prime \left( \mathrm{int}\right) }+\gamma \hat{b}%
_{1}^{\prime \left( \mathrm{int}\right) } &=&\partial _{\mu }\hat{\rho}%
_{1}^{\mu },  \label{i53b} \\
\delta \hat{b}_{1}^{\prime \left( \mathrm{int}\right) }+\gamma \hat{b}%
_{0}^{\prime \left( \mathrm{int}\right) } &=&\partial _{\mu }\hat{\rho}%
_{0}^{\mu },  \label{i53c}
\end{eqnarray}%
so we can take
\begin{equation}
\hat{b}_{2}^{\prime \left( \mathrm{int}\right) }=\hat{b}_{1}^{\prime \left(
\mathrm{int}\right) }=\hat{b}_{0}^{\prime \left( \mathrm{int}\right) }=0
\label{b'hat0}
\end{equation}%
and incorporate the `homogeneous' solution into the first-order deformation $%
\hat{S}_{1}^{\left( \mathrm{int}\right) }$ (see (\ref{S1intcoll})) through a
suitable redefinition of the parameterizing constants. At this point we act
like in sections \ref{caseI} and \ref{analysiscaseI}. Replacing (\ref{b'hat0}%
) and (\ref{caseIcollfull}) into (\ref{S1intcoll}), (\ref{newint2}), (\ref%
{S1PFcollfin}), (\ref{S2PFcollfin}), and (\ref{izw})--(\ref{izy}) and
regrouping the terms from (\ref{S1coll}) and (\ref{S2dec}) with the help of (%
\ref{notS2intcoll}) and (\ref{i32}), we find that there are no
cross-couplings among different gravitons intermediated by the vector field.
The vector field gets coupled to a single graviton (the first one in our
convention) and the resulting interactions fit the rules prescribed by
General Relativity.

The Lagrangian formulation of the coupled model can be completed by imposing
some gauge-fixing conditions similar to (\ref{wzz1}), one for each graviton
sector. If in addition we make the convention%
\begin{equation}
f_{1}=1=-p_{1},  \label{convf1p1}
\end{equation}%
then the fully deformed solution to the master equation
\begin{equation}
\hat{S}^{(\mathrm{I})}=\bar{S}^{\prime }+k\hat{S}_{1}^{(\mathrm{I})}+k^{2}%
\hat{S}_{2}^{(\mathrm{I})}+\cdots ,  \label{SfullcollcaseI}
\end{equation}%
where $\bar{S}^{\prime }$ is the\textquotedblleft free\textquotedblright\
solution (\ref{i11}), leads to a Lagrangian action in which a \emph{single}
graviton ($A=1$) couples to the vector field $V_{\mu }$ according to the
standard coupling from General Relativity, while each of the other gravitons
($B=\overline{2,n}$) interacts \emph{only} with itself according to an
Einstein-Hilbert action (or possibly a Pauli-Fierz action if $f_{B}=0$) with
a cosmological term. Accordingly, in case I we obtain the Lagrangian action
\begin{eqnarray}
&&\hat{S}^{\mathrm{L(I)}}\left[ h_{\mu \nu }^{A},V_{\mu }\right] =\int d^{D}x%
\left[ \frac{2}{k^{2}}\sqrt{-g^{1}}\left( R^{1}-2k^{2}\Lambda _{1}\right)
\right.  \notag \\
&&-\frac{1}{4}\sqrt{-g^{1}}g^{1\mu \nu }g^{1\rho \lambda }\bar{F}_{\mu \rho
}^{1}\bar{F}_{\nu \lambda }^{1}+k\left( q_{1}\delta _{3}^{D}\varepsilon
^{1\mu _{1}\mu _{2}\mu _{3}}\bar{V}_{\mu _{1}}^{1}\bar{F}_{\mu _{2}\mu
_{3}}^{1}\right.  \notag \\
&&\left. \left. +q_{2}\delta _{5}^{D}\varepsilon ^{1\mu _{1}\mu _{2}\mu
_{3}\mu _{4}\mu _{5}}\bar{V}_{\mu _{1}}^{1}\bar{F}_{\mu _{2}\mu _{3}}^{1}%
\bar{F}_{\mu _{4}\mu _{5}}^{1}\right) \right]  \notag \\
&&+\sum_{B=2}^{n}\left[ \int d^{D}x\frac{2}{k_{B}^{2}}\sqrt{-g^{B}}\left(
R^{B}-2kk_{B}\Lambda _{B}\right) \right]  \notag \\
&\equiv &\hat{S}^{\mathrm{L(I)}}\left[ g_{\mu \nu }^{1},\bar{V}_{\mu }^{1}%
\right] +\sum_{B=2}^{n}\hat{S}^{\mathrm{L(E-H)}}\left[ g_{\mu \nu }^{B}%
\right] ,  \label{lagactcollcaseI}
\end{eqnarray}%
where $\bar{V}_{\mu }^{1}$ and $\bar{F}_{\mu \nu }^{1}$ are `curved' with
the vielbein fields from the first graviton sector%
\begin{eqnarray}
\bar{V}_{\mu }^{1} &=&e_{\mu }^{1a}V_{a},\qquad \bar{F}_{\mu \nu
}^{1}=\partial _{\lbrack \mu }\left( e_{\nu ]}^{1a}V_{a}\right) ,
\label{VbarFbar} \\
\varepsilon ^{1\mu _{1}\mu _{2}\ldots \mu _{D}} &=&\sqrt{-g^{1}}%
e_{a_{1}}^{1\mu _{1}}\cdots e_{a_{D}}^{1\mu _{D}}\varepsilon ^{a_{1}\ldots
a_{D}}.  \label{epsilon1}
\end{eqnarray}%
The notations $R^{A}$ and $g^{A}$ ($A=\overline{1,n}$) denote the full
scalar curvature and the determinant of the metric tensor $g_{\mu \nu
}^{A}=\sigma _{\mu \nu }+k_{A}h_{\mu \nu }^{A}$ (without summation over $A$)
from the $A$-th graviton sector respectively, while $k_{B}=kf_{B}$, $B=%
\overline{2,n}$. The final conclusion is that \emph{in the first case there
is no cross-interaction among different gravitons to all orders in the
coupling constant.}

\subsubsection{Case II: no-go results for the new couplings in $D=3$\label%
{exoticcoll}}

The second case is subject to the conditions%
\begin{equation}
\left( p_{A}\right) _{A=\overline{1,n}}=0,\quad \left( f_{\bar{A}}\right) _{%
\bar{A}=\overline{1,m}}=0,\quad \left( y_{3A^{\prime }}\delta
_{3}^{D}\right) _{A^{\prime }=\overline{m+1,n}}=0=\left( y_{2A^{\prime
}}\right) _{A^{\prime }=\overline{m+1,n}},  \label{caseIIcollfull}
\end{equation}%
so the deformed solution to the master equation is maximally parameterized
by $\left( f_{A^{\prime }}\right) _{A^{\prime }=\overline{m+1,n}}$, $\left(
y_{3\bar{A}}\delta _{3}^{D}\right) _{\bar{A}=\overline{1,m}}$, $\left( y_{2%
\bar{A}}\right) _{\bar{A}=\overline{1,m}}$, $\left( \Lambda _{A}\right) _{A=%
\overline{1,n}}$, $q_{1}\delta _{3}^{D}$, and $q_{2}\delta _{5}^{D}$.
Substituting (\ref{caseIIcollfull}) into (\ref{chihat0}), it follows that%
\begin{eqnarray}
\hat{\chi}_{0} &=&-4q_{1}\delta _{3}^{D}y_{2\bar{A}}\left( D-2\right)
\varepsilon _{\mu \nu \rho }F^{\mu \nu }\eta ^{\bar{A}\rho }-6q_{2}\delta
_{5}^{D}y_{2\bar{A}}\varepsilon _{\mu \nu \rho \alpha \beta }F^{\mu \nu
}F^{\rho \alpha }\eta ^{\bar{A}\beta }  \notag \\
&&-\left( \sum_{\bar{A}=1}^{m}y_{2\bar{A}}\Lambda ^{\bar{A}}\right) 4D\eta .
\label{i54}
\end{eqnarray}%
Reasoning exactly like in the case of formulas (\ref{chicaseII}) and (\ref%
{chicaseIII}), we deduce that equation (\ref{i36}) demands an equation of
the type (\ref{condchi0}), $\hat{\chi}_{0}=\delta \hat{\varphi}_{1}+\gamma
\hat{\omega}_{0}+\partial _{\mu }\hat{l}_{0}^{\mu }$, which cannot be
satisfied for local $\hat{\varphi}_{1}$, $\hat{\omega}_{0}$, and $\hat{l}%
_{0}^{\mu }$ unless%
\begin{equation}
\hat{\chi}_{0}=0,  \label{chihat00caseII}
\end{equation}%
which further requires%
\begin{equation}
\left( q_{1}\delta _{3}^{D}y_{2\bar{A}}\right) _{\bar{A}=\overline{1,m}%
}=0,\quad \left( q_{2}\delta _{5}^{D}y_{2\bar{A}}\right) _{\bar{A}=\overline{%
1,m}}=0,\quad \sum_{\bar{A}=1}^{m}\left( y_{2\bar{A}}\Lambda ^{\bar{A}%
}\right) =0.  \label{condsupcaseIIIcoll}
\end{equation}%
Clearly, there are two distinct solutions to the above equations%
\begin{eqnarray}
q_{1}\delta _{3}^{D} &=&0=q_{2}\delta _{5}^{D},\qquad \sum_{\bar{A}%
=1}^{m}\left( y_{2\bar{A}}\Lambda ^{\bar{A}}\right) =0,  \label{caseII.1coll}
\\
y_{2\bar{A}} &=&0,\qquad \bar{A}=\overline{1,m},  \label{caseII.2coll}
\end{eqnarray}%
deserving separate analyses. In each subcase (\ref{chihat10}) and (\ref%
{chihat00caseII}) hold, such that equations (\ref{i34})--(\ref{i36}) take
the `homogeneous' form (\ref{i53a})--(\ref{i53c}), whose solution can be
taken of the form (\ref{b'hat0}).

\paragraph{Subcase II.1}

From (\ref{caseIIcollfull}) and (\ref{caseII.1coll}) we observe that the
deformed solution to the master equation is maximally parameterized in this
situation by $\left( f_{A^{\prime }}\right) _{A^{\prime }=\overline{m+1,n}}$%
, $\left( y_{3\bar{A}}\delta _{3}^{D}\right) _{\bar{A}=\overline{1,m}}$, $%
\left( y_{2\bar{A}}\right) _{\bar{A}=\overline{1,m}}$, and $\left( \Lambda
_{A}\right) _{A=\overline{1,n}}$, where in addition the first $m$
cosmological constants are restricted to satisfy the condition%
\begin{equation}
\sum_{\bar{A}=1}^{m}\left( y_{2\bar{A}}\Lambda ^{\bar{A}}\right) =0.
\label{condadcaseII.1coll}
\end{equation}%
Consequently, the first- and second-order deformations of the solution to
the master equation, (\ref{S1coll}) and (\ref{S2dec}), read as%
\begin{eqnarray}
&&\hat{S}_{1}^{\left( \mathrm{II.1}\right) }=\sum_{A^{\prime
}=m+1}^{n}\left\{ \int d^{D}x\left\{ f_{A^{\prime }}\left[ \frac{1}{2}\eta
^{\ast A^{\prime }\mu }\eta ^{A^{\prime }\nu }\partial _{\lbrack \mu }\eta
_{\nu ]}^{A^{\prime }}+h^{\ast A^{\prime }\mu \rho }\left( \left( \partial
_{\rho }\eta ^{A^{\prime }\nu }\right) h_{\mu \nu }^{A^{\prime }}\right.
\right. \right. \right.  \notag \\
&&\left. \left. \left. \left. -\eta ^{A^{\prime }\nu }\partial _{\lbrack \mu
}h_{\nu ]\rho }^{A^{\prime }}\right) +\hat{a}_{0}^{(\mathrm{EH-cubic}%
)A^{\prime }}\right] -2\Lambda _{A^{\prime }}h^{A^{\prime }}\right\} \right\}
\notag \\
&&+\sum_{\bar{A}=1}^{m}\left\{ \int d^{D}x\left[ y_{2\bar{A}}\left( h^{\ast
\bar{A}}\eta +\left( D-2\right) \left( -V^{\ast \lambda }\eta _{\lambda }^{%
\bar{A}}+V^{\lambda }\partial _{\lbrack \mu }h_{\lambda ]}^{\bar{A}\ \mu
}\right) \right) \right. \right.  \notag \\
&&\left. \left. +y_{3}^{\bar{A}}\delta _{3}^{D}\varepsilon _{\mu \nu \rho
}\left( V^{\ast \mu }\partial ^{\lbrack \nu }\eta _{\bar{A}}^{\rho
]}+F^{\lambda \mu }\partial ^{\lbrack \nu }h_{\bar{A}\ \lambda }^{\rho
]}\right) -2\Lambda _{\bar{A}}h^{\bar{A}}\right] \right\} ,
\label{S1fincollII.1}
\end{eqnarray}%
\begin{eqnarray}
&&\hat{S}_{2}^{\left( \mathrm{II.1}\right) }=\sum_{A^{\prime
}=m+1}^{n}\left\{ f_{A^{\prime }}\left[ f_{A^{\prime }}S_{2}^{(\mathrm{%
EH-quartic})A^{\prime }}+\Lambda _{A^{\prime }}\int d^{D}x\left(
h^{A^{\prime }\mu \nu }h_{\mu \nu }^{A^{\prime }}\right. \right. \right.
\notag \\
&&\left. \left. \left. -\frac{1}{2}\left( h^{A^{\prime }}\right) ^{2}\right) %
\right] \right\} +\sum_{\bar{A},\bar{B}=1}^{m}\left\{ \int d^{D}x\left[ y_{2%
\bar{A}}y_{2\bar{B}}\frac{\left( D-2\right) ^{2}}{4}\left( h^{\bar{A}}h^{%
\bar{B}}-h^{\bar{A}\mu \nu }h_{\mu \nu }^{\bar{B}}\right) \right. \right.
\notag \\
&&\left. \left. +y_{2\bar{A}}y_{3}^{\bar{B}}\delta _{3}^{D}\left( D-2\right)
\varepsilon _{\mu \nu \rho }h_{\ \ \ \lambda }^{\bar{A}\mu }\left( \partial
^{\lbrack \nu }h_{\bar{B}}^{\rho ]\lambda }\right) +y_{3}^{\bar{A}}y_{3\bar{B%
}}\delta _{3}^{D}\left( \partial ^{\lbrack \nu }h_{\bar{A}}^{\rho ]\lambda
}\right) \partial _{\lbrack \nu }h_{\rho ]\lambda }^{\bar{B}}\right] \right\}
\notag \\
&&+\frac{1}{2}\left( D-2\right) \left( D-1\right) \left[ \sum_{\bar{A}%
=1}^{m}\left( y_{2\bar{A}}\right) ^{2}\right] \int d^{D}x\left( V_{\mu
}V^{\mu }\right)  \label{S2fincollII.1}
\end{eqnarray}%
respectively. The third-order deformation results from the equation%
\begin{equation}
\left( \hat{S}_{1}^{\left( \mathrm{II.1}\right) },\hat{S}_{2}^{\left(
\mathrm{II.1}\right) }\right) +s\hat{S}_{3}^{\left( \mathrm{II.1}\right) }=0.
\label{S3collcaseII.1}
\end{equation}%
If we make the notations%
\begin{eqnarray}
S_{1}^{(\mathrm{EH-\Lambda })A^{\prime }} &\equiv &\int d^{D}x\left\{
f_{A^{\prime }}\left[ \frac{1}{2}\eta ^{\ast A^{\prime }\mu }\eta
^{A^{\prime }\nu }\partial _{\lbrack \mu }\eta _{\nu ]}^{A^{\prime
}}+h^{\ast A^{\prime }\mu \rho }\left( \left( \partial _{\rho }\eta
^{A^{\prime }\nu }\right) h_{\mu \nu }^{A^{\prime }}\right. \right. \right.
\notag \\
&&\left. \left. \left. -\eta ^{A^{\prime }\nu }\partial _{\lbrack \mu
}h_{\nu ]\rho }^{A^{\prime }}\right) +\hat{a}_{0}^{(\mathrm{EH-cubic}%
)A^{\prime }}\right] -2\Lambda _{A^{\prime }}h^{A^{\prime }}\right\} ,
\label{S1EHA'}
\end{eqnarray}%
\begin{eqnarray}
S_{2}^{(\mathrm{EH-\Lambda })A^{\prime }} &\equiv &f_{A^{\prime }}\left[
f_{A^{\prime }}S_{2}^{(\mathrm{EH-quartic})A^{\prime }}\right.  \notag \\
&&\left. +\Lambda _{A^{\prime }}\int d^{D}x\left( h^{A^{\prime }\mu \nu
}h_{\mu \nu }^{A^{\prime }}-\frac{1}{2}\left( h^{A^{\prime }}\right)
^{2}\right) \right] ,  \label{S2EHA'}
\end{eqnarray}%
then we observe that $S_{1}^{(\mathrm{EH-\Lambda })A^{\prime }}$ and $%
S_{2}^{(\mathrm{EH-\Lambda })A^{\prime }}$ are nothing but the first- and
second-order components respectively (in the coupling constant) of the
solution to the master equation corresponding to the full Einstein-Hilbert
theory in the presence of a cosmological constant for the graviton $%
A^{\prime }$. Therefore,
\begin{equation}
\left( \sum_{A^{\prime }=m+1}^{n}S_{1}^{(\mathrm{EH-\Lambda })A^{\prime
}},\sum_{B^{\prime }=m+1}^{n}S_{2}^{(\mathrm{EH-\Lambda })B^{\prime
}}\right) =-s\left[ \sum_{A^{\prime }=m+1}^{n}S_{3}^{(\mathrm{EH-\Lambda }%
)A^{\prime }}\right] ,  \label{S3EHA'}
\end{equation}%
where $S_{3}^{(\mathrm{EH-\Lambda })A^{\prime }}$ is the third-order
component of the solution to the master equation associated with the full
Einstein-Hilbert theory with a cosmological term in the graviton sector $%
A^{\prime }$. By direct computation we then infer that%
\begin{eqnarray}
&&\left( \hat{S}_{1}^{\left( \mathrm{II.1}\right) },\hat{S}_{2}^{\left(
\mathrm{II.1}\right) }\right) =s\left[ \frac{4}{D-2}\left( \sum_{\bar{A}%
=1}^{m}y_{2\bar{A}}y_{3}^{\bar{A}}\right) \left( \sum_{\bar{B}=1}^{m}y_{3%
\bar{B}}\delta _{3}^{D}h^{\ast \bar{B}}\eta \right) \right.  \notag \\
&&\left. -\sum_{A^{\prime }=m+1}^{n}S_{3}^{(\mathrm{EH-\Lambda })A^{\prime }}%
\right] +\left( \sum_{\bar{A}=1}^{m}\left( y_{2\bar{A}}\right) ^{2}\right)
\left( D-2\right) \left( D-1\right) \times  \notag \\
&&\times \left\{ \sum_{\bar{B}=1}^{m}\left[ \int d^{D}x\left( \frac{\left(
D-2\right) }{2}y_{2\bar{B}}\left( h^{\bar{B}}\eta -2V_{\lambda }\eta ^{\bar{B%
}\lambda }\right) \right. \right. \right.  \notag \\
&&\left. \left. \left. +y_{3\bar{B}}\delta _{3}^{D}\varepsilon _{\mu \nu
\rho }F^{\mu \nu }\eta ^{\bar{B}\rho }\right) \right] \right\} ,  \label{i55}
\end{eqnarray}%
such that the existence of local solutions to equation (\ref{S3collcaseII.1}%
) demands that $(h^{\bar{B}}\eta -2V_{\lambda }\eta ^{\bar{B}\lambda })$ and
$\varepsilon _{\mu \nu \rho }F^{\mu \nu }\eta ^{\bar{B}\rho }$ are $s$-exact
modulo $d$ quantities from local functions for each $\bar{B}=\overline{1,m}$%
. It is easy to show that none of them has this property, so we must set%
\begin{eqnarray}
\left( \sum_{\bar{A}=1}^{m}\left( y_{2\bar{A}}\right) ^{2}\right) y_{2\bar{B}%
} &=&0,\qquad \bar{B}=\overline{1,m},  \label{condcollII.1a} \\
\left( \sum_{\bar{A}=1}^{m}\left( y_{2\bar{A}}\right) ^{2}\right) y_{3\bar{B}%
}\delta _{3}^{D} &=&0,\qquad \bar{B}=\overline{1,m}.  \label{condcollII.1b}
\end{eqnarray}%
The solution to these equations,%
\begin{equation}
y_{2\bar{B}}=0,\qquad \bar{B}=\overline{1,m},  \label{solsupplcollcaseII.1}
\end{equation}%
solves in addition equation (\ref{condadcaseII.1coll}). Substituting (\ref%
{solsupplcollcaseII.1}) into (\ref{i55}) and then in (\ref{S3collcaseII.1})
we find the equivalent equation%
\begin{equation}
s\left( \hat{S}_{3}^{\left( \mathrm{II.1}\right) }-\sum_{A^{\prime
}=m+1}^{n}S_{3}^{(\mathrm{EH-\Lambda })A^{\prime }}\right) =0,  \label{i56}
\end{equation}%
whose solution can be chosen, without loss of generality, of the form%
\begin{equation}
\hat{S}_{3}^{\left( \mathrm{II.1}\right) }=\sum_{A^{\prime }=m+1}^{n}S_{3}^{(%
\mathrm{EH-\Lambda })A^{\prime }}.  \label{S3fincollII.1}
\end{equation}%
We recall that $S_{3}^{(\mathrm{EH-\Lambda })A^{\prime }}$ gathers the
contributions of order three in the coupling constant from the solution of
the master equation corresponding to the full Einstein-Hilbert action with a
cosmological constant for the graviton $A^{\prime }$.

Putting together the results expressed by formulas (\ref{caseIIcollfull}), (%
\ref{caseII.1coll}), and (\ref{solsupplcollcaseII.1}) we conclude that in
subcase II.1 the consistency of the deformed solution to the master equation
requires the conditions%
\begin{eqnarray}
\left( p_{A}\right) _{A=\overline{1,n}} &=&0=\left( y_{2A}\right) _{A=%
\overline{1,n}},\qquad \left( f_{\bar{A}}\right) _{\bar{A}=\overline{1,m}}=0,
\label{condfincollcaseII.1a} \\
\left( y_{3A^{\prime }}\delta _{3}^{D}\right) _{A^{\prime }=\overline{m+1,n}%
} &=&0,\qquad q_{1}\delta _{3}^{D}=0=q_{2}\delta _{5}^{D}.
\label{condfincollcaseII.1b}
\end{eqnarray}%
The full deformed solution to the master equation $\hat{S}^{\left( \mathrm{%
II.1}\right) }$ reads as%
\begin{equation}
\hat{S}^{\left( \mathrm{II.1}\right) }=\bar{S}^{\prime }+k\hat{S}%
_{1}^{\left( \mathrm{II.1}\right) }+k^{2}\hat{S}_{2}^{\left( \mathrm{II.1}%
\right) }+k^{3}\hat{S}_{3}^{\left( \mathrm{II.1}\right) }+\cdots ,
\label{i58}
\end{equation}%
(with $\bar{S}^{\prime }$ the solution of the master equation for the free
model, (\ref{i11})) and it is maximally parameterized by $\left(
f_{A^{\prime }}\right) _{A^{\prime }=\overline{m+1,n}}$, $\left( y_{3\bar{A}%
}\delta _{3}^{D}\right) _{\bar{A}=\overline{1,m}}$, and the cosmological
constants $\left( \Lambda _{A}\right) _{A=\overline{1,n}}$. Taking into
account relations (\ref{i11}), (\ref{S1fincollII.1}), (\ref{S2fincollII.1}),
(\ref{S3fincollII.1}) and notations (\ref{S1EHA'})--(\ref{S2EHA'}), we can
decompose $\hat{S}^{\left( \mathrm{II.1}\right) }$ as a sum between two
basic parts%
\begin{equation}
\hat{S}^{\left( \mathrm{II.1}\right) }=\left( \sum_{A^{\prime }=m+1}^{n}S^{(%
\mathrm{EH-\Lambda })A^{\prime }}\right) +\hat{S}^{(\mathrm{special})}
\label{SfincollcaseII.1}
\end{equation}%
that are independent one of the other. The first part decomposes into $%
\left( n-m\right) $ components that are all series in the constant coupling $%
k$
\begin{equation*}
S^{(\mathrm{EH-\Lambda })A^{\prime }}=\bar{S}^{\prime A^{\prime }}+kS_{1}^{(%
\mathrm{EH-\Lambda })A^{\prime }}+k^{2}S_{2}^{(\mathrm{EH-\Lambda }%
)A^{\prime }}+k^{3}S_{2}^{(\mathrm{EH-\Lambda })A^{\prime }}+\cdots ,
\end{equation*}%
with%
\begin{equation}
\bar{S}^{\prime A^{\prime }}\equiv \int d^{D}x\left[ \mathcal{L}_{0}^{\left(
\mathrm{PF}\right) }\left( h_{\mu \nu }^{A^{\prime }},\partial _{\lambda
}h_{\mu \nu }^{A^{\prime }}\right) +h^{\ast A^{\prime }\mu \nu }\partial
_{(\mu }\eta _{\nu )}^{A^{\prime }}\right]  \label{notcollPF}
\end{equation}%
and $\mathcal{L}_{0}^{\left( \mathrm{PF}\right) }\left( h_{\mu \nu
}^{A^{\prime }},\partial _{\lambda }h_{\mu \nu }^{A^{\prime }}\right) $ the
Pauli-Fierz Lagrangian for the graviton $A^{\prime }$. Each $S^{(\mathrm{%
EH-\Lambda })A^{\prime }}$ represents a copy of the solution to the master
equation for the full Einstein-Hilbert theory with a cosmological constant
associated with the graviton field $h_{\mu \nu }^{A^{\prime }}$ ($A^{\prime
}=\overline{m+1,n}$), so \emph{they cannot produce couplings among different
gravitons}. We emphasize that \emph{none of the }$\left( n-m\right) $\emph{\
gravitons gets coupled to the vector field }$V_{\mu }$. Let us analyze in
more detail the second part. It stops at order two in the coupling constant
\begin{eqnarray}
&&\hat{S}^{(\mathrm{special})}=\sum_{\bar{A}=1}^{m}\left\{ \int d^{D}x\left[
\mathcal{L}_{0}^{\left( \mathrm{PF}\right) }\left( h_{\mu \nu }^{\bar{A}%
},\partial _{\lambda }h_{\mu \nu }^{\bar{A}}\right) -2k\Lambda _{\bar{A}}h^{%
\bar{A}}+h^{\ast A\mu \nu }\partial _{(\mu }\eta _{\nu )}^{A}\right] \right\}
\notag \\
&&+\int d^{D}x\left\{ -\frac{1}{4}F_{\mu \nu }F^{\mu \nu }+V^{\ast \mu
}\partial _{\mu }\eta +k\sum_{\bar{A}=1}^{m}\left[ y_{3}^{\bar{A}}\delta
_{3}^{D}\varepsilon ^{\mu \nu \rho }\left( V_{\mu }^{\ast }\partial
_{\lbrack \nu }\eta _{\rho ]}^{\bar{A}}\right. \right. \right.  \notag \\
&&\left. \left. \left. +F_{\lambda \mu }\partial _{\lbrack \nu }h_{\rho ]}^{%
\bar{A}\ \lambda }\right) \right] +k^{2}\sum_{\bar{A},\bar{B}=1}^{m}\left[
y_{3}^{\bar{A}}y_{3}^{\bar{B}}\delta _{3}^{D}\left( \partial _{\lbrack \nu
}h_{\rho ]\lambda }^{\bar{A}}\right) \partial _{\lbrack \nu ^{\prime
}}h_{\rho ^{\prime }]}^{\bar{B}\ \lambda }\sigma ^{\nu \nu ^{\prime }}\sigma
^{\rho \rho ^{\prime }}\right] \right\}  \label{Sspecial}
\end{eqnarray}%
and in $D=3$ spacetime dimensions seems to mix different spin-two
fields via
the terms from the last (double) sum in the right-hand side of (\ref%
{Sspecial}) with $\bar{A}\neq \bar{B}$.

In order to focus in more detail on (\ref{Sspecial}) we take the limit
situation $m=n$ (so $\bar{A}\rightarrow A$) in the conditions (\ref%
{condfincollcaseII.1a})--(\ref{condfincollcaseII.1b}) and work in $D=3$,
such that the entire deformed solution to the master equation, $\hat{S}%
^{\left( \mathrm{II.1}\right) }$, consistent to all orders in the coupling
constant, reduces to (\ref{Sspecial}). We can express $\hat{S}^{(\mathrm{%
special})}$ in a nicer form by acting in a manner similar to that followed
in Section \ref{speccoupl}. Based on the observation that the deformed
solution to the master equation is unique up to addition of $s$-exact terms,
in the sequel we work with%
\begin{eqnarray}
&&\left. \hat{S}^{(\mathrm{special})}\right\vert _{\substack{ m=n  \\ D=3}}%
-s\left\{ 2k^{2}\sum_{A,B=1}^{n}\left[ \int d^{3}x\,y_{3}^{A}y_{3}^{B}\left(
h^{\ast A\mu \nu }h_{\mu \nu }^{B}+\eta ^{\ast A\mu }\eta _{\mu }^{B}\right) %
\right] \right\}  \notag \\
&=&\int d^{3}x\left\{ -\frac{1}{4}F_{\mu \nu }F^{\mu \nu }+V_{\mu }^{\ast
}\partial ^{\mu }\eta +\sum_{A=1}^{n}\left[ \mathcal{L}_{0}^{\left( \mathrm{%
PF}\right) }\left( h_{\mu \nu }^{A},\partial _{\lambda }h_{\mu \nu
}^{A}\right) \right. \right.  \notag \\
&&\left. -2k\Lambda _{A}h^{A}+h^{\ast A\mu \nu }\partial _{(\mu }\eta _{\nu
)}^{A}+ky_{3}^{A}\varepsilon ^{\mu \nu \rho }\left( V_{\mu }^{\ast }\partial
_{\lbrack \nu }\eta _{\rho ]}^{A}-F_{\mu \nu }\partial _{\lbrack \theta
}h_{\rho ]}^{A\ \theta }\right) \right]  \notag \\
&&\left. +2k^{2}\sum_{A,B=1}^{n}\left[ y_{3}^{A}y_{3}^{B}\left( \partial
_{\lbrack \mu }h_{\rho ]}^{A\ \mu }\right) \partial _{\lbrack \nu
}h_{\lambda ]}^{B\ \nu }\sigma ^{\rho \lambda }\right] \right\} .
\label{SspecialD=3}
\end{eqnarray}%
The part of antighost number zero gives the Lagrangian action of the coupled
model%
\begin{eqnarray}
\hat{S}^{\mathrm{L(II.1)}}[h_{\mu \nu }^{A},V^{\mu }] &=&\int d^{3}x\left\{ -%
\frac{1}{4}F_{\mu \nu }F^{\mu \nu }+\sum_{A=1}^{n}\left[ \mathcal{L}%
_{0}^{\left( \mathrm{PF}\right) }\left( h_{\mu \nu }^{A},\partial _{\lambda
}h_{\mu \nu }^{A}\right) \right. \right.  \notag \\
&&\left. -2k\Lambda _{A}h^{A}-ky_{3}^{A}\varepsilon ^{\mu \nu \rho }F_{\mu
\nu }\partial _{\lbrack \theta }h_{\rho ]}^{A\ \theta }\right]  \notag \\
&&\left. +2k^{2}\sum_{A,B=1}^{n}\left[ y_{3}^{A}y_{3}^{B}\left( \partial
_{\lbrack \mu }h_{\rho ]}^{A\ \mu }\right) \partial _{\lbrack \nu
}h_{\lambda ]}^{B\ \nu }\sigma ^{\rho \lambda }\right] \right\}
\label{Lagspecial}
\end{eqnarray}%
and the terms of antighost one provide its gauge symmetries%
\begin{equation}
\delta _{\epsilon }^{\mathrm{(II.1)}}h_{\mu \nu }^{A}=\partial _{(\mu
}\epsilon _{\nu )}^{A},\qquad \delta _{\epsilon }^{\mathrm{(II.1)}}V^{\mu
}=\partial ^{\mu }\epsilon +k\sum_{A=1}^{n}\left( y_{3}^{A}\varepsilon ^{\mu
\nu \rho }\partial _{\lbrack \nu }\epsilon _{\rho ]}^{A}\right) .
\label{gaugespecial}
\end{equation}%
This Lagrangian action can be brought to a simpler form by redefining the
field strength of the vector field as%
\begin{equation}
\hat{F}^{\mu \nu }=F^{\mu \nu }+2k\sum_{A=1}^{n}\left( y_{3}^{A}\varepsilon
^{\mu \nu \rho }\partial _{\lbrack \theta }h_{\rho ]}^{A\ \theta }\right) ,
\label{fieldnewspec}
\end{equation}%
in terms of which%
\begin{equation}
\hat{S}^{\mathrm{L(II.1)}}[h_{\mu \nu }^{A},V^{\mu }]=\int d^{3}x\left[
\sum_{A=1}^{n}\left( \mathcal{L}_{0}^{\left( \mathrm{PF}\right) }\left(
h_{\mu \nu }^{A},\partial _{\lambda }h_{\mu \nu }^{A}\right) -2k\Lambda
_{A}h^{A}\right) -\frac{1}{4}\hat{F}_{\mu \nu }\hat{F}^{\mu \nu }\right] .
\label{actSII.1coll}
\end{equation}%
The absence of terms of antighost number strictly greater than one
indicates that the deformed gauge symmetries (\ref{gaugespecial})
are independent and Abelian (their commutators close everywhere in
the space of field histories). We remark that this case corresponds
to the situation from Section \ref{speccoupl} (in the absence of
internal Pauli-Fierz indices), where we obtained a result
complementary to the usual couplings prescribed by General
Relativity. The gauge symmetries of the vector field are modified by
terms proportional with the antisymmetric first-order derivatives of
the Pauli-Fierz gauge parameters, while the gravitons keep their
original
gauge symmetries. The invariance of $\hat{S}^{\mathrm{L(II.1)}}$ under (\ref%
{gaugespecial}) is ensured by the gauge invariance of the deformed field
strength, $\delta _{\epsilon }^{\mathrm{(II.1)}}\hat{F}_{\mu \nu }=0$.

Unfortunately, action (\ref{actSII.1coll}) does not describe in fact
cross-couplings between different spin-two fields. In order to make this
observation clear, let us denote by $Y$ the matrix of elements $%
y_{3}^{A}y_{3}^{B}$. It is simple to see that the rank of $Y$ is equal to
one. By an orthogonal transformation $M$ we can always find a matrix $\hat{Y}
$ of the form%
\begin{equation}
\hat{Y}=M^{T}YM,  \label{w1}
\end{equation}%
with $M^{T}$ the transposed of $M$, such that
\begin{equation}
\hat{Y}^{11}=\sum_{A=1}^{n}\left( y_{3}^{A}\right) ^{2}\equiv \lambda
^{2},\qquad \hat{Y}^{1A^{\prime }}=\hat{Y}^{B^{\prime }1}=\hat{Y}^{A^{\prime
}B^{\prime }}=0,\qquad A^{\prime },B^{\prime }=\overline{2,n}.  \label{w2}
\end{equation}%
If we make the notation%
\begin{equation}
\hat{y}^{A}=M^{AC}y_{3}^{C},  \label{w3}
\end{equation}%
then relation (\ref{w2}) implies%
\begin{equation}
\hat{y}^{A}=\lambda \delta _{1}^{A}.  \label{w4}
\end{equation}%
Now, we make the field redefinition%
\begin{equation}
h_{\mu \nu }^{A}=M^{AC}\hat{h}_{\mu \nu }^{C},  \label{w5}
\end{equation}%
with $M^{AC}$ the elements of $M$. This transformation of the spin-two
fields leaves $\sum_{A=1}^{n}\mathcal{L}_{0}^{\left( \mathrm{PF}\right)
}\left( h_{\mu \nu }^{A},\partial _{\lambda }h_{\mu \nu }^{A}\right) $
invariant and, moreover, based on the above results, we obtain%
\begin{equation}
\sum_{A,B=1}^{n}\left[ y_{3}^{A}y_{3}^{B}\left( \partial _{\lbrack \mu
}h_{\rho ]}^{A\ \mu }\right) \partial _{\lbrack \nu }h_{\lambda ]}^{B\ \nu
}\sigma ^{\rho \lambda }\right] =\lambda ^{2}\left( \partial _{\lbrack \mu }%
\hat{h}_{\rho ]}^{1\ \mu }\right) \partial _{\lbrack \nu }\hat{h}_{\lambda
]}^{1\ \nu }\sigma ^{\rho \lambda },  \label{w6}
\end{equation}%
\begin{equation}
\sum_{A=1}^{n}\left( y_{3}^{A}\varepsilon ^{\mu \nu \rho }F_{\mu \nu
}\partial _{\lbrack \theta }h_{\rho ]}^{A\ \theta }\right) =\lambda
\varepsilon ^{\mu \nu \rho }F_{\mu \nu }\partial _{\lbrack \theta }\hat{h}%
_{\rho ]}^{1\ \theta },  \label{w7}
\end{equation}%
such that (\ref{actSII.1coll}) becomes%
\begin{equation}
\hat{S}^{\mathrm{L(II.1)}}[\hat{h}_{\mu \nu }^{A},V^{\mu }]=\int d^{3}x\left[
\sum_{A=1}^{n}\left( \mathcal{L}_{0}^{\left( \mathrm{PF}\right) }\left( \hat{%
h}_{\mu \nu }^{A},\partial _{\lambda }\hat{h}_{\mu \nu }^{A}\right) -2k\hat{%
\Lambda}_{A}\hat{h}^{A}\right) -\frac{1}{4}\hat{F}_{\mu \nu }^{\prime }\hat{F%
}^{\prime \mu \nu }\right] ,  \label{w8}
\end{equation}%
where
\begin{equation}
\hat{\Lambda}_{A}=\Lambda _{B}M^{BA},\qquad \hat{F}^{\prime \mu \nu }=F^{\mu
\nu }+2k\lambda \varepsilon ^{\mu \nu \rho }\partial _{\lbrack \theta }\hat{h%
}_{\rho ]}^{1\ \theta }.  \label{w9}
\end{equation}%
Action (\ref{w8}) is invariant under the gauge transformations%
\begin{equation}
\delta _{\hat{\epsilon}}^{\mathrm{(II.1)}}\hat{h}_{\mu \nu }^{A}=\partial
_{(\mu }\hat{\epsilon}_{\nu )}^{A},\qquad \delta _{\epsilon }^{\mathrm{(II.1)%
}}V^{\mu }=\partial ^{\mu }\epsilon +k\lambda \varepsilon ^{\mu \nu \rho
}\partial _{\lbrack \nu }\hat{\epsilon}_{\rho ]}^{1},  \label{w10}
\end{equation}%
where%
\begin{equation}
\hat{\epsilon}_{\mu }^{A}=\epsilon _{\mu }^{B}M^{BA}.  \label{w11}
\end{equation}%
We observe that action (\ref{w8}) decouples into action (\ref{LagactcaseII})
(derived in Section \ref{speccoupl}) for the first spin-two field ($A=1$)
and a sum of Pauli-Fierz actions with cosmological terms for the remaining $%
\left( n-1\right) $ spin-two fields. In conclusion, \emph{we cannot couple
different spin-two fields even outside the framework of General Relativity}.

\paragraph{Subcase II.2}

Now, we start from conditions (\ref{caseIIcollfull}) and (\ref{caseII.2coll}%
), such that the deformed solution to the master equation is maximally
parameterized in this situation by $\left( f_{A^{\prime }}\right)
_{A^{\prime }=\overline{m+1,n}}$, $\left( y_{3\bar{A}}\delta _{3}^{D}\right)
_{\bar{A}=\overline{1,m}}$, $\left( \Lambda _{A}\right) _{A=\overline{1,n}}$%
, $q_{1}\delta _{3}^{D}$, and $q_{2}\delta _{5}^{D}$. Without entering
unnecessary details, we only mention that this case is similar to subcase
II.1.2 in the absence of Pauli-Fierz internal indices, briefly discussed in
the final part of Section \ref{caseII}. The consistency of the deformed
solution to the master equation goes on unobstructed up to order five in the
coupling constant, where the existence of a local $\hat{S}_{5}^{\left(
\mathrm{II.1.2}\right) }$ requires a condition of the type $%
y_{3}^{3}q_{1}^{2}=0$, namely
\begin{equation}
q_{1}^{2}\left( \sum_{\bar{A}=1}^{m}\left( y_{3\bar{A}}\right) ^{2}\right)
y_{3\bar{B}}\delta _{3}^{D}=0,\qquad \bar{B}=\overline{1,m}.
\label{casecollII.2}
\end{equation}%
There are two main possibilities, none of them leading to cross-couplings
between different spin-two fields. Thus, if we take $D\neq 3$, then no
couplings among different gravitons are allowed since the Lagrangian of the
interacting model is a sum of independent Einstein-Hilbert Lagrangians with
cosmological terms for the last $\left( n-m\right) $ gravitons (none of them
coupled to the vector field), a sum of Pauli-Fierz Lagrangians plus simple
cosmological terms $-2k\Lambda _{\bar{A}}h^{\bar{A}}$ for the first $m$
gravitons and the Maxwell Lagrangian supplemented by the generalized Abelian
Chern-Simons density $kq_{2}\delta _{5}^{D}\varepsilon ^{\mu \nu \lambda
\alpha \beta }V_{\mu }F_{\nu \lambda }F_{\alpha \beta }$. If $D=3$, then
either $q_{1}=0$, in which situation we re-obtain the case from the previous
section, described by formula (\ref{SfincollcaseII.1}), where we have shown
that there are no cross-couplings between different gravitons, or $\left(
y_{3\bar{A}}\right) _{\bar{A}=\overline{1,m}}=0$, such that again no
cross-couplings are permitted and the resulting Lagrangian is like in the
above for $D\neq 3$ (after formula (\ref{casecollII.2})) up to replacing the
density $kq_{2}\delta _{5}^{D}\varepsilon ^{\mu \nu \lambda \alpha \beta
}V_{\mu }F_{\nu \lambda }F_{\alpha \beta }$ with the standard Abelian
Chern-Simons term $kq_{1}\varepsilon ^{\mu \nu \lambda }V_{\mu }F_{\nu
\lambda }$.

\section{Generalization to an arbitrary $p$-form\label{comm}}

The results obtained so far in the presence of a massless vector
field can be generalized to the case of deformations for one or
several gravitons and an arbitrary $p$-form gauge field.

In the case of a single graviton the starting point is the sum between the
Pauli-Fierz action and the Lagrangian action of an Abelian $p$-form with $%
p>1 $%
\begin{equation}
S_{0}^{\mathrm{L}}[h_{\mu \nu },V_{\mu _{1}\ldots \mu _{p}}]=\int
d^{D}x\left( \mathcal{L}_{0}^{\left( \mathrm{PF}\right) }-\frac{1}{2\cdot
\left( p+1\right) !}F_{\mu _{1}\ldots \mu _{p+1}}F^{\mu _{1}\ldots \mu
_{p+1}}\right) ,  \label{aa1}
\end{equation}%
in $D\geq p+1$ spacetime dimensions, with $F_{\mu _{1}\ldots \mu _{p+1}}$
the Abelian field strength of the $p$-form gauge field $V_{\mu _{1}\ldots
\mu _{p}}$
\begin{equation}
F_{\mu _{1}\ldots \mu _{p+1}}=\partial _{\lbrack \mu _{1}}V_{\mu _{2}\ldots
\mu _{p+1}]}.  \label{abfstr}
\end{equation}%
This action is known to be invariant under the gauge transformations%
\begin{equation}
\delta _{\epsilon }h_{\mu \nu }=\partial _{(\mu }\epsilon _{\nu )},\qquad
\delta _{\epsilon }V_{\mu _{1}\ldots \mu _{p}}=\partial _{\lbrack \mu
_{1}}\epsilon _{\mu _{2}\ldots \mu _{p]}}^{(p)}.  \label{aa2}
\end{equation}%
Unlike the Maxwell field ($p=1$), the gauge transformations of the
$p$-form for $p>1$ are off-shell reducible of order $\left(
p-1\right) $. This property has strong implications at the level of
the BRST complex and of the BRST cohomology in the form sector: a
whole tower of ghosts of ghosts and of antifields will be required
in order to incorporate the reducibility, only the ghost of maximum
pure ghost number, $p$, will enter $H\left( \gamma \right) $, and
the local characteristic cohomology will be richer in the
sense that (\ref{a33}) and (\ref{a35}) become~\cite{knaep1}%
\begin{equation}
H_{J}\left( \delta |d\right) =0=H_{J}^{\mathrm{inv}}\left( \delta |d\right)
,\qquad J>p+1.  \label{aa3}
\end{equation}%
In spite of these new cohomological ingredients, which complicate the
analysis of deformations, the results from Sections \ref{analysiscaseI} and %
\ref{speccoupl} can still be generalized.

Thus, two complementary cases are revealed. One describes the
standard graviton-$p$-form interactions from General Relativity and
leads to a Lagrangian action similar to (\ref{LagactcaseI}) up to
replacing $\left( 1/4\right) g^{\mu \nu }g^{\rho \lambda
}\bar{F}_{\mu \rho }\bar{F}_{\nu \lambda }$ with the expression
$\left( 2\cdot \left( p+1\right) !\right) ^{-1}g^{\mu _{1}\nu
_{1}}\cdots g^{\mu _{p+1}\nu _{p+1}}\bar{F}_{\mu _{1}\ldots \mu
_{p+1}}\bar{F}_{\nu _{1}\ldots \nu _{p+1}}$ and, if $p$ is odd, also
the terms containing $\delta _{3}^{D}\varepsilon ^{\mu _{1}\mu
_{2}\mu _{3}}$ and $\delta _{5}^{D}\varepsilon ^{\mu _{1}\mu _{2}\mu
_{3}\mu _{4}\mu _{5}}$ with some densities involving $\delta
_{2p+1}^{D}\varepsilon ^{\mu _{1}\ldots \mu _{2p+1}}$ and $\delta
_{3p+2}^{D}\varepsilon ^{\mu _{1}\ldots \mu _{3p+2}}$ respectively
(if $p$ is even, the terms proportional with either $q_{1}$ or
$q_{2}$ must be suppressed). The other case emphasizes that \emph{it
is possible to construct some new deformations in }$D=p+2$\emph{,
describing a spin two-field coupled to a }$p$\emph{-form and having (\ref%
{aa1}) and (\ref{aa2}) as a free limit, which are consistent to all orders
in the coupling constant and are not subject to the rules of General
Relativity}. Their source is a generalization of the terms proportional with
$y_{3}$ from the first-order deformation (\ref{a60int})%
\begin{equation}
S_{1}^{\left( \mathrm{int}\right) }\left( y_{3}\right) =y_{3}\varepsilon
_{\mu _{1}\ldots \mu _{p}\nu \rho }\int d^{p+2}x\left( V^{\ast \mu
_{1}\ldots \mu _{p}}\partial ^{\lbrack \nu }\eta ^{\rho ]}+\frac{1}{p!}%
F^{\lambda \mu _{1}\ldots \mu _{p}}\partial ^{\lbrack \nu }h_{\ \ \lambda
}^{\rho ]}\right) .  \label{aa4}
\end{equation}%
Performing the necessary computations, we find the Lagrangian action%
\begin{eqnarray}
S^{\mathrm{L}}[h_{\mu \nu },V_{\mu _{1}\ldots \mu _{p}}] &=&\int
d^{p+2}x\left( \mathcal{L}_{0}^{\left( \mathrm{PF}\right) }-2k\Lambda
h\right.  \notag \\
&&\left. -\frac{1}{2\cdot \left( p+1\right) !}F_{\mu _{1}\ldots \mu
_{p+1}}^{\prime }F^{\prime \mu _{1}\ldots \mu _{p+1}}\right) ,  \label{aa5}
\end{eqnarray}%
where the field strength of the $p$-form is deformed as%
\begin{equation}
F_{\mu _{1}\ldots \mu _{p+1}}^{\prime }=F_{\mu _{1}\ldots \mu
_{p+1}}+2\left( -\right) ^{p+1}ky_{3}\varepsilon _{\mu _{1}\ldots \mu
_{p+1}\rho }\partial ^{\lbrack \theta }h_{\ \ \theta }^{\rho ]}.  \label{aa6}
\end{equation}%
This action is fully invariant under the original Pauli-Fierz gauge
transformations and%
\begin{equation}
\bar{\delta}_{\epsilon }V_{\mu _{1}\ldots \mu _{p}}=\partial _{\lbrack \mu
_{1}}\epsilon _{\mu _{2}\ldots \mu _{p]}}^{(p)}+ky_{3}\varepsilon _{\mu
_{1}\ldots \mu _{p}\nu \rho }\partial ^{\lbrack \nu }\epsilon ^{\rho ]}.
\label{aa7}
\end{equation}%
The gauge algebra remains Abelian and the reducibility of (\ref{aa7}) is not
affected by these couplings: the associated functions and relations are the
initial ones.

It is important to notice that all the standard hypotheses imposed
to consistent deformations are fulfilled. Indeed, in the free limit
($k=0$) the field strength (\ref{aa6}) is restored to its original
form (\ref{abfstr}), the cosmological term $-2k\Lambda h$ is
destroyed, and the Pauli-Fierz gauge
parameters $\epsilon ^{\rho }$ are discarded from the gauge transformations $%
\bar{\delta}_{\epsilon }V_{\mu _{1}\ldots \mu _{p}}$, leaving us with the
original action (\ref{aa1}) and initial gauge transformations (\ref{aa2}).
Also, the spacetime locality, Lorentz covariance, and Poincar\'{e}
invariance of action (\ref{aa5}) are obvious. Likewise, the smoothness of
the deformed theory in the coupling constant is ensured by the polynomial
behaviour of (\ref{aa5}) and (\ref{aa7}) with respect to $k$: the action is
a polynomial of order two and the gauge transformations are polynomials of
order one. Furthermore, the differential order of the coupled field
equations is preserved with respect to that of the free equations
(derivative order assumption), being equal to two, as it can be observed
from the concrete form of the Euler-Lagrange derivatives of action (\ref{aa5}%
)
\begin{eqnarray}
\frac{\delta S^{\mathrm{L}}[h_{\mu \nu },V_{\mu _{1}\ldots \mu _{p}}]}{%
\delta h_{\mu \nu }} &=&\frac{\delta S_{0}^{\mathrm{L}}[h_{\mu \nu },V_{\mu
_{1}\ldots \mu _{p}}]}{\delta h_{\mu \nu }}-2k\Lambda \sigma ^{\mu \nu }
\notag \\
&&-\frac{ky_{3}}{\left( p+1\right) !}\left[ \left( -\right) ^{p}\varepsilon
^{\mu _{1}\ldots \mu _{p+1}(\mu }\partial ^{\nu )}F_{\mu _{1}\ldots \mu
_{p+1}}^{\prime }\right.  \notag \\
&&\left. +2\sigma ^{\mu \nu }\varepsilon ^{\mu _{1}\ldots \mu
_{p+2}}\partial _{\mu _{1}}F_{\mu _{2}\ldots \mu _{p+2}}^{\prime }\right]
\notag \\
&\equiv &\square h^{\mu \nu }+\left( 1+4k^{2}y_{3}^{2}\right) \partial ^{\mu
}\partial ^{\nu }h-\left( 1+2k^{2}y_{3}^{2}\right) \partial ^{(\mu }\partial
_{\theta }h^{\nu )\theta }  \notag \\
&&+\left( 1+4k^{2}y_{3}^{2}\right) \sigma ^{\mu \nu }\left( \partial _{\rho
}\partial _{\theta }h^{\rho \theta }-\square h\right) -2k\Lambda \sigma
^{\mu \nu }  \notag \\
&&+\left( -\right) ^{p+1}\frac{ky_{3}}{\left( p+1\right) !}\varepsilon ^{\mu
_{1}\ldots \mu _{p+1}(\mu }\partial ^{\nu )}F_{\mu _{1}\ldots \mu _{p+1}},
\label{eqshp}
\end{eqnarray}%
\begin{eqnarray}
\frac{\delta S^{\mathrm{L}}[h_{\mu \nu },V_{\mu _{1}\ldots \mu _{p}}]}{%
\delta V_{\mu _{1}\ldots \mu _{p}}} &=&\frac{1}{p!}\partial _{\nu }F^{\prime
\nu \mu _{1}\ldots \mu _{p}}  \notag \\
&=&\frac{\delta S_{0}^{\mathrm{L}}[h_{\mu \nu },V_{\mu _{1}\ldots \mu _{p}}]%
}{\delta V_{\mu _{1}\ldots \mu _{p}}}-\frac{ky_{3}}{p!}\varepsilon ^{\mu
_{1}\ldots \mu _{p}\nu \rho }\partial _{\lbrack \nu }\partial ^{\theta
}h_{\rho ]\theta }  \notag \\
&\equiv &\frac{1}{p!}\left( \partial _{\nu }F^{\nu \mu _{1}\ldots \mu
_{p}}-ky_{3}\varepsilon ^{\mu _{1}\ldots \mu _{p}\nu \rho }\partial
_{\lbrack \nu }\partial ^{\theta }h_{\rho ]\theta }\right) .  \label{eqsp}
\end{eqnarray}%
It is truly remarkable that these new couplings comply with the derivative
order assumption.

Let us analyze the main physical consequences of these new couplings. First,
we investigate some direct outcomes of the field equations, obtained by
equating (\ref{eqshp}) and (\ref{eqsp}) to zero. By taking the trace of the
field equations for the graviton, $\sigma _{\mu \nu }\delta S^{\mathrm{L}%
}/\delta h_{\mu \nu }=0$, we infer the equivalent equation%
\begin{equation}
K=\frac{2k\Lambda \left( p+2\right) }{p+\left( p+1\right) 4k^{2}y_{3}^{2}},
\label{modcurv}
\end{equation}%
where $K$ is the linearized scalar curvature, $K=\partial _{\rho }\partial
_{\theta }h^{\rho \theta }-\square h$. Due to the presence of the
cosmological constant, the linearized scalar curvature is a non-vanishing
constant. The field equation of the $p$-form, $\delta S^{\mathrm{L}}/\delta
V_{\mu _{1}\ldots \mu _{p}}=0$, is nothing but a nontrivial conservation law
of order $\left( p+1\right) $, $\partial _{\nu }F^{\prime \nu \mu _{1}\ldots
\mu _{p}}=0$, where the associated current is precisely the deformed field
strength (\ref{aa6}). It is not a usual conservation law because it results
from some rigid symmetries of the solution to the master equation for the
coupled theory. The main difference between the free theory (\ref{aa1}) and
the coupled one is that the $\left( p+1\right) $-order conservation law of
the latter contains a nontrivial component from the Pauli-Fierz sector, $%
2\left( -\right) ^{p+1}ky_{3}\varepsilon ^{\nu \mu _{1}\ldots \mu _{p}\rho
}\partial _{\lbrack \theta }h_{\rho ]}^{\ \ \theta }$. Another interesting
observation is that, unlike the free limit (\ref{aa1}), the field equations
of the coupled model admit to be written in a compact form. Indeed, it can
be shown that both field equations, $\delta S^{\mathrm{L}}/\delta h_{\mu \nu
}=0$ and $\delta S^{\mathrm{L}}/\delta V_{\mu _{1}\ldots \mu _{p}}=0$, are
\emph{completely equivalent} with the following expression of the
first-order derivatives of the Abelian field strength (\ref{abfstr})%
\begin{eqnarray}
\partial ^{\nu }F_{\mu _{1}\ldots \mu _{p+1}} &=&\frac{\left( -\right) ^{p}}{%
2}\varepsilon _{\mu _{1}\ldots \mu _{p+1}\rho }\left\{ 2ky_{3}\partial
^{\lbrack \nu }\partial _{\theta }h^{\rho ]\theta }+\frac{1}{ky_{3}}\left[
2k\Lambda \sigma ^{\nu \rho }-\square h^{\nu \rho }\right. \right.   \notag
\\
&&-\left( 1+4k^{2}y_{3}^{2}\right) \partial ^{\nu }\partial ^{\rho }h+\left(
1+2k^{2}y_{3}^{2}\right) \partial ^{(\nu }\partial _{\theta }h^{\rho )\theta
}  \notag \\
&&\left. \left. -\left( 1+4k^{2}y_{3}^{2}\right) \sigma ^{\nu \rho }\left(
\partial _{\lambda }\partial _{\theta }h^{\lambda \theta }-\square h\right)
\right] \right\} .  \label{equivfeqs}
\end{eqnarray}%
The direct as well as the converse implication results from simple algebraic
manipulations of the coupled field equations or respectively of (\ref%
{equivfeqs}) and also by means of the identity%
\begin{equation}
\varepsilon ^{\mu \nu \mu _{1}\ldots \mu _{p}}\partial ^{\rho }F_{\rho \mu
_{1}\ldots \mu _{p}}=\frac{\left( -\right) ^{p+1}}{p+1}\varepsilon ^{\mu
_{1}\ldots \mu _{p+1}[\mu }\partial ^{\nu ]}F_{\mu _{1}\ldots \mu _{p+1}},
\label{idD=p+2}
\end{equation}%
valid in $D=p+2$.

Regarding a collection of spin-two fields and a $p$-form, it can be used a
line similar to that employed in Section \ref{manyspintwoem}. Thus, it can
be shown that two complementary cases are again unfolded. One is similar to
the situation discussed in Section \ref{nogocollGR} and the other with the
result from Section \ref{exoticcoll}. In both cases there are no
cross-couplings among different spin-two fields intermediated by a $p$-form
gauge field: the $p$-form couples to a single spin-two field.

\section{Conclusion\label{conc}}

To conclude with, in this paper we have investigated the couplings
between a single spin-two field or a collection of such fields
(described in the free limit by a sum of Pauli-Fierz actions) and a
massless $p$-form using the powerful setting based on local BRST
cohomology. Under the hypotheses of locality, smoothness in the
coupling constant, Poincar\'{e} invariance, Lorentz covariance, and
preservation of the number of derivatives on each field (plus
positivity of the metric in the internal space in the case of a
collection of spin-two fields), we found two complementary
situations. One submits to the well-known prescriptions of General
Relativity, but the other situation discloses some new type of
couplings in $\left( p+2\right) $ spacetime dimensions, which only
modify the gauge symmetries of the $p$-form. It is remarkable that
these $\left( p+2\right) $-dimensional cross-couplings comply with
the derivative order assumption, unlike other situations from the
literature. Unfortunately, in the case of a collection of spin-two
fields none of these coupled theories allows for (indirect)
cross-couplings between different gravitons.

\section*{Acknowledgment}

The authors are partially supported by the European Commission FP6 program
MRTN-CT-2004-005104 and by the type A grant 305/2004 with the Romanian
National Council for Academic Scientific Research (C.N.C.S.I.S.) and the
Romanian Ministry of Education and Research (M.E.C.). The authors thank the
referee for his/her valuable comments and suggestions.

\end{document}